\documentclass[reqno]{tran-l}

\usepackage{amsmath}
\usepackage{amssymb}
\usepackage{amsthm}

\allowdisplaybreaks[3]

 \newtheorem{thm}{Theorem}[section]
 \newtheorem{cor}[thm]{Corollary}
 \newtheorem{lem}[thm]{Lemma}
 \newtheorem{prop}[thm]{Proposition}
 \theoremstyle{definition}
 
 \theoremstyle{remark}
 \newtheorem{rem}[thm]{Remark}
 \numberwithin{equation}{section}
 \DeclareMathOperator{\RE}{Re}
 \DeclareMathOperator{\IM}{Im}

 \newcommand{\h}{\mathcal{H}}

 \newcommand{\W}{\mathcal{W}}

 \newcommand{\Real}{\mathbb{R}}
 \newcommand{\Complex}{\mathbb{C}}

 \newcommand{\abs}[1]{\left\vert#1\right\vert}
 \newcommand{\set}[1]{\left\{#1\right\}}
 
 \newcommand{\norm}[1]{\left\Vert#1\right\Vert}
 
\begin{document}

\title[Energy and Momentum Conservation for Stochastic Processes]
 {Energy and Momentum Conservation for Diffusion -
 A Stochastic Mechanics Approximation - Part I}

\author{ Johan G.B. Beumee}

\address{FP Financial, The Tile House,
Liphook, Hampshire GU30 7JE, United Kingdom}

\email{johan.beumee@btinternet.com}

\thanks{This work was completed with support from FP Financial.}

\thanks{Rights Reserved FP Financial - 28 September 2006}

\thanks{The author thanks Mark Davis and Chris Rogers for comments.}


\date{}

\keywords{stochastic processes, quantum mechanics, heatbath,
diffusion}


\commby{James F. & Fransina I. Beumee}


\begin{abstract}
This paper models the classical diffusion of a main particle
through a heatbath by means of a pre-limit microscopic
representation of its drifted momentum and energy transfers at
collision times. The collision point linear interpolated path
can be approximated by the solution to the "inscribed"
continuous stochastic differential equation using the same drift
function. Employing results from stochastic mechanics it is then
shown that the combined main particle/heatbath system does not
exchange or radiate energy if the probability distribution for
the position of the main particle is derived from
Schr\"{o}dinger's equation. Furthermore it is shown that the
main particle distance traveled between collisions and the mean
inter-collision time must satisfy a type of Minkowski invariant.
Hence if there is a correlation between the pre- and
post-collision velocities of the main particle through a
collision point then the mean distance traveled can be related
to the mean inter-particle collision times via a Lorentz
transformation. The last Section shows that this approach can be
applied to all elastic main particle/heatbath particle
collisions either via direct calculation involving modeling the
collision scattering or by altering the properties of the
heatbath.
\end{abstract}

\maketitle

\section*{Introduction}
\smallskip
Ever since Einstein's introduction of the molecular-kinetic
theory of heat in 1905 the Brownian motion/Markovian formalism
has been applied to a large variety of topics including
classical particle diffusion and stochastic mechanics.  The
first subject is more the domain of statistical mechanics and
focusses on thermodynamic properties,
Goldstein~\cite{GOLDSTEIN1}, Isothermal Flows
Garbaczewksi~\cite{GARBA1}, transport equations (e.g. Master
equation, Boltzmann's equation or Kramer's equation) and Markov
Chains, for instance Posilicano~\cite{POSILICANO1}, van
Kampen~\cite{KAMPEN1} or Gamba~\cite{GAMBA1}.  The second topic
falls under the interpretation of quantum mechanics see for
instance the recent review by Carlen~\cite{ECARL2} or
Nelson~\cite{ENELSON1}. For a historical view on the development
of Brownian Motion consult Nelson~\cite{ENELSON2}. The present
paper represents a (larger) particle (the "main" particle)
diffusing through a heatbath of smaller particles employing a
"finite-energy" Markovian difference equation and investigates
various energy conditions using results that have been developed
in stochastic mechanics.

The simplifying assumption for diffusion is that the frequent
energy exchanges between the main and heatbath particles induce
a continual acceleration and deceleration that make the motion
of the main particle look macroscopically as if it has no memory
of its previous whereabouts.  Mathematically the Brownian motion
process is the result of a limiting procedure involving an
infinite amount of collisions exchanging an infinite amount of
energy.  The original derivation of the associated diffusion
equation for the density distribution is due to
Einstein~\cite{EINST1}. As a result of the limiting process the
resulting Markovian process is not differentiable and therefore
no energy or momentum can be ascribed to the main particle.
Brownian motion is very attractive because of its "random step"
intuition and because of the tractability of the associated
diffusion equation.

Many ways of associating momentum and energy with the motion of
a diffusing particle have been suggested in the literature.  A
straightforward and transparent approach is to augment the
position process for the diffusing particle with a momentum
process and investigate the position process in the limit that
the momentum becomes extremely large. Modeling both position and
momentum Nelson~\cite{ENELSON2} showed that the position of the
main particle converges in probability to the solution of a
stochastic differential equation if the associated momentum
process contains an ever larger mean reversion (and variance).
The more typical - and mathematical - construction of Brownian
motion is based on successive probabilistic additions of
orthogonal functions, see Rogers \& Williams~\cite{ROGERS1} or
Karatzas \& Shreve~\cite{SHREVE1}. These two constructions do
not consider the underlying particle collision process or the
implied energy exchange between the main particle and heatbath
environment.

This paper uses a direct approach by creating a pre-limit
microscopic representation of a main particle by modeling its
momentum and energy transfers at collision times. The assumption
is that the collisions occur at stopping times $t_j$, $j \geq 0$
with $t_{j+1}-t_j=\tau_j$, $j \geq 0$, so that the main particle
will have corresponding positions
$x_j=x\left(t_j,\beta\right)\in \Real^n$, $j\geq 0$,
$\beta=2/\overline{\tau}$ at these collision times.  Here
$\beta$ denotes the main particle path dependence on the mean
particle inter-collision time $\overline{\tau}$. The particle
position at the collision point $x_{j+1}$ depends on the
position of the previous collision point $x_j$, a drift term
$b^+\left(x_j,t_j\right)$ and a Gaussian random shock.  Since
the (main) particle drift over the interval
$\left[t_j,t_{j+1}\right],j \geq 0,$ is a function of $x_j$,
$t_j,j \geq 0,$ the position process of the main particle $x_j,j
\geq 0,$ reduces to a discrete Markovian stochastic process. The
crucial assumption in this paper is that the inter-collision
times $\tau_j$,$j \geq 0$, can be represented by a second (or
higher) order gamma distribution. The main reason for this
assumption is that the energy and momentum of the main particle
between collisions are then well defined and finite with
probability one.

The set of collision points
$\set{x_j=x\left(t_j,\beta\right)\vert j=0,1,...}$ does not
prescribe the position of the main particle at an arbitrary time
$t\nsubseteq \set{t_j\vert j=0,1,...}$. A reasonable estimate
for the position of particle at time $t$ is to define
$x\left(t,\beta\right),t\geq 0$, as the linear interpolation on
the collision positions.  So $x\left(t,\beta\right)$, is
linearly interpolated from $x\left(t_j,\beta\right)$ and
$x\left(t_{j+1},\beta\right)$ where $t_{j}\leq t < t_{j+1}$
($t_0=0$, the origin). The expectation is that if the number of
collisions increases the process $x\left(t,\beta\right)\in
\Real^n$, approaches the continuous stochastic process
$x\left(t\right)\in \Real^n$, $t \geq 0$, which satisfies an
appropriate stochastic differential equation with drift
$b^+(x,t)$.  This obviously depends on the properties of the
drift function, the variance function in the stochastic shock
and the existence of collision points. Section 1 introduces a
set of conditions for which the discrete process $x(t,\beta)$
converges a.e. to the strong solution of a stochastic
differential equation.

By design, if the particle experiences a collision at time $t$
with $x(t,\beta)=x$ then the next collision will occur at time
$t+\tau_2$ while the last collision of the main particle
occurred at time $t-\tau_1$. Hence the main particle has a
(forward) momentum (and energy) towards the next collision equal
$v_2=\left(x(t+\tau_2)-x\right)/\tau_2$. The forward momentum is
a random variable depending only on the current position of the
main particle due to the fact that the stochastic process
$x(t_j),j\geq 0$, is Markovian.  Similarly, the main particle
will have a backward (or incoming) momentum (and energy) from
the last collision equal $v_1=\left(x-x(t-\tau_1)\right)/\tau_1$
since $t-\tau_1$ is the collision time previous to $t$.  The
backward momentum is a random variable that will have to be
conditioned on the fact that it is now known that $x(t)=x$.  At
least in principle, the distribution for $x(t-\tau_1)$ follows
from the distribution of the main particle for $x(t)$ employing
Bayes' theorem.  The forward and backward time step perspective
for continuous stochastic processes can be gleaned from the
extensive work on stochastic processes by
Nelson~\cite{ENELSON1}, Carlen~\cite{ECARL1}, ~\cite{ECARL3},
Guerra~\cite{GUERRA1}, ~\cite{GUERRA2} and see the references in
Carlen~\cite{ECARL2}.

Now that pre- and post-collision momenta and energies are determined
for the main particle the collision process and energy exchange can
be investigated for the main and heatbath particle. Section 2
investigates the consequences of an elastic collision and introduces
the canonical linear relationship between the pre- and post
collision velocities of the main and heatbath particle.  This linear
relationship is parameterized by a random anti-symmetric matrix $Z$
(manufactured from a random unitary matrix $U$) which incorporates
the random center of mass line and collision impact angle
information. This matrix will be referred to as the collision
scattering matrix and collision energy exchanges with $Z\equiv 0$
are referred to as 'simple' collisions.  Notice that in one
dimension all collisions are simple.

Section 2 shows that the combined kinetic energy $\h_k$ of the main
and heatbath particle for simple elastic collisions can be expressed
in the form of a simple quadratic combination of the forward and
backward momenta/velocities of the main particle. In one dimension
this relationship is exact while in more than one dimension the
total kinetic energy $\h_k$ of the colliding system refers to the
energy of the motion along the center of mass line of the collision.
The remaining kinetic energy of the main and heatbath particles is
embedded in motion perpendicular to the collision center of mass
line and remains invariant under the collision.  Section 2 only
investigates the canonical case where $\h_k=\h_k\left(Z\equiv
0\right)$ while Section 4 shows that the case
$\h_k=\h_k\left(Z\right),Z\neq 0$, can be reduced to the simple
collision case.

Using this result Section 2 then shows that if the total kinetic
energy of the system is not conserved then either the main
particle is radiating energy into the heatbath or the main
particle is absorbing energy from its heatbath surroundings.
Such an exchange is possible as the result of external forces in
the form of a potential $\Phi_p$ but in that case the combined
kinetic energy $\h_k$ and $\Phi_p$ together must be a conserved
quantity, i.e. its expected value over all paths and positions
must be a constant in time.  The main result of the paper is
that the only probability density that renders the total energy
$E\left[\h_k+\Phi_p\right]$ time invariant is the squared
modulus of the wave function obtained from Schr\"{o}dingers
equation. Appropriate conditions and examples for the potential
will presented in Sections 2 and 4. This is a purely classical
representation and Planck's constant is now replaced by a
constant depending on the variance of the underlying stochastic
process $\sigma^2$ and the main particle/heatbath particle mass
ratio $\gamma$. This Section also shows (using collision
elasticity) that the backward and forward momenta for the
heatbath particle are tightly correlated if the main particle
follows a Markovian path.

The main particle/heatbath particle elastic collision representation
also provides a similar quadratic expression showing that the
forward and backward velocities of the main particle are directly
related to the (forward and backward) velocities of the heatbath
particle.  This relationship will often be referred to as the
"momentum" constraint.  Section 3 will show that if the heatbath
particles are in energetic equilibrium with the main particle then
the drift of the main particle and the correlation between the
forward and backward velocities of the colliding heatbath particles
must depend on the average inter-collision time. The relationship
between the mean inter-particle collision time and the particle
energy can be expressed in the form of the geometric Minkovski
invariant.  A consequence is that the mean distance traveled for the
main particle and its mean inter-particle collision time can then be
related via a Lorentz transformation.

The last Section focusses on the non-simple elastic collisions and
qualifies the main and heatbath particle total energy dependence on
the anti-symmetric scattering matrix $Z=Z(U)$. The same conservation
of total energy employed in Section 2 now produces an equation which
also contains the dynamics of the $Z$ matrix.  The conservation
result is discussed and some examples presented but no full solution
for this case can be derived.  However, the final results in this
Section show that the total energy conservation and "momentum"
constraint can be reformulated in terms of a transformed heatbath
with heatbath particles that have a higher kinetic energy and a
different correlation structure. It is possible therefore to
transform the scattering matrix $Z$ away by subsuming it into the
heatbath. Conveniently then all the results of Section 2 and 3
become valid again for the transformed heatbath.

The work is organized as follows.  Section 1 introduces the
details of the main particle collision representation and shows
that the collision point linear interpolate converges a.e. to
the solution of a continuous stochastic differential equation.
Section 2 employs the fact that the particle collisions are
elastic to show that a non-radiation condition demands that the
only acceptable probability density for the position of the main
particle is derived from Schr\"{o}dinger's equation. Section 3
uses the "momentum" constraint to show that the mean
inter-particle collision time and inter-particle distance
traveled satisfy a geometric Minkowski invariant.  The last
Section shows that the results of Sections 2 and 3 can be
extended to almost all types of elastic collisions. All proofs
have been delegated to the Appendices to make the paper more
readable.

The notation employed in this this paper uses $E[.]$ or
$E[.\,|.]$ for expectation or conditional expectation
respectively.  Typically $E[.]$ indicates an expectation over
all variables between the brackets including the random
scattering matrix $Z$ present and the forward and backward
velocities. also $b^\pm (x,t):\Real^n\times
[0,\infty)\longmapsto \Real^n$ are referred to as the forward
and backward instantaneous drifts of the main particle and
$\sigma(x,t):\Real^n\times [0,\infty)\longmapsto \Real^n\times
\Real^n$ are the corresponding variance terms. Moreover $\tau$
refers to the inter-particle collision time,
$\overline{\tau}=E\left[\tau\right]$ and
$\beta=2/\overline{\tau}$. Often the reference to the time $t$
implicitly assumes that this time point is a collision time for
the main particle.  Clearly for finite mean collision times a
distinction must be made between the collision points
$t\in\set{t_j,j=0,1,...}$ and the remaining time points but the
paper does not always complete the analysis.  The norm
$\norm{.}$ indicates the usual distance norm in $\Real^n$.

\section{Diffusion and Energy}

This Section introduces the details of the
$x_j=x\left(t_j,\beta\right)$, $j=0,1,...$, collision
representation, the converged process $x(t),t\geq 0$, and the
distribution for the inter-particle collision times
$\tau_j,j\geq 0$.  It is shown that $\lim_{\beta \rightarrow
\infty} x(t,\beta)$ exists under certain smoothness conditions
and this Section also investigates the correlation between the
forward and backward main particle velocities.  Some results
from Nelson and Carlen will be quoted without proof.

Following the Introduction let the collision points for the main
particle be $x(t_j)\in\Real^n$, $j=0,1,...$, at collision time
$t\in\set{t_j,j=0,1,...}$, and assume that the next collision
will occur in $x(t+\tau_2)$ at stopping time $t+\tau_2$. The
collision previous to time $t$ occurred at $t-\tau_1$ when the
main particle was in position $x(t-\tau_1)$.  The collisions
must be assumed "real" in the sense that the main particle
actually interchanges energy with a colliding heatbath particle.
The inter-collision times are modeled as independent random
variables distributed with a second order Gamma distribution so
that $f_\beta (t)=\beta^2t e^{-\beta t},t\geqq 0$.  As a result
the time to the next collision $\tau_2$ and the time since the
previous collision $\tau_1$ has the following moments
\begin{gather}
\label{COLL1}
\begin{split}
&E[\tau_2]=E[\tau_1]=\int_{0}^{\infty}
\beta^2 t^2 e^{-\beta t} dt=\frac{2}{\beta}=\overline{\tau},\\
&var(\tau_2)=var(\tau_1)=\frac{2}{\beta^2}=\frac{1}{2}\overline{\tau}^2.
\end{split}
\end{gather}
Most importantly for this distribution it is also true that
\begin{align*}
&E\left[\frac{1}{\tau_2}\right]=E\left[\frac{1}{\tau_1}\right]
=\int_{0}^{\infty} \beta^2 e^{-\beta t}
dt=\beta=\frac{2}{\overline{\tau}},
\\
&E\left[\frac{1}{\sqrt{\tau_2}}\right]
=E\left[\frac{1}{\sqrt{\tau_1}}\right] =\int_{0}^{\infty}\beta^2
\sqrt{t}e^{-\beta t} dt=\sqrt{\frac{2\pi}{\overline{\tau}}}.
\end{align*}
Higher order Gamma distributions could have been employed for
the inter-collision times or in fact any distribution such that
$E\left[\tau^{-1}\right]<\infty$ would have been suitable but
there seems little fundamental difference in the analysis.
Interestingly, the ubiquitous exponential distribution is
excluded due to the last restriction but notice that the second
order Gamma distribution is the distribution of the sum of two
exponential random variables.

By assumption above if $t$ is a collision time with main
particle position $x(t,\beta)=x\in \Real^n$ then the previous
collision occurred at $t-\tau_1$.  In this case $\tau_1$ is a
random variable conditional on the fact that a collision
occurred in the future $t$ with the particle in position
$x(t,\beta)=x\in \Real^n$. Unfortunately, the distribution for
$\tau_1$ conditional on this event is no longer a second order
Gamma distribution, see Feller~\cite{FELLER1} for a more
complete discussion.  The future conditioning alters the
distribution of $\tau_1$ which is easy to see since any
collision time $t-\tau_1$ previous to $t$ must satisfy $0\leqq
t-\tau_1\leqq t$. Hence $0\leqq \tau_1\leqq t$ counter to the
domain definition of a Gamma distribution. Moreover, there is
always the possibility that no previous collision occurred so
the real density for $\tau_1$ is in fact defective. However, for
$t\gg\overline{\tau}$ it is reasonable to assume that $\tau_1$
is at least approximately a second order Gamma distribution with
$E\left[\tau_1\right]=E\left[\tau_2\right]$ and
$E\left[\tau^{-1}_1\right]=E\left[\tau^{-1}_2\right]$.

The forward and backward step for the position process
$x(t,\beta)$ at collision time $t$ are now defined as follows
\begin{subequations}
\begin{align}
\label{l:DRFTEQ1}
&x(t+\tau_2,\beta)-x(t,\beta)=\Delta^+
x(t,\beta)=b^+(x(t),\beta,t)\tau_2+\sigma\Delta^+z,
\\
\label{l:DRFTEQ6}
&x(t,\beta)-x(t-\tau_1,\beta)=\Delta^-
x(t,\beta)= b^-(x(t,\beta),t)\tau_1+\sigma\Delta^-z,
\end{align}
where $\tau_1$, $\tau_2$ are the inter-collision stopping times
with (independent) second order Gamma distributions with $b^\pm
(x,t):\Real^n\times [0,\infty)\longmapsto \Real^n$ and
$\sigma(x,t):\Real^n\times [0,\infty)\longmapsto \Real^n\times
\Real^n$ as the drift vector and variance matrix respectively.
Here $\beta=2/\overline{\tau}=E\left[\tau_2^{-1}\right]=
E\left[\tau_1^{-1}\right]$, see equation \eqref{COLL1}, and
$\Delta^+z=z(t_{j+1})-z(t)$ where $z(t)\in\Real^n,j\geqq 0$, is
a Gaussian process and $\Delta^-z$ is a Gaussian increment
independent of $\Delta^+z$ with
$E\left[\left(\Delta^-z_j\right)^2\right]=\tau_1$.

Due to the aforementioned issue with the distribution for $\tau_1$
the postulated form for the backward velocity in \eqref{l:DRFTEQ6}
cannot be correct. For fixed $\tau_2$ and $\tau_1$ Bayes' Theorem
implies that $\Delta^- x(t,\beta)$ must have a Gaussian distribution
but not necessarily with the same variance matrix
$\sigma=\sigma(x,t,\beta)$. As $\tau_1$ and $\tau_2$ are random
variables the forward increment is no longer Gaussian and the
distribution of the backward velocity distribution is unclear.
Therefore equation \eqref{l:DRFTEQ6} should have been written as
\begin{align}
\label{l:DRFTEQ6a} x(t,\beta)-x(t-\tau_1,\beta)=\Delta^-
x(t,\beta)= b_\beta^-(x(t,\beta),t)\tau_1+\sigma_\beta\Delta^-z,
\end{align}
with
\begin{align}
\label{l:DRFTEQ6b}
\begin{split}
&b_\beta^-(x(t,\beta),t)=\frac{E\left[\Delta^- x(t,\beta)\vert
x(t,\beta)\right]}{\tau_1},
\\
&\sigma_\beta\Delta^-z=\Delta^-
x(t,\beta)-b_\beta^-(x(t,\beta),t)\tau_1,
\end{split}
\end{align}
\end{subequations}
where again $E\left[\Delta^-z\right]=\tau_1$. Furthermore for
finite $\beta$ the increment $\Delta^-z$ is not a Gaussian
increment. However the expectation is that for progressively
smaller average inter-particle collision times $(\overline{\tau}
\rightarrow 0)$ that $b_\beta^-(x,t) \rightarrow b^-(x,t)$,
$\sigma_\beta \rightarrow \sigma$ and $\Delta ^-z$ becomes
approximately Gaussian. In other words $b_\beta^-(x,t)$
approaches the backward drift used in stochastic mechanics and
for sufficiently large $\beta$ (small $\overline{\tau}$)
equation \eqref{l:DRFTEQ6b} approaches \eqref{l:DRFTEQ6}.
Equation \eqref{l:DRFTEQ1} is therefore a definition while
\eqref{l:DRFTEQ6} is only true in the limit of sufficiently
small time steps.

Due to \eqref{l:DRFTEQ1}, \eqref{l:DRFTEQ6} the forward and
backward velocities can now be written as
\begin{gather}
\label{l:DRFTDEF0}
\begin{split}
v_2\left(x(t),t,\beta\right)=
\frac{x(t+\tau_2,\beta)-x(t,\beta)}{\tau_2}=
b^+(x(t),t)+\frac{1}{\tau_2}\sigma\Delta^+z,
\\
v_1\left(x(t),t,\beta\right)=
\frac{x(t,\beta)-x(t-\tau_1,\beta)}{\tau_1}=
b^-(x(t),t)+\frac{1}{\tau_1}\sigma\Delta^-z,
\end{split}
\end{gather}
which is properly defined because the diffusion shocks
$\sigma\Delta^+z/\tau_2$ and $\sigma\Delta^-z/\tau_1$ are proper
random variables with distribution
$f_{\frac{\Delta^+z}{\tau_2}}(v)\sim
\left(\beta+\frac{v^2}{2}\right)^{-5/2}$. In fact it is
straightforward to show that
\begin{gather*}
\begin{split}
&E\left[v_2\left(x(t),t,\beta\right)\right\vert x(t)] = b^+(x(t),t),
\\
&E\left[v_1\left(x(t),t,\beta\right)\right \vert x(t)]= b^-(x(t),t),
\\
&Cov\left[v_1\right\vert x(t)]=Cov\left[v_2\right\vert x(t)]=
\frac{4}{\overline{\tau}}\sigma\sigma^T,
\end{split}
\end{gather*}
using the fact that $E\left[\tau_2^{-1}\right]=
E\left[\tau_1^{-1}\right]=2/\overline{\tau}$ from equation
(\ref{COLL1}).  The expectations and covariances were calculated
conditional on the collision occurring at time $t$.

To define the solution to the difference equation properly, fix
a time interval $[0,T]$ and consider the collision set in this
time interval $x(t_j,\beta),j=1,...,N$, for a random variable
$N$ such that $t_N\leq T$ and $t_{N+1}>T$.  Hence $N$ is the
last collision in the time interval $[0,T]$. Define
$N(t)=\max_{j\in{\{0,1,...\}}} \{ t_j|t_j<t \}$ with
$t_{mn}=t_{N(t)},t_{mx}=t_{N(t)+1}$ then
\begin{gather}
\label{l:DRFTEQ9}
x(t,\beta)=
\begin{cases}
\begin{matrix}
\alpha_l
x\left(t_{mn},\beta\right)+(1-\alpha_l)x\left(t_{mx},\beta\right),
\\
\alpha_l=\frac{(t_{mx}-t)}{(t_{mx}-t_{mn})}, \end{matrix}
&\text{if $N(t)\geqq 0$},
\\
x(0),&\text{if $N(t)=\infty$},
\end{cases}
\end{gather}
where $N=\infty$ is an event that has probability zero as the
particle inter-collision times are all finite and identically
distributed. Hence $P\left[N(t)<\infty\right]=1$ so that
$t_{mx}$ and $t_{mn}$ are well defined with probability 1.  An
alternative expression used by the Theorem below is the
following martingale representation
$x(t,\beta)=x_N+b\left(x_N,t_N\right)(t-t_N)+\sigma\Delta(t-t_N)$
which differs from \eqref{l:DRFTEQ9} only by the proportionality
on the Gaussian shock.

The point of the Theorem below is to show under what
circumstances $x(t,\beta)\rightarrow x(t), a.e.$, where the
position process $x(t)$ is a strong solution to the stochastic
differential equation corresponding to \eqref{l:DRFTEQ1} with
drift $b^+(x(t),t)$ and variance matrix $\sigma=\sigma(x(t),t)$.

\begin{thm}
\label{a:THEOREM1}
Let the forward drift terms $b^+
(x,t):\Real^n\times [0,T]\longmapsto \Real^n$ and variance
matrix $\sigma(x,t):\Real^n\times [0,T]\longmapsto \Real^n\times
\Real^n$ satisfy the following growth and global Lipschitz
condition
\begin{gather}
\label{l:DRFTDEF3}
\begin{split}
&\norm{b^+\left(x,s\right)-b^+\left(y,s\right)}+
\norm{\sigma\left(x,s\right)-\sigma\left(y,s\right)}\leq
K\norm{x-y},
\\
&\norm{b^+\left(x,s\right)}^2+ \norm{\sigma\left(x,s\right)}^2\leq
K^2\left(1+\norm{x}^2\right),
\end{split}
\end{gather}
for all $x,y \in \Real^n$ and $s \in [0,T]$.  Let the drift function
also satisfy a Lipschitz condition in time such that a constant
$M_f$ exists so that
\begin{gather}
\label{l:DRFTDEF4}
\norm{b^+\left(x,s\right)-b^+\left(x,t\right)}\leq M_f\abs{s-t},
\end{gather}
for all $x \in \Real^n$, and all $s,t \in [0,T]$.  Let
$x(t,\beta)$ be the position of the main particle at time $t$
due to a finite set of collisions as specified in
\eqref{l:DRFTEQ1} and \eqref{l:DRFTDEF0}, i.e.
\begin{align}
\label{DRFTEQ2}
\begin{split}
x(t,\beta) &=x_0+\sum_{j=0}^{j=N}v_2 \left(x(t_j,\beta)\right)
\tau_j
\\
&=x_0+\sum_{j=0}^{j=N}\left[
b^+(x(t_j,\beta),t_j)\tau_j+\sigma\Delta^+z_j\right],
\end{split}
\end{align}
where $\tau_j=t_{j+1}-t_j,0\leq j< N(t)$,
$\tau_{N(t)}=t-t_{N(t)-1}$, $t_0=0$,$t_{N(t)+1}=t$ (not a collision
point) and where $\Delta^+z_j=z(t_{j+1})-z(t_j)$,
$z(t)\in\Real^n,j=0,1,...,N(t)$ are the Gaussian pulses.

Let $x_0=x(\beta,0)=x(0)$ be a random variable such that
$E[x_0^2]<\infty$ then $x(t,\beta)\rightarrow x(t)$ almost
everywhere as $\beta\uparrow \infty$ for the process $x(t)$
satisfying the stochastic differential equation
\begin{gather}
\label{l:DRFTDEF1}
x(t)=x_0+\int_0^t
b^+\left(x(s),s\right)ds+\int_0^t\sigma
 dz(t).
\end{gather}
\end{thm}

\begin{rem}
Condition \eqref{l:DRFTDEF3} together with $E[x_0^2]<\infty$ insure
that there is a strong solution $x(t),t\in\Re,$ to equation
\eqref{l:DRFTDEF1}. This means that $x(t)\in\Real^n$ is a continuous
process adapted to the filtration $\{\mathfrak{F}_t;0\leq
t<\infty\}$ of the probability space
$\left(\Omega,\mathfrak{F},P\right)$ and that
\begin{gather}
\label{l:DRFTDEF2}
P\left[\int_0^t\left[\abs{b_j^+\left(x(s),s\right)}+
\sigma_{ij}^2\left(x(s),s\right)\right] ds < \infty\right]=1,
\end{gather}
for all $i,j=1,...,n$. Condition \eqref{l:DRFTDEF4} is not required
for a strong solution but is an essential condition for the
convergence. The filtration $\{\mathfrak{F}_t;0\leq t<\infty\}$
encompasses the $\sigma-$algebra of events associated with the
initial condition $x_0$ (a random variable) and the $\sigma$-algebra
associated with the stochastic process $z(t),t\geq 0$.  The solution
is unique (any two solutions are equal with probability one) and can
be written as a functional of the initial condition and a
realization of the Gaussian process $z(t);0\leq t<\infty$.
\end{rem}
\begin{rem}
A strong solution for \eqref{l:DRFTDEF1} implies a growth condition
on the second moment of $x(t),t>0$ such that
\begin{gather*}
E\left[x(t)^2\right] \leq C\left(1+E\left[x_0^2\right]^2\right)
e^{Ct},
\end{gather*}
for some appropriate constant $C>0$.  In the case of few
interactions no collisions occur before $t=t_N$ ($\beta\downarrow
0$) so then \eqref{DRFTEQ2} reduces to
\begin{gather*}
x(t,0)=x_0+\left[ b^+(x_0,0)t+\sigma\Delta^+z_0\right],
\end{gather*}
with $\Delta^+z_0=z(t)-z(0)$.  Then
\begin{gather*}
E\left[x(t,0)^2\right]=E\left[x_0^2\right]+E\left[
b^+(x_0,t_j)^2t^2+\sigma^2t\right]
\\
\leq E\left[x_0^2\right]+K^2E\left[x_0^2\right]t^2+\sigma^2t <
\infty,
\end{gather*}
from the growth condition \eqref{l:DRFTDEF3}.
\end{rem}

\begin{proof}
For the sake of convenience the variance matrix will be assumed to
be constant which does not materially alter the proof.  Fix $N$ then
\begin{gather}
\label{a:THM10}
\begin{split}
\norm{x(t,\beta)-x(t)}=
\norm{\sum_{j=0}^{j=N}\int_{t_j}^{t_j+1}\left[
b^+\left(x(t_j,\beta),t_j\right)-b^+\left(x(s),s\right)\right]ds}
\\
\leq \sum_{j=0}^{j=N}\int_{t_j}^{t_j+1}\norm{
b^+\left(x(t_j,\beta),t_j\right)-b^+\left(x(s),s\right)}ds.
\end{split}
\end{gather}
Because of \eqref{l:DRFTDEF3} and \eqref{l:DRFTDEF4} the
integrand can be majorized as
\begin{align*}
&\norm{ b^+\left(x(t_j,\beta),t_j\right)-b^+\left(x(s),s\right)}
\\
&= \norm{ b^+\left(x(t_j,\beta),t_j\right)-b^+\left(x(s),t_j\right)
+b^+\left(x(s),t_j\right)-b^+\left(x(s),s\right)}
\\
&\leqq
\norm{b^+\left(x(t_j,\beta),t_j\right)-b^+\left(x(s),t_j\right)}
+\norm{b^+\left(x(s),t_j\right)-b^+\left(x(s),s\right)}
\\
&\leqq\norm{b^+\left(x(t_j,\beta),t_j\right)-b^
+\left(x(s),t_j\right)}+ M_f\abs{t_j-s},
\end{align*}
and applying this to \eqref{a:THM10} results in
\begin{align}
\label{DRIFTDEF4}
\begin{split}
&\norm{x(t,\beta)-x(t)}
\\
&\leq
\sum_{j=0}^{j=N}\int_{t_j}^{t_j+1}\norm{
b^+\left(x(t_j,\beta),t_j\right)-b^+\left(x(s),t_j\right)}ds
\\
&+M_f\sum_{j=0}^{j=N}\int_{t_j}^{t_j+1}\abs{ t_j-s}ds
\\
&\leq K\sum_{j=0}^{j=N}\sup_{t_j\leq s\leq t_{j+1}}\norm{
x(t_j,\beta)-x(s)}\tau_j +\frac{M_f}{2}\sum_{j=0}^{j=N}\tau_j^2.
\end{split}
\end{align}
and so
\begin{align*}
&\sup_{t_N\leq s\leq t}\norm{x(t_N,\beta)-x(s)}
\\
&\leq K\sum_{j=0}^{j=N}\sup_{t_j\leq s\leq t_{j+1}}\norm{
x(t_j,\beta)-x(s)}\tau_j +\frac{M_f}{2}\sum_{j=0}^{j=N}\tau_j^2.
\end{align*}
To extract the growth of the $\sup_{t_j\leq s<t_{j+1}}
\norm{x(t_j,\beta)-x(t_j)}$ term from this inequality the following
discrete version of Gronwall's Inequality will be applied, see
Shreve~\cite{SHREVE1}.
\begin{lem}
Let $a_n>0,n=0,1,...$ be a set of numbers such that
\begin{gather}
\label{GRONWOLL1} a_n\leq
\sum_{j=0}^{j=n-1}\left(P_ja_j\right)+Q_n,\quad n=1,...,
\end{gather}
for positive numbers $P_j,Q_j,j=0,1,...$.  Then for all $n>0$
\begin{gather}
\label{DRIFTDEF5} a_n \leq a_0P_0\frac{\Pi_{n-1}}
{\Pi_0}+\left(Q_n+\sum_{k=1}^{k=n-1} \frac{
\Pi_{n-1}}{\Pi_k}P_kQ_k\right),\quad n=1,...,
\end{gather}
where $\Pi_n=\prod_{j=0}^{j=n}\left(1+P_j\right),n=0,1,...$.
\end{lem}
\begin{proof}
From \eqref{GRONWOLL1} it follows that
\begin{gather}
\label{GRONWALL2}
a_n= \sum_{j=0}^{j=n-1}P_ja_j+Q_n-z_n,\quad
n=1,...,
\end{gather}
for some positive sequence of numbers $z_n>0,n=1,...$.  Let
$\beta_n=\sum_{j=0}^{n}P_ja_j,n=0,1,...,$ then \eqref{GRONWALL2} can
be written as
\begin{gather*}
\beta_n=\left(1+P_n\right)\beta_{n-1}+P_n\left(Q_n-z_n\right),\quad
n=1,...,
\end{gather*}
with initial condition $\beta_0=P_0a_0$.  The solution to this
equation equals
\begin{gather*}
\beta_n=\beta_0\frac{\Pi_n}{\Pi_0}+\sum_{k=1}^{k=n} \frac{
\Pi_n}{\Pi_k}P_k\left( Q_k-z_k\right),
\end{gather*}
using $\Pi_n=\prod_{j=0}^{j=n}\left(1+P_j\right)$.  Finally then
\begin{align*}
&P_na_n=\beta_n-\beta_{n-1}
\\
&=\beta_0\left(\frac{\Pi_n-\Pi_{n-1}} {\Pi_0}\right)+P_nQ_n
+\left(\Pi_n-\Pi_{n-1}\right)\sum_{k=1}^{k=n-1} \Pi_k^{-1}P_kQ_k
\\
&-P_nz_n-\left(\Pi_n-\Pi_{n-1}\right)\sum_{k=1}^{k=n-1}
\Pi_k^{-1}P_kz_k
\\
&=\beta_0P_n\frac{\Pi_{n-1}} {\Pi_0}+P_n\left(Q_n+\sum_{k=1}^{k=n-1}
\frac{\Pi_{n-1}}{\Pi_k}P_kQ_k\right)
\\
&-P_n\left(z_n+\Pi_{n-1}\sum_{k=1}^{k=n-1}
\Pi_k^{-1}P_kz_k\right),
\end{align*}
since
$\Pi_n-\Pi_{n-1}=\Pi_{n-1}\left(1+P_n\right)-\Pi_{n-1}=P_n\Pi_{n-1}$.
Now $\beta_0=P_0a_0$ and $z_n\geq 0,n\geq 1$, so dividing by $P_n$
it follows that
\begin{gather}
\label{GRONWALL3} a_n \leq a_0P_0\frac{\Pi_{n-1}}
{\Pi_0}+\left(Q_n+\sum_{k=1}^{k=n-1} \frac{
\Pi_{n-1}}{\Pi_k}P_kQ_k\right).
\end{gather}
If all $P_n,Q_n,n=0,1,...$ are equal so that
$P_n=P,Q_n=Q,n=0,1,...$ then \eqref{GRONWALL3} reduces to
\begin{gather*}
a_n\leq (Pa_0+Q)(1+P)^{n-1},\quad n=1,....
\end{gather*}
Notice that the righthand side of \eqref{GRONWALL3} increases
monotonically with $n$.
\end{proof}

Applying \eqref{DRIFTDEF5} to \eqref{DRIFTDEF4} using
$\Pi_n=\prod_{j=0}^{j=n}\left(1+K\tau_j\right)$,$P_n=K\tau_n,n\geq
0$ and $Q_n=\frac{M_f}{2}\sum_{j=0}^{j=n}\tau_j^2,n\geq 0$, with
the last remark in the Lemma, it follows that
\begin{gather}
\label{DRIFTDEF6}
\sup_{t_N\leq s\leq t} {\norm{x(t_N,\beta)-x(s)}}
\leq \norm{x(0,\beta)-x(0)}K\tau_0\frac{\Pi_{N}} {\Pi_0}
\\
+\frac{M_f}{2}\left(\sum_{j=0}^{j=N}\tau_j^2 +\sum_{k=1}^{k=N}
\frac{ \Pi_{N}}{\Pi_k}K\tau_l \sum_{j=0}^{j=k-1}\tau_j^2\right),
\end{gather}
Now $x(0,\beta)=x(0)$ and also
\begin{gather*}
\Pi_N=\prod_{j=0}^{j=N}\left(1+K\tau_j\right) =
e^{\sum_{j=0}^{j=N}log\left(1+K\tau_j\right)} \leq e^{Kt},
\end{gather*}
hence \eqref{DRIFTDEF6} reduces to
\begin{gather*}
\max_{0\leq j \leq N}\sup_{t_j\leq s\leq
t_{j+1}}\norm{x(s,\beta)-x(s)}
\\
\leq \frac{M_f}{2}\left(\sum_{j=0}^{j=N}\tau_j^2\right)\left(1+
KTe^{KT}\right),
\end{gather*}
so that finally
\begin{gather*}
E\left[\sup_{0\leq t\leq T}\norm{x(t,\beta)-x(t)}\right] \leq
\frac{L}{\beta},
\end{gather*}
where
\begin{gather*}
L=3M_fT\left(1+ KTe^{KT}\right).
\end{gather*}
This is the result of the fact that
$E\left[\left(\sum_{j=0}^{j=N}\tau_j^2\right)\right]
=3T\overline{\tau}$ and the fact that $t_{N+1}=t$.  From
Chebychev then
\begin{gather*}
Pr\left[\sup_{0\leq t\leq T}\norm{x(t,2^{2k})-x(t)}
\geq\frac{1}{2^k}\right] \leq \frac{L}{2^k},
\end{gather*}
so that by Borel-Cantelli $x(t,\beta)\longrightarrow x(t)$
almost everywhere on the interval [0,T] if
$\overline{\tau}\downarrow 0$. This completes the proof.
\end{proof}
\begin{rem}
This result justifies posing equation \eqref{l:DRFTEQ1} to
specify the discrete dynamics of the main particle and use the
solution $x(t),t\geq 0,$ \eqref{l:DRFTDEF1} as an approximation
of $x(t,\beta),t\geq 0$. Clearly the discrete subordinate
process that solves \eqref{l:DRFTEQ1} is difficult to determine.
The theorem above does not address the validity of equation
\eqref{l:DRFTEQ6} nor does it provide a hint as to the form of
the backward drift $b^-(x,t)$.
\end{rem}
\begin{rem}
Unfortunately, the types of solutions that are most of interest
have drift terms $b^+(x,t)$ that can be singular and therefore
do not satisfy conditions \eqref{l:DRFTDEF3} and
\eqref{l:DRFTDEF4}. However, Carlen showed in ~\cite{ECARL3}
that for a wide array of interesting drift functions a weak
solution exists for $x(t),t\in\Real^n$. Hence a filtration
$\{\mathfrak{F}_t;0\leq t<\infty\}$ exists for which
$\{z(t);0\leq t<\infty\}$ is a Brownian motion and for which
conditions \eqref{l:DRFTDEF1} and \eqref{l:DRFTDEF2} are
satisfied. In this case uniqueness does not necessarily hold (in
the same fashion) and no (measurable) functional exists to map
the solution given the initial condition and the Brownian path.
It is not clear at all how Theorem \eqref{a:THEOREM1} can be
generalized to appropriate weak solutions of equations
\eqref{l:DRFTDEF1} and \eqref{l:DRFTDEF2}.  For more discussion
on strong and weak solutions in the present context, see
Shreve~\cite{SHREVE1}, Rogers~\cite{ROGERS1} or
Carlen~\cite{ECARL1}.
\end{rem}

Following Nelson~\cite{ENELSON1} and Carlen~\cite{ECARL1} it can
be shown that the backward increment for a continuous stochastic
process is equivalent to a time reversed Markovian process with
a related drift and Gaussian increment. Specifically the
following applies.

\begin{thm}
\label{a:THM1}
If conditions \eqref{l:DRFTDEF3} and
\eqref{l:DRFTDEF4} apply, let $\sigma\equiv\sigma I$ so that the
variance matrix $\sigma$ is constant. Then, for a diffusion
process $x(t) \in \Real^n$ the forward motion of the particle
$x(t+\tau_2)-x(t)$ and the backward step $x(t)- x(t-\tau_1)$
conditional on $x(t)=x$ are given by
\begin{subequations}
\begin{align}
\label{l:DRFT2a}
&x(t+\tau_2)-x(t)=\Delta^+
x(t)=b^+(x(t),t)\tau_2+\sigma\Delta^+z,
\\
\label{l:DRFT2b}
&x(t)-x(t-\tau_1)=\Delta^- x(t)=
b^-(x(t),t)\tau_1+\sigma\Delta^-z,
\\
\label{l:DRFT2c}
&b^-(x,t)=b^+(x,t)-\sigma^2 \frac{\nabla
\rho(x,t)}{\rho(x,t)},
\end{align}
\end{subequations}
where $\rho(x,t)\in\Real,x\in\Real^n$ is the probability density
for finding the main particle at position $x$ at time $t$. The
shocks $\Delta^+z$ and $\Delta^-z$ are independent Gaussian
increments with mean zero and variances $\tau_2$ and $\tau_1$
respectively. From this it follows that
\begin{subequations}
\begin{gather}
\label{l:DRFT3a}
\rho(x,t)_t= - \nabla\left(b^+(x,t)\rho(x,t)\right)
+\Delta_x\rho(x,t),
\\
\label{l:DRFT3b}
\rho(x,t)_t= -\nabla\left(b^-(x,t)\rho(x,t)\right)
- \Delta_x \rho(x,t),
\end{gather}
so that
\begin{gather}
\label{l:DRFT3c}
\rho(x,t)_t= -
\nabla\left(\left(\frac{b^+(x,t)+b^-(x,t)}{2}\right)\rho(x,t)\right),
\end{gather}
which is the continuity equation.
\end{subequations}
Here $\nabla=\left(\frac{\partial}{\partial
x_1},...,\frac{\partial}{\partial x_n}\right)$,
$\Delta_x=\left(\frac{\partial^2}{\partial
x^2_1},...,\frac{\partial^2}{\partial x^2_n}\right)$  and
$\left(.\right)_t$ denotes the time derivative.
\end{thm}

\begin{proof}
See references Nelson ~\cite{ENELSON1} and Carlen ~\cite{ECARL1} and
the references therein for proofs where the collision times are
infinitesimally small.  For the justification of \eqref{l:DRFT2a},
\eqref{l:DRFT2b} and \eqref{l:DRFT2c} see remark \eqref{l:DRFT26}
below.
\end{proof}

\begin{rem}
From this point onward no distinction will be made between
$x(t,\beta)$ and $x(t)$ as the latter can be obtained as the
limit of the former.  This is not only to relieve notation but
also to emphasize the fact that the results that are to follow
are only true in the limit of $\beta\uparrow \infty$.
\end{rem}

\begin{rem}
\label{l:DRFT26} Equation \eqref{l:DRFT2a} is exact by
definition and its solution approaches $x(t),t\geq 0$, which has
a probability density $\rho(x,t)$ and a forward drift
$b^+(x,t)$.  Therefore the backward drift $b^-(x,t)$ can be
obtained as a result of \eqref{l:DRFT2c}.  As a result equation
\eqref{l:DRFT2b} is approximately correct for small discrete
time steps.
\end{rem}

The Corollary below uses Ito's Lemma and \eqref{l:DRFTDEF1} to find
an approximation for a function of the position $x(t)$.
\begin{cor}
For any function
\begin{align}
\label{l:DRFT4}
\begin{split}
\Delta^+f&=f(x(t+\tau_2),t+\tau_2)-f(x(t),t)
\\
&=(f_t+\sigma^2\Delta_xf)\tau_2+\nabla f.\Delta^+x(t)+O(\tau_2)
\end{split}
\end{align}
and
\begin{align}
\label{l:DRFT5}
\begin{split}
\Delta^-f&=f(x(t),t)-f(x(t-\tau_1),t-\tau_1) \\
&=(f_t-\sigma^2\Delta_xf)\tau_1+\nabla f.\Delta^-x(t)+O(\tau_1),
\end{split}
\end{align}
where $\Delta^+$, $\Delta^-$,$\tau_1$ and $\tau_2$ are introduced in
Theorem \eqref{a:THM1} above.
\end{cor}
\begin{proof}
Equation \eqref{l:DRFT4} follows from Ito's Lemma, while for
\eqref{l:DRFT5} it is clear that
\begin{align*}
&f(x(t-\tau_1),t-\tau_1)=f(x(t)-\Delta^-x,t-\tau_1)
\\
&=f(x(t),t)-f_t \tau_1-f_x\Delta^-x+
\frac{1}{2}f_{xx}\left(\Delta^-x\right)^2+O(\tau_1),
\end{align*}
so
\begin{align*}
&\Delta^-f=f(x(t),t)-f(x(t-\tau_1),t-\tau_1)
\\
&= f_t
\tau_1+f_x\Delta^-x-\frac{1}{2}f_{xx}\left(\Delta^-x\right)^2+O(\tau_1)
\\
&=(f_t-\sigma^2f_{xx})\tau_1+f_x\Delta^-x(t)+O(\tau_1),
\end{align*}
which settles the proof.
\end{proof}

\begin{rem}
Consider the one-dimensional case where $x(t) \in \Real$ and
assume that the variance term depends on $x(t)$ and $t$
explicitly so that $\sigma=\sigma(x(t),t)$.  If the variance
process is at least once differentiable then the backward and
forward representation can still be obtained as follows. Let
$y=f(x,t)=\int_{-\infty}^x\frac{\sigma_r}{\sigma(p,t)}dp$ and
let $\rho_f(y)$ be the density function for $y$ at time $t$ then
using Ito's formula
\begin{align*}
&\Delta^+y=F^+(y,t)dt+\sigma_r \Delta^+z
\\
&=\left(f_t+b^+\nabla f+\frac{\sigma^2}{2}\Delta_x f
\right)dt+\sigma_r \Delta^+z,
\\
&\Delta^-y=F^-(y,t)dt+\sigma_r \Delta^-z,
\end{align*}
where
\begin{gather*}
F^-(y,t)=F^+(y,t)-\sigma_r^2\frac{\rho_f(y)}{\partial y},
\end{gather*}
and where $\Delta^+z$ and $\Delta^-z$ are the increments as
defined above. Inverting the function above $x(y)=f^{-1}(y,t)$
and writing $\frac{\partial \rho}{\partial x}=\rho_x$, the
probability density for $y$ can then be translated back.
\end{rem}

Returning to the backward and forward representation of the motion
of the main particle in \eqref{l:DRFTEQ1}, \eqref{l:DRFTEQ1} the
following is now straightforward.
\begin{lem}
\label{l:ENERG8}
 Let the position process $x(t) \in \Real^n$ and the
inter-collision times be 2nd order Gamma distributed.  Then at
the collision time $t$
\begin{align}
\label{l:DRFT6}
\begin{split}
&E\left[\frac{\Delta^+x(t)}{\tau_2}\Bigg\vert
x(t)\right]=b^+(x(t),t), E\left[\frac{\Delta^-
x(t)}{\tau_1}\Bigg\vert x(t)\right]=b^-(x(t),t),
\\
&\qquad E\left[\left(\frac{\Delta^+
x(t)}{\tau_2}\right)^2\Bigg\vert x(t)\right]=b^+(x(t),t)^2+
\frac{2\sigma^2}{\overline{\tau}},
\\
&\qquad E\left[\left(\frac{\Delta^-
x(t)}{\tau_1}\right)^2\Bigg\vert
x(t)\right]=b^-(x(t),t)^2+\frac{2\sigma^2}{\overline{\tau}},
\end{split}
\end{align}
where $\overline{\tau}=2/\beta$ is the mean inter-particle collision
time and $\epsilon=\sigma^2/M$ is the diffusion per unit mass.
\end{lem}
\begin{proof}
The expectations are straightforward as the terms on the
righthand side of the second and third equation in
(\ref{l:DRFTDEF0}) are finite curtesy of the fact that
\begin{gather*}
E\left[\left(\frac{\Delta^+z}{\tau_2}\right)^2\right]=
E\left[\frac{1}{\tau_2}\right]=\frac{2}{\overline{\tau}},\\
E\left[\left(\frac{\Delta^-z}{\tau_1}\right)^2\right]=
E\left[\frac{1}{\tau_1}\right]=\frac{2}{\overline{\tau}}.
\end{gather*}
The equality in the first equation in (\ref{l:DRFT5}) is due to
the fact that
\begin{gather*}
\int_{-\infty}^{\infty}(b^+(x,t)-b^-(x,t))\rho(x,t)=
\int_{-\infty}^{\infty}\left(\frac{\rho_x(x,t)}{\rho(x,t)}\right)\rho(x,t)=0,
\end{gather*}
so that the average velocity of the main particle is the same
whether a forward or backward view is developed.
\end{proof}
\begin{rem}
As a result of \eqref{l:DRFT6} the forward and backward energy
for the main particle can be defined as
\begin{align}
\label{l:DRFT27}
\begin{split}
&\h^+_M(x(t),t)=\frac{M}{2}E\left[\left(\frac{\Delta^+
x(t)}{\tau_2}\right)^2\Bigg\vert x(t)\right]=
\frac{M}{2}b^+(x(t),t)^2+\frac{\epsilon}{\overline{\tau}},
\\
&\h^-_M(x(t),t)=\frac{M}{2}E\left[\left(\frac{\Delta^-
x(t)}{\tau_2}\right)^2\Bigg\vert x(t)\right]=
\frac{M}{2}b^-(x(t),t)^2+\frac{\epsilon}{\overline{\tau}},
\end{split}
\end{align}
because
$\frac{M}{2}2\sigma^2/\overline{\tau}=\epsilon/\overline{\tau}$.
\end{rem}
\begin{rem}
Recall that it was assumed that $t\gg\overline{\tau}$ otherwise
$\tau_1$ is not sufficiently close to a 2nd order Gamma
distribution.
\end{rem}
\begin{rem}
In the typical stochastic mechanics setting, $\sigma^2=\hbar/M$
so that the backward and forward energies have a drift component
and depend on a constant equal to $\hbar/\overline{\tau}$.
Typically in molecular applications at room temperatures
$Mb^+(x(t),t)^2>>\hbar/\overline{\tau}$ and only at lower
temperatures $Mb^+(x(t),t)^2 \sim\hbar/\overline{\tau}$. However
in other applications like astrophysics or economics it remains
to be seen what the drift energy is in proportion to the
diffusion energy. In the molecular-kinetic theory of heat the
diffusion coefficient equals $\sigma^2\sim kT/\eta_v$ so the
mean versus diffusion energy ratio then depends on the
temperature $T$ and the viscosity $\eta_v$.
\end{rem}

The following example demonstrates a consequence of Theorem
\eqref{a:THM1} and equations \eqref{l:DRFTEQ1} and
\eqref{l:DRFT2c}. Consider a Brownian particle that arrived in
$x(t)=x$ and assume that the collisions occur closely in time so
$\tau_2$ and $\tau_1$ are small. From \eqref{l:DRFT2c} follows
that the forward and drift for Brownian motion are determined as
$b^+(x,t)=0$ and $b^-(x,t)=-x(t)/t$. The approximate energy gain
or loss for the main particle can now be approximated from the
outgoing and incoming energies around the collision point so in
one dimension
\begin{align*}
&2\left(\h^+_M\left(x(t),t\right)-\h^-_M\left(x(t),t\right)\right)
\\
=&M\left( \frac{x(t+\tau_2)-x(t)}{\tau_2}\right)^2- M\left(
\frac{x(t)-x(t-\tau_1)}{\tau_1}\right)^2,
\end{align*}
where $\h^+_M\left(x(t),t\right)-\h^-_M\left(x(t),t\right)$ is
the energy gain or loss due to the collision in $x(t)$ at time
$t$.

Averaging over all collision points this reduces to
\begin{gather}
\label{l:DRFT7}
\begin{split}
&2E\left[\h^+_M\left(x(t),t\right)-\h^
-_M\left(x(t),t\right)\right]
\\
&=ME\left[\left(\frac{\Delta^+x(t)}{\tau_2}\right)^2\right]-
ME\left[\left( \frac{\Delta^-x(t)}{\tau_1}\right)^2\right]
\\
&=2M\frac{\sigma^2}{\overline{\tau}}-
ME\left[\frac{x(t)^2}{t^2}\right]
-2M\frac{\sigma^2}{\overline{\tau}}=-\frac{M\sigma^2}{2t}.
\end{split}
\end{gather}
The expectation includes all paths emanating from the origin so
this suggests that the main particle sheds energy continually.
The expression does not depend on the mean inter-particle
collision time or relates in any other way to the
characteristics of the impacting particles. In fact the large
contributions of the diffusion terms in the backward and forward
momenta cancel exactly leaving the time dependent drift
$-M\sigma^2/(2t)$.

Equation (\ref{l:DRFT7}) should have a more complicated
appearance to reflect the discreteness of equation
\eqref{l:DRFTEQ1}. However, for sufficiently small
$\overline{\tau}$ the conclusion must be that a particle
following a Brownian motion path must be flooding its
surrounding heatbath with energy thereby raising its
temperature. In the continuous representation the energy
transfer is infinite at the origin which is an unrealistic
physical anomaly due to the discreteness of equation
\eqref{l:DRFTEQ1}. It is clear that a continual energy exchange
is possible (viz. high energy protons fired into a plasma torus)
but it is unexpected that a Brownian (main) particle is not in
equilibrium with its surroundings.

If not Brownian Motion then the question is what stochastic
process describes the equilibrium motion of the main particle in
a heatbath such that there is no energy exchange between main
particle and heatbath?  To answer this question the next Section
shows that the main particle equilibrium motion is related to
the total energy conservation and investigates the constraints
on the probability density function for the position of the main
particle.

\section{Energy Conservation}
This Section combines the properties of elastic collisions to
show that the main particle plus heatbath particle kinetic
energy equals a quadratic expression of the backward and forward
velocities.  From this it is shown that if the main particle is
in equilibrium with the heatbath then there is only one
acceptable probability density for the position of the main
particle. Another conclusion in this Section is that the forward
and backward velocities of the main particle through a collision
point can only be independent if the heatbath particle motion
through the collision is highly correlated. Examples are
presented in the form of the Gaussian Wave packet and Brownian
motion.

To analyze the energy exchange associated with the collisions
consider an elastic two-particle collision in $n$ (typically 2
or 3) dimensions at time $t$ between the main particle of mass
$M$ and the heatbath particle $m$ where typically $(M> m)$. The
main particle has pre- and post-collision velocities $v_2, v_1
\in \Real^n$ while the heatbath particle has pre- and
post-collision velocities $w_2, w_1 \in \Real^n$ respectively.
By assumption the particles exchange energy and momentum during
the collision hence $v_2\neq v_1$ and $w_2 \neq w_1$.  As there
are no other interactions, the momentum and the energy during
the collision must be conserved so that
\begin{gather}
\label{l:ENERG1}
\begin{split}
&p_2+q_2=Mv_2+mw_2=p_1+q_1=Mv_1+mw_1,
\\
&\frac{1}{2}(M\abs{v_2}^2+m\abs{w_2}^2)
=\frac{\abs{p_2}^2}{2M}+\frac{\abs{w_2}^2}{2m}
\\
&=\frac{1}{2}(M\abs{v_1}^2+m\abs{w_1}^2)
=\frac{\abs{p_1}^2}{2M}+\frac{\abs{w_1}^2}{2m}.
\end{split}
\end{gather}
To obtain a full solution to equation \eqref{l:ENERG1} assume a
linear relationship between the main pre- and post-collision
particle velocities $v_1, v_2 \in \Real^n$ and the incident pre-
and post-collision particle velocities $w_1, w_2 \in \Real^n$ as
follows
\begin{align}
\label{l:GAMMDEF1}
\begin{pmatrix}
v_2 \\
w_2
\end{pmatrix}
=
\begin{pmatrix}
P & Q \\
V & G
\end{pmatrix}
\begin{pmatrix}
v_1 \\
w_1
\end{pmatrix}
=\Gamma
\begin{pmatrix}
v_1 \\
w_1
\end{pmatrix},
\end{align}
where $P=P(x,t),Q=Q(x,t),V=V(x,t)$ and $G=G(x,t)$ are
$\Real^n\times\Real^n$ matrices so that
$\Gamma=\Gamma\left(x,t\right)$ is a
$\Real^{2n}\times\Real^{2n}$ matrix. The following theorem
determines the form of the matrices $P,Q,V$ and $G$ under the
conservation of energy and momentum constraints
\eqref{l:ENERG1}.
\begin{thm}
\label{l:ENERG16}
Assume that the $\Gamma$ matrix in
\eqref{l:GAMMDEF1} can be decomposed as $\Gamma=
\left(\begin{smallmatrix}
P & Q \\
V & G
\end{smallmatrix}\right)$
then the energy and momentum conservation in \eqref{l:ENERG1} is
equivalent to
\begin{align}
\label{l:COLLMAT2}
\begin{split}
&\Gamma^T
\begin{pmatrix}
M \\
m
\end{pmatrix}
=
\begin{pmatrix}
M \\
m
\end{pmatrix}, \\
&\Gamma^T
\begin{pmatrix}
M & 0 \\
0 & m
\end{pmatrix}
\Gamma=
\begin{pmatrix}
M & 0 \\
0 & m
\end{pmatrix},
\end{split}
\end{align}
or equivalently
\begin{subequations}
\begin{align}
\label{l:COLLMAT3a}
&P^TM+R^Tm=M,P^TMP+V^TmV=M,
\\
\label{l:COLLMAT3b}
&Q^TM+S^Tm=m,Q^TMQ+G^TmG=M,
\\
\label{l:COLLMAT3c}
&P^TMQ+V^TmG=0.
\end{align}
\end{subequations} If $M=MI,m=mI$ ($I$ being the unit matrix)
then these equations can be solved to yield
\begin{align}
\label{l:COLLMAT1}
\begin{split}
&P=\frac{\sin(\theta)}{2\gamma}\left(I-\gamma^2U\right),
Q=\frac{\gamma \sin\theta}{2}\left(I+U\right),
\\
&V=\frac{sin\theta}{2\gamma}\left(I+U\right) ,
G=\frac{\gamma\sin(\theta)}{2}\left(I-\frac{1}{\gamma^2}U\right),
\\
&U^TU=I,
\end{split}
\end{align}
where $U$ is an arbitrary $n\times n$ unitary matrix $U^TU=I$.
\end{thm}
\begin{proof}
A straightforward but lengthy proof for this can be found in
Appendix B.
\end{proof}
\begin{rem}
\label{l:REM1} Notice that $Q=\gamma^2 V$, $P+Q=I$ and $V+G=I$
so the collision matrix $\Gamma(x,t)$ can be easily expressed in
terms of the matrix $Q$.
\end{rem}
\begin{rem}
The matrix $\Gamma(x,t)$ in \eqref{l:GAMMDEF1} consists of all
center of mass collision information to translate the
pre-collision velocities $v_1$, $w_1$ information into the
post-collision $v_2$, $w_2$ configuration.  In one dimension
this matrix is naturally absent and in higher dimensions the
matrix varies randomly from collision to collision.  The matrix
depends on $\Gamma=\Gamma\left(x,t,Z\right)$ where $Z$ is an
asymmetric matrix $Z=I-2\left(I+U\right)^{-1}$ with $U$ defined
in \eqref{l:COLLMAT1}.  The matrix $Z$ will be referred to as
the collision scattering matrix.  With elastic collisions this
matrix is independent of the energy pre-collision energy $v_!$
and $w_1$ and typically reflects the physical circumstances of
the collision event.
\end{rem}

For $U=I, Z\equiv 0$, equation \eqref{l:COLLMAT1} reduces to a
simpler set of linear equations which shall be referred to as
the "simple" elastic collision
\begin{align}
\label{l:ENERG2}
\begin{split}
\begin{pmatrix}
v_2 \\
w_2
\end{pmatrix}
&=
\begin{pmatrix}
\cos(\theta)I & \gamma\sin(\theta)I
\\
\frac{\sin(\theta)}{\gamma}I & -\cos(\theta)I
\end{pmatrix}
\begin{pmatrix}
v_1
\\
w_1
\end{pmatrix}
= \Omega
\begin{pmatrix}
v_1 \\
w_1
\end{pmatrix},
\end{split}
\end{align}
with obvious definition of the matrix $\Omega$ and $I$ being the
unit matrix in n dimensions.  Here $\gamma=\frac{m}{M}$,
$\cos(\theta)=\frac{1-\gamma^2}{1+\gamma^2}$,
$\sin(\theta)=\frac{2\gamma}{1+\gamma^2}$.  In one dimension
this solution is unique as in that case $Z\equiv 0$.  This
expression also applies if $\gamma = 1$ ($m=M$) in which case
the particles simply exchange velocities $v_2=w_1, w_2=v_1$. The
matrix $\Omega$ is a reflection transformation with
$det(\Omega)=-1$, $\Omega^T\Omega=I$, Trace$(\Omega)=0$ and has
eigenvectors
\begin{align*}
\begin{pmatrix}
\cos\left(\frac{\theta}{2}\right) \\
\sin\left(\frac{\theta}{2}\right)
\end{pmatrix}
,
\begin{pmatrix}
\sin\left(\frac{\theta}{2}\right) \\
-\cos\left(\frac{\theta}{2}\right)
\end{pmatrix},
\end{align*}
with eigenvalues $1,-1$.  Notice also that
$\cos(\theta/2)=\frac{1}{1+\gamma^2}$ and
$\sin(\theta/2)=\frac{\gamma}{1+\gamma^2}$.

\begin{rem}
\label{l:ENERG15} Equation \eqref{l:ENERG1} has a unique
solution in the sense that the main and heatbath particle
exchange momentum and energy along the center of mass line at
the moment of collision while other components of the motion
remain invariant.  This means that in higher dimensions equation
\eqref{l:ENERG2} applies to the pre - and post-collision
velocity components of $v_2,w_2,v_1,w_1$ along the center of
mass line. The components of $v_1$ and $w_1$ orthogonal to the
collision center of mass line remain invariant under an elastic
collision. In this Section no real distinction is made between
the properties of the heatbath particle and the properties of
its center of mass line components.
\end{rem}

A simple Theorem now describes how the energy conservation for
the motion for the heatbath particle $m, w_1$ (post-collision
$m,w_2$) and the main particle $M, v_1$ (post-collision $M,v_2$)
can be expressed exclusively in terms of $v_2$ and $v_1$ if
\eqref{l:ENERG2} holds.
\begin{thm}
\label{l:THMENERG}
Let the momentum of the main particle and
interacting particle be presented as $p_1=Mv_1$ (post-collision
$p_2=Mv_2$) and $q_1=Mw_1$ (post-collision $q_2=Mw_2$) with
$v_1,v_2,w_1,w_2\in\Real^n$. Then the total energy
$\h_T=\frac{1}{2}(Mv^2_2+mw^2_2)=\frac{1}{2}(Mv^2_1+mw^2_1)$ is
related to the pre - and post  collision of momenta of the main
particle as follows:
\[
\frac{M\gamma^4}{1+\gamma^2}\h_k=
\frac{1}{2}\abs{\frac{q_2+q_1}{2}}^2+\frac{\gamma^2}{2}
\abs{\frac{q_2-q_1}{2}}^2,
\]
for the heatbath particle, while for the main particle
\[
\frac{M}{1+\gamma^2}\h_k=\frac{1}{2}
\abs{\frac{p_2+p_1}{2}}^2+\frac{1}{2\gamma^2}
\abs{\frac{p_2-p_1}{2}}^2.
\]
In terms of the velocities these equations become
\begin{subequations}
\begin{align}
\label{l:DRFT8}
\begin{split}
&\frac{\h_k}{M_T}=
\frac{1}{2}\abs{\frac{w_2+w_1}{2}}^2+\frac{\gamma^2}{2}
\abs{\frac{w_2-w_1}{2}}^2,
\\
&\frac{\h_k}{M_T}\frac{1}{2}=
\abs{\frac{v_2+v_1}{2}}^2+\frac{1}{2\gamma^2}
\abs{\frac{v_2-v_1}{2}}^2,
\end{split}
\end{align}
with $M_T=M+m=M\left(1+\gamma^2\right)$.  Also a more direct
relationship between the momenta can be derived as follows
\begin{align}
\label{l:DRFT9}
\abs{\frac{v_2+v_1}{2}}^2+\frac{1}{\gamma^4}
\abs{\frac{v_2-v_1}{2}}^2=\frac{1}{2}\abs{w_1}^2+\abs{w_2}^2.
\end{align}
\end{subequations}
\end{thm}
\begin{proof}
See Appendix A.
\end{proof}
\begin{rem}
The previous remark \eqref{l:ENERG15} implies that in higher
dimensions equations \eqref{l:DRFT8} and \eqref{l:DRFT9} apply
to the center of mass line components of the main and heatbath
particles.  In other words if $n>1$ the kinetic energy $\h_k$
incorporates only the parts of the motion of the heatbath
particle that are altered due to the collision.  The motion
perpendicular to the center of mass line remains invariant hence
this energy component remains the same and must be added to
$\h_k$ in order to derive the full energy of the heatbath and
main particle combined.  In Section 4 equations \eqref{l:DRFT8}
and \eqref{l:DRFT9} will be investigated for the case where
$Z\neq 0$ so there $\h_k$ will represent the combined kinetic
energy.
\end{rem}

\begin{rem}
It deserves mentioning that equations \eqref{l:DRFT8} and
\eqref{l:DRFT9} only hold for a collision point and are in fact
incorrect for any intermediate points.
\end{rem}

Conceptually equation \eqref{l:DRFT8} suggests that the
fracturing of the path of the main particle provides a direct
indication of the amount of energy that is involved in the main
and heatbath particle system combined. Moreover, condition
\eqref{l:DRFT9} is independent of the energy constraint
\eqref{l:DRFT8} and tends to constrain the momentum exchange
between the particles involved.  For this reason \eqref{l:DRFT8}
is referred to as the "energy" balance/constraint and
\eqref{l:DRFT9} as the "momentum" constraint.

The agenda for further developments is now to employ the
collision representation of the Markovian process describing the
path of the main particle in equation \eqref{l:DRFTEQ1} and
substitute the respective forward and backward velocities
\eqref{l:DRFTDEF0} into \eqref{l:DRFT8}.  Due to the
distribution of the inter-collision stopping times the
expectation of the total energy components in the collision
process will be well defined.  The mean inter-particle collision
is small so the use of equation \eqref{l:DRFTEQ6} is well
justified. The total energy as a function of time and position
can then be investigated.

First an important consequence of equation \eqref{l:GAMMDEF1}
must be noted.  Rewrite equation (\ref{l:ENERG2}) so as to
represent the pre- and post-collision velocities of the main
particle as a function of the velocities of the colliding
heatbath particle. This shows that
\begin{align}
\label{l:ENERG3}
\begin{split}
\begin{pmatrix}
v_2 \\
v_1
\end{pmatrix}
&=
\frac{\gamma}{\sin(\theta)}
\begin{pmatrix}
\cos(\theta)I & I \\
I & \cos(\theta)I
\end{pmatrix}
\begin{pmatrix}
w_2
\\
w_1
\end{pmatrix},
\end{split}
\end{align}
hence the statistical properties of $v_2,v_1$ are generated by the
behavior of  $w_2,w_1$ and vice versa.

Specifically equations \eqref{l:DRFTEQ1} and \eqref{l:DRFTEQ6}
demand that the motion of the heatbath particle must be of the
following form
\begin{gather*}
\begin{split}
w_2=g^+(x(t),t)+\omega\frac{\Delta^+_w}{\tau_2},\\
w_1=g^+(x(t),t)+\omega\frac{\Delta^+_w}{\tau_1},
\end{split}
\end{gather*}
which substituted into equation \eqref{l:ENERG3} yields
\begin{align*}
\left(b^+(x(t),t)+\sigma\frac{\Delta^+z}{\tau_2}\right)
=&\cos(\theta)\left(b^-(x(t),t)+\sigma\frac{\Delta^-z}{\tau_1}\right)
\\
&+\gamma\sin(\theta) \left(g^-(x(t),t)+\omega\frac{\Delta^+_w}{\tau_1}\right),
\\
\left(g^+(x(t),t)+\sigma\frac{\Delta^+_w}{\tau_2}\right)
=&\frac{\sin(\theta)}{\gamma}\left(b^-(x(t),t)+\sigma\frac{\Delta^-z}{\tau_1}\right)
\\
&-\cos(\theta)\left(g^-(x(t),t)+\omega\frac{\Delta^-_w}{\tau_1}\right),
\end{align*}
or equivalently
\begin{subequations}
\begin{align}
\label{DELTA1a}
\begin{pmatrix}
b^+(x(t),t) \\
b^-(x(t),t)
\end{pmatrix}
&= \frac{\gamma}{\sin(\theta)}
\begin{pmatrix}
\cos(\theta) & 1 \\
1 & \cos(\theta)
\end{pmatrix}
\begin{pmatrix}
g^+(x(t),t) \\
g^-(x(t),t)
\end{pmatrix},
\\
\label{DELTA1b}
\begin{pmatrix}
\sigma\frac{\Delta^+z}{\tau_2} \\
\sigma\frac{\Delta^-z}{\tau_1}
\end{pmatrix}
&= \frac{\gamma}{\sin(\theta)}
\begin{pmatrix}
\cos(\theta) & 1 \\
1 & \cos(\theta)
\end{pmatrix}
\begin{pmatrix}
\omega\frac{\Delta^+_w}{\tau_2} \\
\omega\frac{\Delta^+_w}{\tau_1}
\end{pmatrix}.
\end{align}
\end{subequations}

Hence the motion of the heatbath particle must be driven by
backward and forward drifts $g^\pm(x,t)$ and corresponding
random impulses $\omega\frac{\Delta^+_w}{\tau_2}$,
$\omega\frac{\Delta^-_w}{\tau_1}$.  As subsequent collisions
involve different heatbath particles the realizations from the
heatbath particles $w_1$, $w_2$ should be the result of a time
dependent random field.  Specifically the post-collision
heatbath velocity $w_2(x(t),t)$ for the collision at $x(t),t$ is
not equal to the pre-collision velocity $
w_1(x(t+\tau_2),t+\tau_2)$ as they involve different particles.
No continuity equation therefore applies to $g^\pm(x,t)$.
However equations \eqref{DELTA1a} and \eqref{DELTA1b} show that
the heatbath realizations are constrained by the motion of the
main particle.

The other important consequence of \eqref{DELTA1b} is that the
main particle forward and backward velocities $\Delta^+z$ and
$\Delta^-z$ can only be independent if the heatbath particle
forward and backward velocities $\Delta^+_w$ and $\Delta^-_w$
are appropriately correlated. The following Theorem details the
autocorrelation in the heatbath particle motion and also shows
that the random terms $\Delta^+_w$ and $\Delta^-_w$ must be
cross correlated with $\Delta^+z$ and $\Delta^-z$.

\begin{thm}
\label{a:THM9}
If the main particle follows a Markovian path i.e
if $\Delta^+z$ and $\Delta^-z$ in equation \eqref{DELTA1b} are
uncorrelated, then $\omega^2=\frac{\sigma^2}{2\alpha^2}$, where
$\alpha^2=\gamma^4/(1+\gamma^4)$.  Also the $\Delta_w^+$ and
$\Delta_w^-$ must be correlated as
\begin{gather*}
corr\left(\frac{\Delta_w^+}{\tau_2},\frac{\Delta_w^-}{\tau_1}
\right)=-\left(1-2\alpha^2\right).
\end{gather*}
If $m<<M$ then $\alpha$ is very small hence the variance term
$\omega$ must be large and the correlation between the
pre-collision and post-collision velocities of the heatbath
particle must be very high.

In addition the forward and backward velocities of the main and
incident particle have covariances
\begin{align*}
&E\left[ \frac{\Delta^+_w\Delta^+z}{\tau_2^2}\right]=E\left[
\frac{\Delta^-_w\Delta^-z}{\tau^2_1}\right]=
-\frac{\sqrt{2}\alpha}{\gamma\sin(\theta)}\cos(\theta),
\\
&E\left[ \frac{\Delta^+_w\Delta^-z}{\tau_2\tau_1}\right]=E\left[
\frac{\Delta^-_w\Delta^+z}{\tau_2\tau_1}\right]=
\frac{\sqrt{2}\alpha}{\gamma\sin(\theta)}.
\end{align*}
\end{thm}
\begin{proof}
Inverting equation \eqref{DELTA1b} yields
\begin{align}
\label{Delta2}
\begin{pmatrix}
\omega\frac{\Delta^+_w}{\tau_2} \\
\omega\frac{\Delta^-_w}{\tau_1}
\end{pmatrix}
= \frac{1}{\gamma\sin(\theta)}
\begin{pmatrix}
-\cos(\theta) & 1 \\
1 & -\cos(\theta)
\end{pmatrix}
\begin{pmatrix}
\sigma\frac{\Delta^+z}{\tau_2} \\
\sigma\frac{\Delta^-z}{\tau_1}
\end{pmatrix},
\end{align}
and if the increments $\sigma\frac{\Delta^+z}{\tau_2}$ and
$\sigma\frac{\Delta^-z}{\tau_1}$ are independent then
\begin{align*}
&\omega^2 E\left[
\begin{pmatrix}
\frac{\Delta^+_w}{\-\tau_2} \\
\frac{\Delta^-_w}{\-\tau_1}
\end{pmatrix}
\begin{pmatrix}
\frac{\Delta^+_w}{\-\tau_2} & \frac{\Delta^-_w}{\-\tau_1}
\end{pmatrix}
\right]
\\
&=
\frac{2\sigma^2}{\overline{\tau}\gamma^2\sin^2(\theta)}
\begin{pmatrix}
-\cos(\theta) & 1 \\
1 & -\cos(\theta)
\end{pmatrix}^2
\\
&=\frac{\sigma^2}{\overline{\tau}\alpha^2}
\begin{pmatrix}
1 & -\left(1-2\alpha^2\right) \\
-\left(1-2\alpha^2\right) & 1
\end{pmatrix},
\end{align*}
where $\alpha^2=\gamma^4/(1+\gamma^4)$.  If it is assumed that
\begin{align*}
E\left[\left(\frac{\Delta^+_w}{\tau_2}\right)^2\right]=
E\left[\left(\frac{\Delta^-_w}{\tau_1}\right)^2\right]=
\frac{2}{\overline{\tau}},
\end{align*}
then $\omega=\sigma/\left(\alpha\sqrt{2}\right)$ and the
correlation follows directly.

Using \eqref{Delta2} again it is also clear that
\begin{align*}
&\omega\sigma E\left[
\begin{pmatrix}
\frac{\Delta^+_w}{\-\tau_2} \\
\frac{\Delta^-_w}{\-\tau_1}
\end{pmatrix}
\begin{pmatrix}
\frac{\Delta^+z}{\-\tau_2} & \frac{\Delta^-z}{\-\tau_1}
\end{pmatrix}
\right]
\\
&= \frac{2\sigma^2}{\overline{\tau}\gamma\sin(\theta)}
\begin{pmatrix}
-\cos(\theta) & 1 \\
1 & -\cos(\theta)
\end{pmatrix},
\end{align*}
so that
\begin{align*}
&E\left[
\begin{pmatrix}
\frac{\Delta^+_w\Delta^+z}{\tau_2^2} &
\frac{\Delta^+_w\Delta^-z}{\tau_2\tau_1}
\\
\frac{\Delta^-_w\Delta^+z}{\tau_2\tau_1}
&\frac{\Delta^-_w\Delta^-z}{\tau_1^2}
\end{pmatrix}
\right]
\\
&= \frac{\sqrt{2}\alpha}{\gamma\sin(\theta)}
\begin{pmatrix}
-\cos(\theta) & 1 \\
1 & -\cos(\theta)
\end{pmatrix},
\end{align*}
which concludes the proof.
\end{proof}

\begin{rem}
If $\gamma<<1$ then the $\omega^2$ variance term becomes very
large and the correlation between
$\sigma\frac{\Delta_w^+}{\tau_2}$ and
$\sigma\frac{\Delta_w^-}{\tau_1}$ becomes tighter.  In the limit
where $\gamma\approx 0$, e.g. a football interacting with
molecules or a planet interacting with cosmic particles (or
photons) a small diffusion constant $\sigma$ for the main object
is associated with an enormous heatbath particle momentum
proportional to $\sigma/\left(\alpha\sqrt{2}\right)$. In this
case the post-collision heatbath particle velocity size is
identical to the pre-collision heatbath velocity but moving in
the opposite direction.
\end{rem}

The next issue to investigate is the total kinetic energy $\h_k$
in \eqref{l:DRFT8} as a function of time while the main particle
travels from collision to collision.  By assumption the
collision is elastic and conserves the total energy so if $w_1$
and $w_2$ refer to the same particle (the first one before the
collision and the second one after the collision with the main
particle) then by definition
\begin{align*}
\h_k&=\h_k(x(t),t)=\frac{1}{2}M\abs{v_2}^2+\frac{1}{2}m\abs{w_2}^2
\\
&= \frac{1}{2}M\abs{v_1}^2+\frac{1}{2}m\abs{w_1}^2.
\end{align*}
However, for the subsequent collision the main particle moving with
momentum $Mv_2$ particle will meet a different heatbath particle at
time $t+\tau_2$ in a new position
$x(t+\tau_2)=x(t)+b^+(x,t)\tau_2+\sigma\Delta^+z(t)/\tau_2$. This
new heatbath particle will have a new momentum $mq'$ which is chosen
randomly from the ensemble. Obviously the exiting total kinetic
energy at $x(t)$ and the combined "new" total kinetic energy at
$x(t+\tau_2)$ will not be equal since
\begin{align*}
H'_k(x(t+\tau_2),t+\tau_2)&=\frac{1}{2}M\abs{v_2}^2+\frac{1}{2}m\abs{q}'^2
\\
&\neq \frac{1}{2}M\abs{v_2}^2+\frac{1}{2}M\abs{w_2}^2.
\end{align*}

Hence that the combined system (main + heatbath particle) gains or
looses energy $\Delta \h_k(x(t),t)$ equal to
\begin{align*}
\Delta \h_k(x(t),t)&=\h_k(x(t+\tau_2),t+\tau_2)-\h_k(x(t),t)
\\
&= \frac{m}{2}\left(\abs{q}'^2-\abs{w_2}^2\right).
\end{align*}
On average the amount of energy exchanged equals
\begin{align}
\label{l:POTNT2} E\left[\Delta \h_k(x(t),t)\right]=
\frac{m}{2}E\left[ \abs{q}'^2\right]-\frac{m}{2}E\left[
\abs{w_2}^2\right]
\end{align}
where the expectation $\frac{m}{2}E\left[ \abs{q}'^2\right]$ is
the average energy of the new incoming heatbath particle at
collision time $t+\tau_2$ and $\frac{m}{2}E\left[
\abs{w_2}^2\right]$ is the post collision energy of the previous
heatbath particle at the previous collision time $t$. The
expectation here is over all paths, heatbath interactions and
all positions $x(t)$.

If \eqref{l:POTNT2} is positive the original post collision
energy of the heatbath particle at $t$ has a lower energy than
the new heatbath particle colliding at $t+\tau_2$ and if this
quantity is negative the collision accelerated the heatbath
particle coming out of $x(t)$ in comparison to the typical
heatbath particle.  In the first case the colliding heatbath
particle returned to the heatbath with less energy than the
average heatbath particle and in the second case the collision
accelerated the colliding heatbath particle beyond the heatbath
average. If \eqref{l:POTNT2} is positive the heatbath puts
energy into the main particle lowering its temperature and if
\eqref{l:POTNT2} is negative the main particle radiates energy
into the heatbath thereby heating it up.  Changes in the
expected total energy along the path of the main particle
therefore show the amount of energy that is being exchanged
between main particle and heatbath.

The important result is that if the main particle resides in
heatbath equilibrium the expectation \eqref{l:POTNT2} should be
zero. In other words if the main particle does not radiate
energy into the heatbath, if the main particle movement has
adopted the heatbath temperature then the total kinetic energy
$\h_k$ must be a conserved quantity.  If radiation occurs in the
form of an energy exchange between main particle and heatbath
then there must a potential or explanatory term describing the
additional physical process.  In this case it must be assumed
that the total kinetic energy and the potential term together
are conserved.

Assume therefore a static potential $\Phi_p\in\Real$, $\Phi_p:\Real
\times [0,\infty] \longmapsto\Real$ such that the (average) energy
exchange can be related to \eqref{l:POTNT2} as
\begin{align}
\label{l:POTNT1}
\begin{split}
&E\left[\Phi_p(x(t+\tau_2),t+\tau_2)\right]
-E\left[\Phi_p(x(t),t)\right]
\\
&=-\frac{m}{2}E\left[ \abs{q}'^2\right]+\frac{m}{2}E\left[
\abs{w_2}^2\right],
\end{split}
\end{align}
then
\begin{gather*}
E\left[\h_k(x(t+\tau_2),t+\tau_2)-\h_k(x(t),t)\right]
\\
=-\left( E\left[\Phi_p(x(t+\tau_2),t+\tau_2)\right]
-E\left[\Phi_p(x(t),t)\right]\right).
\end{gather*}
Defining the total energy of main and heatbath particle as
$\h_T(x(t),t)=\h_k(x(t),t) +\Phi_p(x(t),t)$ then
\eqref{l:POTNT1} implies that
$E\left[\h_T(x(t+\tau_2),t+\tau_2)\right]=
E\left[\h_T(x(t),t)\right]$.  In other words the Hamiltonian
$E\left[\h_T(x(t),t)\right]$ must be time invariant. The
expectation $E\left[\h_k(x(t),t)\right]$ is the kinetic energy
term associated with the motion (of both the main and incident
particle) and $\Phi_p$ is the static potential energy term that
regulates the energy exchange between heatbath and main
particle.

Using Theorem \eqref{l:THMENERG}, definition \eqref{l:DRFTEQ1}
and approximation \eqref{l:DRFTEQ6} it is possible to provide
more detail on the precise form of the total energy total energy
$\h_T$.
\begin{prop}
\label{l:HAMILT12} Let the average velocity of the main particle
be defined as
\begin{align}
\label{l:HAMILT19}
\overline{v}=\overline{v}(x,t)=\frac{b^+(x(t),t)+b^-(x(t),t)}{2},
\end{align}
then in $n$ dimensions the expectation of the total energy
$E\left[\h_T\right]$ can be written as
\begin{align}
\label{l:HAMILT1}
\begin{split}
E&\left[\h_T\right]=E\left[\h_k+\Phi_p\right]
\\
=&\frac{M_T}{2}E\left[\abs{\overline{v}(x(t),t)}^2\right]
+\frac{M_T\sigma^4}{8\gamma^2} E\left[\abs{\frac{1}{\rho}\nabla
\rho}^2\right]
\\
&+E\left[\Phi_p(x(t),t)\right]+ \frac{
2n\epsilon}{\sin^2(\theta)\overline{\tau}}.
\end{split}
\end{align}
with $M_T=M+m$, $\sigma^2=\frac{\epsilon}{M}$.  Here $\rho(x,t)$
is the probability density function for $x(t)$ and the potential
$\Phi_p$ is defined in \eqref{l:POTNT1}.
\end{prop}
\begin{proof}
The terms that requires an explanation are the expectation of
the kinetic energy term $\h_k$ and the resident constant.  Using
\eqref{l:DRFT8} this expectation reduces to
\begin{align*}
E&\left[\frac{\h_k}{M_T}\right]= \frac{1}{2}
E\left[\abs{\frac{v_2+v_1}{2}}^2+ \frac{1}{\gamma^2}
\abs{\frac{v_2-v_1}{2}}^2\right]
\\
=& \frac{1}{2} E\left[\abs{\overline{v}(x(t),t)+
\frac{1}{2}\sigma\frac{\Delta^+z}{\tau_2}+\frac{1}{2}
\sigma\frac{\Delta^-z}{\tau_1} }^2\right]
\\
&+\frac{1}{2\gamma^2}
E\left[\abs{\frac{b^+(x(t),t)-b^-(x(t),t)}{2}
+\frac{1}{2}\sigma\frac{\Delta^+z}{\tau_2}-\frac{1}{2}
\sigma\frac{\Delta^-z}{\tau_1} }^2\right],
\end{align*}
which can be simplified to
\begin{align}
\label{l:HAMILT3}
\begin{split}
E&\left[\frac{\h_k}{M_T}\right]=\frac{1}{2}
E\left[\abs{\overline{v}(x(t),t)}^2\right]
\\
&+\frac{1}{2\gamma^2}
E\left[\abs{\frac{b^+(x(t),t)-b^-(x(t),t)}{2}}^2\right]
+\frac{n}{2}\left(1+\frac{1}{\gamma^2}
\right)\frac{\sigma^2}{\overline{\tau}}
\\
=&\frac{1}{2} E\left[\abs{\overline{v}(x(t),t)}^2\right]
+\frac{\sigma^4}{8\gamma^2} E\left[\abs{\frac{1}{\rho}\nabla
\rho}^2\right]
\\
&+ \frac{n\left(1+\gamma^2\right)\epsilon}{2 m\overline{\tau}},
\end{split}
\end{align}
if $\sigma^2=\frac{\epsilon}{M}$.  Again here $\rho(x,t)$ is the
probability density for $x(t)$ used in equation \eqref{l:DRFT8}
to find an expression for the
$b^+\left(x(t),t\right)-b^-\left(x(t),t\right)$ term.

The constant in $E\left[\h_k\right]/M_T$ is the result of the
fact that
\begin{align*}
&\frac{1}{2}\left(1+\frac{1}{\gamma^2}\right)
\frac{\sigma^2}{\overline{\tau}}=\frac{1}{2}
\left(1+\frac{1}{\gamma^2}\right)
\frac{\epsilon}{M\overline{\tau}}
\\
&=\frac{\left(1+\gamma^2\right)\epsilon}{2\gamma^2
M\overline{\tau}}=\frac{\left(1+\gamma^2\right)
\epsilon}{2m\overline{\tau}},
\end{align*}
a constant depending only on the properties of the incident
particles.  This means that
\begin{align*}
E\left[\h_T\right]\thicksim nM_T\frac{\left(1+\gamma^2\right)
\epsilon}{2m\overline{\tau}}= \frac{n\left(1+\gamma^2\right)^2
\epsilon}{2\gamma^2\overline{\tau}}=\frac{
2n\epsilon}{\sin(\theta)^2\overline{\tau}}
\end{align*}
which explains the constant in \eqref{l:HAMILT1}.  This
reconciles all the terms and the Theorem is proved.
\end{proof}

\begin{rem}
Comparing the diffusion energy term $\epsilon/\overline{\tau}$
in \eqref{l:DRFT27} it is clear that the diffusion energy for
the total kinetic energy $\frac{
2n\epsilon}{\sin(\theta)^2\overline{\tau}}$ in Proposition
\eqref{l:HAMILT12} above is much larger if $\gamma<<1$. This is
due to the presence of the heatbath particle diffusion energy.
\end{rem}

The radiation requirement introduced above imposes a restriction
on the distribution density for the position of the main
particle. This result is summarized in the following Theorem.
\begin{thm}
\label{l:SCHR1}
Let the potential $\Phi_p:\Real \times
[0,\infty] \longmapsto\Real$ defined in equation
\eqref{l:HAMILT1} be such that
\begin{align}
\label{l:SCHROD3}
\frac{d}{dt}E\left[\Phi_p\left(x(t),t\right)\right]=
E\left[\left(\frac{b^++b^-}{2}\right). \nabla\phi\right],
\end{align}
for a suitable potential
$\phi:\Real\times[0,\infty)\longmapsto\Real$ which is at least once
differentiable.  Then the only probability density distribution
$\rho(x,t)$ for the main particle position process $x(t)$ that
allows the total energy
$E\left[\h_T\right]=E\left[\h_k+\Phi_p\right]$ defined in
Proposition \eqref{l:HAMILT12} to be time invariant is derived from
the wave function $\psi(x,t),x\in\Real,t\in[0,\infty)$, $\psi
:\Real\times[0,\infty)\longmapsto\Complex$ such that
$\rho(x,t)=\,|\psi(x,t)\,|^2$, where the wave function satisfies
Schr\"{o}dinger's equation
\begin{gather}
\label{l:SCHROD1}
i\chi\psi(x,t)_t=-\frac{\chi^2}{2M_T}\Delta_x\psi(x,t)+\phi(x,t)\psi(x,t),
\end{gather}
with $\chi=M_T\eta=M_T\frac{\sigma^2}{\gamma}=
\left(\gamma+\frac{1}{\gamma}\right)\epsilon=2\epsilon/\sin(\theta)$
and $\Delta_x=\left(\frac{\partial^2}{\partial
x^2_1},...,\frac{\partial^2}{\partial x^2_n}\right)$.  If the
wave function is written as
$\psi=\psi(x,t)=e^{\frac{R(x,t)+iS(x,t)}{\chi}}$ then the
forward and backward drift can be written as
\begin{align}
\begin{split}
\label{l:DRFT15}
b^\pm(x,t)&=\frac{1}{M_T}\left(\nabla S \pm \gamma
\nabla R\right)
\\
&= \frac{\chi}{M_T}\left(\IM \pm \gamma
\RE\right)\frac{\nabla\psi(x,t)}{\psi(x,t)}.
\end{split}
\end{align}
The constant total energy now equals
\begin{align}
\label{l:HAMILT4}
\begin{split}
E\left[\h_T\right]=\frac{\chi^2}{2M_T}E\left[ \abs{\nabla
\psi}^2\right]+ E\left[\Phi_p\right]+\frac{
2n\epsilon}{\sin^2(\theta)\overline{\tau}}.
\end{split}
\end{align}
\end{thm}
\begin{proof}
The proof here is part of a more general result presented in
Appendix \ref{l:APP5}.  Nelson~\cite{ENELSON2} showed the
relationship between the wave function and an energy functional
similar to \eqref{l:HAMILT1}.  In Nelson ~\cite{ENELSON1} this
result was extended to incorporate a potential term as
introduced in \eqref{l:POTNT1} though condition
\eqref{l:SCHROD3} is different. The proof in Appendix
\ref{l:APP5} is for a more general energy expression
incorporating the presence of the collision scattering matrix
$Z$ but runs similar to the presentation in Nelson
~\cite{ENELSON1}. Carlen~\cite{ECARL1} demonstrated that the
stochastic differential equation \eqref{l:DRFTDEF1} admits a
weak solution if the potential $\phi(x,t)$ belongs to a class
Kato-Rellich potential.
\end{proof}

\begin{rem}
Equation \eqref{l:HAMILT4} above incorporates the total energy of
the system, i.e. the energy of the main particle as well as the
energy of the incident particle.  The mass in the equation refers to
the combined mass of the system $M_T=M+m$ rather than the main
particle and it is not specified how this energy is distributed
between the two particles.
\end{rem}
\begin{rem}
Under conditions \eqref{l:DRFTDEF3}, \eqref{l:DRFTDEF4} and
sufficiently small $\overline{\tau}$ the discrete collision
process \eqref{l:DRFTEQ1} can be approximated by the continuous
stochastic differential equation \eqref{l:DRFTDEF1}.  Hence both
the momenta and (forward and backward) energies
\eqref{l:DRFTDEF0} can be approximated using \eqref{l:DRFT15}
see Carlen~\cite{ECARL3}. As a result $\h_T$ in
\eqref{l:HAMILT1} is now a well defined estimate of the combined
energy.
\end{rem}
There is no reference in this approach to quantum mechanics or a
stochastic interpretation of quantum mechanics. Though the proof
in Appendix  \ref{l:APP6} is quite similar to the stochastic
mechanics approach presented in Nelson~\cite{ENELSON1}, Theorem
\eqref{l:SCHR1} above does not reproduce an interpretation of
quantum mechanics at least not perfectly as the scaling here is
different.  Obviously, if Planck's constant were chosen as
$\hbar=\chi$ then \eqref{l:SCHROD1} reduces to the wave-function
for a particle, however, the forward and backward drifts in
\eqref{l:DRFT15} explode if the mass ratio $\gamma$ were allowed
to approach zero.  In addition the mass term $M_T$ incorporates
information on both the main and heatbath particle and the
energy \eqref{l:HAMILT4} refers to the combined kinetic and
potential energy.

\subsection{The Brownian Motion}
This continued example demonstrates what happens to the total
kinetic energy for a particle following a Brownian motion if no
conservation law applies. The Gaussian distribution is the
result of the diffusion equation with a zero forward drift
$b^+(x,t)$ so that the probability density and backward drift
equal
\begin{align*}
&\rho(x,t)=\frac{1}{\sigma\sqrt{2\pi t}}
e^{-\frac{(x-\mu)^2}{2\sigma^2t}},
\\
&b^-(x,t)=\frac{(x-\mu)}{t},
\end{align*}
then equation \eqref{l:HAMILT3} reduces to
\begin{align*}
E[\h_k]&=M_TE\left[\abs{x-\mu}^2\right]\left(\frac{1+\gamma^2}{8\gamma^2t^2}\right)+
\frac{M_T\epsilon}{2\overline{\tau}m}
\\
&=M_T\sigma^2\left(\frac{1+\gamma^2}{8\gamma^2t}\right)+
\frac{M_T\epsilon}{2\overline{\tau}m}
\\
&=\frac{\epsilon}
{4\gamma\sin(\theta)\overline{\tau}}\left(1+4\frac{\overline{\tau}}{nt}\right),
\end{align*}
so the total energy in classical diffusion decreases
continuously though there is no evidence of a potential.
Applying \eqref{l:POTNT2} shows that
\begin{align*}
&E\left[\Delta \h_k(x(t),t)\right]
\\
&=E\left[\h_k(x(t+\tau_2),t+\tau_2) -\h_k(x(t),t)\right]
\\
&= \frac{\epsilon}
{\gamma\sin(\theta)n}E\left[\frac{1}{\left(t+\tau\right)}
-\frac{1}{t}\right]
\\
&\approx-\frac{\epsilon\overline{\tau}}
{\gamma\sin(\theta)nt^2},
\end{align*}
which implies that a particle following a Brownian path always
radiates energy into the heatbath.  Brownian motion seems a
non-equilibrium solution and a main particle can only perform a
Brownian path due to a hidden potential that forces the motion
(a dampened time-dependent oscillator would suffice). The effect
mitigates quickly with time $t$ increasing as a multiple of the
collision time. In a very dense heatbath the inter-collision
time shortens so for macroscopic objects the moment to
equilibrium must be almost instantaneous. Notice that the effect
depends on the mass ratio and is not just a function of the main
particle mass $M$.

To understand why for a Brownian Motion the average total energy
changes consider that classical diffusion is a limiting case
(limit in time and space dimension) of a particle stepping
forward over a lattice grid with equal probability.  If the
process has moved a large distance to position $x$ in a short
amount of time the stochastic path of the particle has a strong
backward drift in the direction of the origin. However, the
forward drift is zero so for these paths the contribution to the
total energy is large and ultimately too large to limit the
average total energy. A path produced by a time invariant
average total energy realizes that the backward drift is large
and adapts the forward drift to point in a similar direction as
the backward drift. Changing the drift will then reduce energy
in the regions where $x$ becomes large by reducing the
acceleration.

For the next example the following Corollary will be useful.
\begin{cor}
\label{l:SCHR2}
The drift for the heatbath motion is given by
\begin{align*}
g^\pm(x,t)&=\frac{1}{M_T}\left(\nabla S\mp \frac{1}{\gamma}\nabla
R\right)
\\
&= \frac{\chi}{M_T}\left(\IM \mp \frac{1}{\gamma}
\RE\right)\frac{\nabla\psi(x,t)}{\psi(x,t)},
\end{align*}
\end{cor}
\begin{proof}
Using equation \eqref{DELTA1a} the relationship between the
heatbath particles and main particle can be written as
\begin{align*}
&\begin{pmatrix}
g^+(x,t) \\
g^-(x,t)
\end{pmatrix}
= \frac{1}{\gamma\sin(\theta)}
\begin{pmatrix}
-\cos(\theta) & 1 \\
1 & -\cos(\theta)
\end{pmatrix}
\begin{pmatrix}
b^+(x,t) \\
b^-(x,t)
\end{pmatrix}
\\
&=\frac{1}{M_T\gamma\sin(\theta)}
\begin{pmatrix}
-\cos(\theta) & 1 \\
1 & -\cos(\theta)
\end{pmatrix}
\begin{pmatrix}
\nabla S+\gamma \nabla R \\
\nabla S-\gamma \nabla R
\end{pmatrix},
\end{align*}
so clearly
\begin{align*}
g^+(x,t)=&\frac{1}{\gamma M_T \sin(\theta)}
\begin{pmatrix}
(1-\cos(\theta))\left(\nabla S\right)
\\-\gamma
(1+\cos(\theta)) \nabla R
\end{pmatrix}
\\
=&\frac{1}{M_T}\left(\nabla S -\frac{1}{\gamma}\nabla R\right),
\\
g^-(x,t)=&\frac{1}{M_T}\left(\nabla S+\frac{1}{\gamma}\nabla
R\right),
\end{align*}
and the Corollary is proved.
\end{proof}

\begin{rem}
\label{l:ENERG9} Analogous to Lemma \eqref{l:ENERG8} and
equation \eqref{l:DRFT27} the average momentum and energy for
the heatbath particle can be determined as
\begin{align}
\label{l:DRFT24}
\begin{split}
&E\left[\frac{\Delta^+w(t)}{\tau_2}\Bigg\vert
x(t)\right]=g^+(x(t),t), E\left[\frac{\Delta^-
w(t)}{\tau_1}\Bigg\vert x(t)\right]=g^-(x(t),t),
\\
&\h^+_m\left(x(t),t\right)=\frac{m}{2}E\left[\left(\frac{\Delta^+
w(t)}{\tau_2}\right)^2\Bigg\vert x(t)\right]=
\frac{m}{2}g^+(x(t),t)^2+\frac{\gamma^2\epsilon}{\alpha^2\overline{\tau}},
\\
&\h^-_m\left(x(t),t\right)=\frac{m}{2}E\left[\left(\frac{\Delta^-
w(t)}{\tau_1}\right)^2\Bigg\vert x(t)\right]=
\frac{m}{2}g^-(x(t),t)^2+\frac{\gamma^2\epsilon}{\alpha^2\overline{\tau}},
\end{split}
\end{align}
since $m\sigma^2/(2\alpha^2\overline{\tau})=
\gamma^2\epsilon/(\alpha^2\overline{\tau})$.
\end{rem}

This Corollary together with \eqref{a:THM9} completes a
description of the behavior of the main and heatbath particles.
If the main particle motion behaves like a martingale where its
energy exchange with the heatbath is derived from a potential
$\Phi_p$ then its probability density and (forward and backward)
drifts are derived from Schr\"{o}dinger's equation
\eqref{l:SCHROD1} and equations \eqref{l:DRFT15}. The colliding
heatbath particle has a drift as specified in Corollary
\eqref{l:SCHR2} and exhibits a high degree of correlation
between its backward and forward momentum as shown in Theorem
\eqref{a:THM9}.  The heatbath particle motion is also highly
correlated with the motion of the main particle and can not be a
martingale process itself.

The following example shows a solution to equations
\eqref{l:SCHROD1} and \eqref{l:DRFT15} for the case where the
path of the main particle can be represented by a generic
Gaussian process.

\subsection{The QM Wave Packet}
\label{l:QM1} Applying Theorem \eqref{l:SCHR1} and Corollary
\eqref{l:SCHR2} to the Gaussian function wave packet in one
dimension shows that the wave function can be represented as a
Gaussian superposition of single momentum solutions to the wave
equation \eqref{l:SCHROD1}. Hence
\begin{align*}
\psi(x,t)=&C\int_{-\infty}^{\infty}
e^{-\frac{\left(p-p_0\right)^2}{2\sigma^2_e}}
e^{\frac{i}{\chi}\left(px-\frac{p^2t}{2M_T}\right)}dp
\\
= &C\int_{-\infty}^{\infty} e^{-\frac{z^2}{2\sigma^2_e}}
e^{\frac{i}{\chi}\left(x(z+p_0)-\frac{(z+p_0)^2t}{2M_T}\right)}dz
\\
=&Ce^{\frac{i}{\chi}xp_0-\frac{i}{\chi}\frac{p^2_0t}{2M_T}}
\int_{-\infty}^{\infty} e^{iz\mu(x,t) -\frac{z^2}{2}\Gamma(t)},
\end{align*}
where
\begin{align*}
&\mu(x,t)= \frac{1}{\chi}\left(x-\frac{p_0t}{M_T}\right),
\\
&\Gamma(t)=\left(\frac{1}{\sigma^2_e}+i\frac{t}{\chi
M_T}\right).
\end{align*}
Hence
\begin{align*}
\psi(x,t) =C'(t)e^{ip_0\mu(x,t)}
e^{-\frac{\mu(x,t)^2\overline{\Gamma(t)}} {2Z_\Gamma(t)}},
\end{align*}
where
\begin{align}
\label{l:DRFT28}
&Z_\Gamma(t)=\Gamma(t)\overline{\Gamma(t)}
=\left(\frac{1}{\sigma^4_e}+\alpha^2
t^2\right),\alpha=\frac{1}{\chi M_T},
\end{align}
and where
$C'=\left(\sqrt{\Gamma(t)\chi\sigma_e\sqrt{\pi}}\right)^{-1}
e^{i\frac{p^2_0t}{2\chi M_T}}$.  This constant is chosen to
insure that $\rho(x,t)=\abs{\psi(x,t)}^2$ is a probability
density.

Rewriting $\Gamma(t)=\sqrt{Z_\Gamma(t)}e^{i\arccos\left(\frac{1}
{\sigma^2_e\sqrt{Z_\Gamma(t)}}\right)}$ the wave function
$\psi(x,t)$ can be represented as
$\psi=\psi(x,t)=e^{\frac{R(x,t)+iS(x,t)}{\chi}}$ where
\begin{align*}
R(x,t)&=-\frac{\left(x-\frac{p_0t}{M_T}\right)^2}
{2\chi^2\sigma^2_eZ_\Gamma(t)}-
\frac{1}{2}\log\left(\chi\sigma_e\sqrt{Z_\Gamma(t)\pi}\right),
\\
S(x,t)&=\frac{\alpha t}{2\chi^2Z_\Gamma(t)
}\left(x-\frac{p_0t}{M_T}\right)^2+\left(
\frac{xp_0}{\chi}-\frac{p^2_0t}{2\chi M_T}\right)
\\
&-\frac{1}{2}\arccos\left(\frac{1}
{\sigma^2_e\sqrt{Z_\Gamma(t)}}\right),
\end{align*}
with $\alpha$ and $Z_\Gamma(t)$ as defined in \eqref{l:DRFT28}.
Finally, then
\begin{align*}
&\frac{\partial}{\partial
x}R(x,t)=-\frac{\left(x-\frac{p_0t}{M_T}\right)}
{\chi^2\sigma^2_eZ_\Gamma(t)},\\
&\frac{\partial}{\partial x}S(x,t)=\frac{\alpha
t}{\chi^2Z_\Gamma(t)} \left(x-\frac{p_0t}{M_T}\right)+p_0.
\end{align*}
Hence $x(t)$ is a (Gaussian) random variable such that $E[x(t)]=
\frac{p_0t}{M_T}$,$var(x(t))=Z_\Gamma(t)\chi^2\sigma^2_e$.

Using \eqref{l:DRFT15} the drift functions become
\begin{align}
\label{l:DRFT23}
&b^\pm(x,t)= \frac{1}{\chi M_T Z_\Gamma(t)}
\left(x-\frac{p_0t}{M_T}\right)\left(\alpha t \mp
\frac{\gamma}{\sigma^2_e}\right) + \frac{p_0}{M_T},
\end{align}
and the total energy becomes
\begin{align}
\label{l:DRFT20}
\begin{split}
E[\h_k]&=\frac{1}{2M_T}E\left[R^2_x+S^2_x\right]+\frac{\left(1+\gamma^2\right)^2
\epsilon}{2\gamma^2\overline{\tau}}
\\
&=\frac{\sigma^2_e}{2M_T}+\frac{p^2_0}{2M_T}+ \frac{
2\epsilon}{\sin(\theta)^2\overline{\tau}}.
\end{split}
\end{align}
Hence the energy consists of a contribution due to the energy
dispersion controlled by $\sigma_e$, the mean kinetic energy term
with $p_0$ and the diffusion term proportional to the variance
$\sigma^2$.

Now the forward and backward energy \eqref{l:DRFT6} equal
\begin{align}
\label{l:DRFT21}
E\left[\h^\pm_M\left(x(t),t\right)\right]=\frac{M\sigma^2_e}{2M^2_T
Z_\Gamma(t)}\left(\alpha t\pm\frac{\gamma}{\sigma^2_e}\right)^2
+M\frac{p^2_0}{2M^2_T}+ \frac{\epsilon}{\overline{\tau}},
\end{align}
so that there is a negative energy transfer
$E\left[\h^+_M\left(x(t),t\right)\right]-
E\left[\h^-_M\left(x(t),t\right)\right]\thickapprox-\frac{M\alpha
t\gamma}{4M^2_T Z_\Gamma(t)} \uparrow 0$ which decreases with
time. This shows that the heatbath is absorbing energy generated
by the main particle initially but this approaches zero
depending on the size of the term $\alpha=1/(\chi M_T)$.  For
large objects $\gamma \thickapprox 0$ so then this heat loss for
the main particle will be negligible.  Since the total energy is
conserved this argument suggests that the heatbath must be
gaining energy as a function of time which indeed will be shown
below. This is an example of a case where the total energy is
conserved as demanded by the non-radiation condition but only
because both component energies change in time. To conserve
$E\left[\h_T\right]$ is therefore not equivalent to demanding
that the backward and forward energies for the main particle are
equal.

The average energy for the main particle equals
\begin{align*}
E\left[\h_{M,avg}\left(x(t),t\right)\right]&=
\frac{1}{2}E\left[\h^+_M\left(x(t),t\right)+\h^-_M\left(x(t),
t\right)\right]
\\
&=\frac{M\sigma^2_e}{2M^2_T Z_\Gamma(t)}\left(\alpha^2
t^2+\frac{\gamma^2}{\sigma^4_e}\right) +M\frac{p^2_0}{2M^2_T}+
\frac{\epsilon}{\overline{\tau}}.
\end{align*}
In the case that $\alpha t\thickapprox 0$ - equivalent to a very
short time step $t \thickapprox 0$ or a very small mass ratio
$\gamma \thickapprox 0$ - the average main particle energy
reduces to
\begin{subequations}
\begin{align}
\label{l:DRFT29a}
\begin{split}
E\left[\h_{M,avg}\left(x(0),0\right)\right]&=
\frac{1}{2}E\left[\h^+_M(x(0),0)+\h^-_M(x(0),0)\right]
\\
&=\frac{M\gamma^2}{2M^2_T Z_\Gamma(t)\sigma^2_e}
+M\frac{p^2_0}{2M^2_T}+ \frac{\epsilon}{\overline{\tau}}
\\
&\thickapprox M\frac{p^2_0}{2M^2_T}+
\frac{\epsilon}{\overline{\tau}}.
\end{split}
\end{align}

On  the other hand from \eqref{l:DRFT21} it is clear that
\begin{align}
\label{l:DRFT29b}
\begin{split}
E\left[\h_{M,avg}\left(x(\infty),\infty\right)\right]
&=\frac{1}{2}E\left[\h^+_M(x(\infty),\infty)+
\h^-_M(x(\infty),\infty)\right]
\\
&= \frac{M \sigma^2_e}{2M^2_T} +M\frac{p^2_0}{2M^2_T}+
\frac{\epsilon}{\overline{\tau}},
\end{split}
\end{align}
\end{subequations}
since $\frac{\alpha^2t}{Z_\Gamma(t)}\rightarrow 1$.  Notice that
the terms $M \sigma^2_e/(2M^2_T)$ and $ Mp^2_0/(2M^2_T)$ in the
limit energies
$E\left[\h_{M,avg}\left(x(\infty),\infty\right)\right]$ and
$E\left[\h_{M,avg}\left(x(0),0\right)\right]$ are close in size
to the first two terms in \eqref{l:DRFT20} as long as $\gamma$
is small. However the diffusion term is \eqref{l:DRFT20} is very
large in comparison to the diffusion term
$\epsilon/\overline{\tau}$ in
$E\left[\h_{M,avg}\left(x(\infty),\infty\right)\right]$ and
$E\left[\h_{M,avg}\left(x(0),0\right)\right]$ see the discussion
below.

To investigate the heatbath apply Corollary \eqref{l:SCHR2} to
determine that for the heatbath particle
\begin{align*}
&g^\pm(x,t)= \frac{1}{\chi M_T Z_\Gamma(t)}
\left(x-\frac{p_0t}{M_T}\right)\left(\alpha t \mp
\frac{1}{\gamma\sigma^2_e}\right) + \frac{p_0}{M_T},
\end{align*}
so that
\begin{align}
\label{l:DRFT22}
E\left[\h^{\pm}_m(x(t),t)\right]=\frac{m\sigma^2_e}{4M^2_T
Z_\Gamma(t)}\left(\alpha t \mp
\frac{1}{\gamma\sigma^2_e}\right)^2 +m\frac{p^2_0}{2M^2_T}+
\frac{m\omega^2}{\overline{\tau}},
\end{align}
and
\begin{align*}
E\left[\h_{m,avg}(x(t),t)\right]=
&\frac{1}{2}E\left[\h^+_m(x(t),t)+\h^+_m(x(t),t)\right]
\\
=&\frac{m\sigma^2_e}{2M^2_T Z_\Gamma(t)}\left(\alpha^2 t^2 +
\frac{1}{\gamma^2\sigma^4_e}\right) +m\frac{p^2_0}{2M^2_T}+
\frac{m\omega^2}{\overline{\tau}}.
\end{align*}
Hence
\begin{subequations}
\begin{align}
\label{l:DRFT30a}
\begin{split}
E\left[\h_{m,avg}(x(0),0)\right]
&=\frac{1}{2}E\left[\h^+_m(x(0),0)+\h^+_m(x(0),0)\right]
\\
&=\frac{M\sigma^2_e}{2M^2_T } +m\frac{p^2_0}{2M^2_T}+
\frac{m\omega^2}{\overline{\tau}},
\end{split}
\end{align}
and for large time $t$ the average heatbath energy becomes
\begin{align}
\label{l:DRFT30b}
\begin{split}
E\left[\h_{m,avg}(x(\infty),\infty)\right]
&=\frac{1}{2}E\left[\h^+_m(x(\infty),\infty)
+\h^+_m(x(\infty),\infty)\right]
\\
&=\frac{m\sigma^2_e}{2M^2_T }+m\frac{p^2_0}{2M^2_T}+
\frac{m\omega^2}{\overline{\tau}},
\end{split}
\end{align}
\end{subequations}
which indeed decreases from $M\sigma_e^2/(2M^2_T)$ to
$m\sigma_e^2/(2M^2_T)$ as time progresses.

It is interesting to compare the variance contributions to the
energies \eqref{l:DRFT20}, \eqref{l:DRFT21} and
\eqref{l:DRFT22}. Recall that
$m\omega^2/\overline{\tau}=m\sigma^2/\left(2\alpha^2\right)$. If
$\gamma << 1$ then $m\omega^2/\overline{\tau} >>
\epsilon/\overline{\tau}$ so the diffusion energy contribution
to the heatbath particle is much larger than the diffusion
energy for the main particle.  The energy due to the dispersion
term $\sigma^2_e$ is at first a small contribution in the main
particle energy in \eqref{l:DRFT21} but then increases (see
\eqref{l:DRFT29a}, \eqref{l:DRFT29b}) while this term in the
heatbath particle energy \eqref{l:DRFT22} does exactly the
opposite (see \eqref{l:DRFT30a}, \eqref{l:DRFT30b}). Both the
main and heatbath particle energy are proportional to the same
kinetic energy contribution weighted with their respective
masses.  This implies that the kinetic energy is almost
exclusively carried by the main particle and for small $\gamma$
very little kinetic energy is carried by the heatbath particle.
In fact in the small $\gamma$ limit the heatbath particles are
all moving through the heatbath with energy $\h_{m,avg} \approx
\sigma^2/\left(2\alpha^2\overline{\tau}\right)\approx
\epsilon/\left(2\gamma^2\overline{\tau}\right)$ without being
affected by the presence of the main particle. After the
collision the heatbath particle emerges with exactly the same
speed and opposite direction.

A curious consequence of this example seems to be that the
statistical characteristics of the heatbath particles are
affected by the energy constraint in a similar fashion as the
main particle. The time dependent term in \eqref{l:DRFT22} shows
a behavior that reflects the main particle and \eqref{l:DRFT22}
carries a kinetic contribution that is is also present in
\eqref{l:DRFT21}.  This is slightly unrealistic as it seems to
imply that the heatbath particle behavior depends additionally
on the main particle kinetic energy rather than external factors
alone. Though the effect becomes very small as the mass ratio
$\gamma$ decreases the only way to render equation
\eqref{l:DRFT22} time-independence is to assume that the average
collision time $\overline{\tau}$ changes as a function of energy
or correlate the drift dependence between pre- and post
collision velocities. The next Section addresses these
constructions.

\section{The Minkowski Invariant}

The example in the previous Section showed that a constant total
kinetic energy can be achieved but the individual main and
heatbath particles energies display time dependent behavior.
This Section attempts to investigate the momentum constraint
\eqref{l:DRFT9} to impose energy conservation embedded in the
heatbath particles through the point of collision.  It will be
shown that the relationship between the inter-particle collision
time and the total energy leads to a type of geometric
invariant.

To obtain an equation similar to \eqref{l:HAMILT1} for the
momentum constraint substitute \eqref{l:DRFTEQ1} into
\eqref{l:DRFT9} to find
\begin{align}
\label{l:REL1}
\begin{split}
&\frac{1}{2}E\left[w^2_1+w^2_2\right] =\frac{1}{2}
E\left[\abs{\overline{v}(x(t),t)}^2\right]
\\
&+\frac{1}{2\gamma^4}
E\left[\abs{\frac{b^+(x(t),t)-b^-(x(t),t)}{2}}^2\right] +
\frac{n\epsilon}{2\alpha^2M\overline{\tau}},
\end{split}
\end{align}
where $b^+=b^+(x(t),t)$ and $b^-=b^-(x(t),t)$ are the usual
forward and backward drifts respectively,
$\overline{v}\left(x(t),t\right)=\left(b^+(x(t),t)
+b^-(x(t),t)\right)/2$ as defined in \eqref{l:HAMILT19} and
$\alpha=\gamma^4/(1+\gamma^4)$. Assuming that the underlying
process is a martingale and applying equation \eqref{l:DRFT2c}
reduces this to
\begin{align}
\label{l:REL2}
\begin{split}
&\frac{1}{2}E\left[\abs{w_1}^2+\abs{w_2}^2\right] =\frac{1}{2}
E\left[\abs{\overline{v}(x(t),t)}^2\right] \\
&+\frac{\sigma^4}{8\gamma^4}
E\left[\left(\frac{1}{\rho}\frac{\partial \rho}{\partial
x}\right)^2\right] +
\frac{n\epsilon}{2\alpha^2M\overline{\tau}},
\end{split}
\end{align}
The constant in this expression is the result of the fact that
in one dimension
\begin{align*}
&\frac{1}{4}\left(1+\frac{1}{\gamma^4}\right)E\left[
\left(\frac{\sigma\Delta^+x(t)}{\tau_2}\right)^2\right]=
\frac{1}{4}\left(1+\frac{1}{\gamma^4}\right)E\left[
\frac{\sigma^2}{\tau_2}\right]
\\
=&\frac{1}{2}
\left(1+\frac{1}{\gamma^4}\right)\frac{\sigma^2}{\overline{\tau}}
=\frac{1}{2}
\left(1+\frac{1}{\gamma^4}\right)\frac{\epsilon}{M\overline{\tau}}
=\frac{1}{2\frac{\gamma^4}{1+\gamma^4}}\frac{\epsilon}{M\overline{\tau}}
=\frac{\epsilon}{2\alpha^2M\overline{\tau}},
\end{align*}
and multiplying with $n$ for the multi-dimensional case yields
the value for the constant in \eqref{l:REL2}.

In a heatbath where the main particle is in equilibrium with the
incident particles it should be expected that the post-collision
energy for the heatbath particle is equal to the pre-collision
when averaged over all paths and positions.  Otherwise there
will be an average heat loss or gain for the main particle. So
if the main particle is in equilibrium it is expected that
$E\left[w^2_2\right]=E\left[w^2_1\right]=c^2$ where $c$ is the
average velocity of the heatbath particles. This means that
$\frac{1}{2}\left(E\left[w^2_2\right]+E\left[w^2_1\right]\right)=c^2$
and defining $\Delta\overline{x}_{\overline{\tau}}$ as
\begin{align}
\label{l:REL11}
\abs{\Delta \overline{x}_{\overline{\tau}}}^2
=\frac{1}{2}
\begin{pmatrix}
E\left[\abs{\overline{v}\left(x(t),t\right)}^2\right]
\\
+\frac{1}{\gamma^4}
E\left[\abs{\frac{b^+(x(t),t)-b^-(x(t),t)}{2}}^2\right]
\end{pmatrix}
\overline{\tau}^2,
\end{align}
reduces equation \eqref{l:REL2} to
\begin{align*}
c^2 = \frac{\abs{\Delta\overline{x}_{\overline{\tau}}}^2}
{\overline{\tau}^2}
+\frac{n\epsilon}{2\alpha^2M\overline{\tau}}.
\end{align*}
Hence
\begin{align}
\label{l:REL12}
c^2\overline{\tau}^2=\abs{\Delta\overline{x}_{\overline{\tau}}}^2+
\frac{n\epsilon\overline{\tau}}{2\alpha^2M}.
\end{align}
The results can now be summarized in the following theorem
\begin{thm}
\label{l:HAMILT18}
Let $x(t) \in \Real^n$ be the coordinate
process of the main particle and define $\abs{\Delta
\overline{x}_{\overline{\tau}}}^2$ as in equation
\eqref{l:REL11}. Assume that the main particle is not radiating
energy or receiving energy so that the backward and forward
velocities of the heatbath particle are equal, e.g.
$E\left[w_2^2\right]=E\left[w_1^2\right]=c^2$. Then the drifts
must be such that $c^2\overline{\tau}^2-\abs{\Delta
\overline{x}_{\overline{\tau}}}^2$ forms an invariant of the
motion, i.e.
\begin{align}
\label{l:REL3}
\sum_{j=0}^{j=N_{\overline{\tau}}}
\left(c^2\overline{\tau}^2-
\abs{\Delta\overline{x}_{\overline{\tau}}}^2\right) =
\frac{n\epsilon T}{2\alpha^2M}=\text{constant},
\end{align}
where $N_{\overline{\tau}}$ is the average number of collisions
such that $E\left[\sum_{j=1}^{j=N_{\tau}}\tau_j\right]
=N_{\overline{\tau}}\overline{\tau}=T$.
\end{thm}
\begin{proof}
The proof is simply that \eqref{l:REL12} implies that
\begin{align*}
\sum_{j=0}^{j=N_{\overline{\tau}}} \left(c^2\overline{\tau}^2-
\abs{\Delta\overline{x}_{\overline{\tau}}}^2\right) =
\sum_{j=0}^{j=N_{\overline{\tau}}}
\frac{n\epsilon\overline{\tau}}{2\alpha^2M} =\frac{n\epsilon
T}{2\alpha^2M},
\end{align*}
since $N_{\overline{\tau}}$ is the (average) number of
inter-particle collisions between time $0$ and time $T$.  The
righthand side is independent of the average inter-particle
collision time.
\end{proof}
\begin{rem}
If $\abs{\Delta \overline{x}_{\overline{\tau}_0}}^2 \approx 0 $
and $c^2$ is large then the solution to equation \eqref{l:REL3}
equals
\begin{align}
\label{l:VELC1} \overline{\tau}_0=
\frac{n\epsilon}{2c^2\alpha^2M},
\end{align}
so there is a reference average inter-particle collision time
for a slow-moving (stationary) main particle in the heatbath.
Combining this and \eqref{l:REL3} shows immediately that
\begin{align*}
\sum_{j=0}^{j=N_{\overline{\tau}}} \left(c^2\overline{\tau}^2-
\abs{\Delta\overline{x}_{\overline{\tau}}}^2\right) =
c^2T\overline{\tau}_0.
\end{align*}
\end{rem}

To gain some insight into this Minkowski type relativistic
invariant \eqref{l:REL3} consider the example of the Gaussian
wave packet \eqref{l:QM1} in the previous Section. From
equations \eqref{l:DRFT23} it is clear that
\begin{align*}
&\frac{1}{2}\left(b^+(x,t)+b^-(x,t)\right)= \frac{\alpha t}{\chi
M_T Z_\Gamma(t)} \left(x-\frac{p_0t}{M_T}\right) +
\frac{p_0}{M_T} \thickapprox \frac{p_0}{M_T},
\\
&\frac{1}{2}\left(b^+(x,t)-b^-(x,t)\right)= \frac{\gamma}{\chi
M_T Z_\Gamma(t)\sigma^2_e} \left(x-\frac{p_0t}{M_T}\right)
\thickapprox 0,
\end{align*}
since $\alpha t/Z_\Gamma(t) \thickapprox 0 $ and $1 /Z_\Gamma(t)
\thickapprox 0 $ for sufficiently large time $t$. As a result
equation \eqref{l:REL3} becomes
\begin{align*}
\sum_{j=0}^{j=N_{\overline{\tau}}}\left(c^2\overline{\tau}^2-
\left(\frac{p_0}{M_T}\right)^2\overline{\tau}^2\right) =
\overline{\tau}\left(c^2- \left(\frac{p_0}{M_T}\right)^2\right)=
\frac{\epsilon T}{\alpha^2M}.
\end{align*}
This suggests that the average inter particle collision time
$\overline{\tau}$ depends on the mean motion of the main
particle specifically $\overline{\tau} \thicksim
\left(c^2-(p_0/M)^2\right)^{-1}$. However this presumes that the
drift $\abs{\Delta\overline{x}_{\overline{\tau}}}$ does not
depend on the average inter-particle collision time
$\overline{\tau}=\overline{\tau}
\left(\abs{\Delta\overline{x}_{\overline{\tau}}}\right)$ which
is unlikely to be reasonable.  In general the mean
inter-collision time and the drift will depend on each other and
the most straightforward approach is to explore a linear
relationship.

There does not seem to be a separate mechanism for introducing a
dependent inter-collision time such that \eqref{l:REL3} holds
except possibly a statistical correlation between the forward
and backward velocities $\Delta^+z$ and $\Delta^-z$.  This
creates a separate relationship between the correlation
$corr\left(\frac{\Delta^+z}{\tau_2},\frac{\Delta^-z}{\tau_1}
\right)=\rho_{\Delta^+z,\Delta^-z}I$ ($I$ being the unit matrix)
of the coordinate process $x(t)$ and the inter-particle
collision time $\overline{\tau}$. Returning to \eqref{l:REL1}
and introducing the correlation term shows that
\begin{align}
\label{l:VELC2}
\begin{split}
c^2 = &\frac{1}{2} E\left[\abs{v\left(x(t),t\right)}^2\right]
\\
&+\frac{1}{8\gamma^4}E\left[\abs{b^+(x(t),t)-b^-(x(t),t)}^2\right]
\\
&+\frac{\sigma^2}{8}E\left[\abs{\frac{\Delta^+z}{\tau_2}
+\frac{\Delta^-z}{\tau_1}}^2\right]
+\frac{\sigma^2}{8\gamma^4}E\left[\abs{\frac{\Delta^+z}{\tau_2}
-\frac{\Delta^-z}{\tau_1}}^2\right]
\\
=&\frac{\abs{\Delta\overline{x}_
{\overline{\tau}}}^2}{\overline{\tau}^2}
+\frac{n\epsilon}{\alpha^2M\overline{\tau}}
+\left(1-\frac{1}{\gamma^4}\right)\frac{n\epsilon
\rho_{\Delta^+z\Delta^-z}}{2M\overline{\tau}}
\\
=&\frac{\abs{\Delta\overline{x}_
{\overline{\tau}}}^2}{\overline{\tau}^2}
+\frac{n\epsilon}{2\alpha^2M\overline{\tau}}\left(
1-\rho^2_v\right),
\end{split}
\end{align}
if $\rho^2_v=\left(1-2\alpha^2\right)\rho_{\Delta^+z\Delta^-z}$.
Multiplying with $\overline{\tau}^2$ and dividing by $\left(
1-\rho^2_v\right)$ this can be written as
\begin{align*}
c^2\tau_v^2 =\abs{\Delta \overline{x}_v}^2
+\frac{n\epsilon\overline{\tau}}{2\alpha^2M},
\end{align*}
where
\begin{align*}
&\overline{\tau}_v=\frac{\overline{\tau}}{\sqrt{1-\rho^2_v}},
\\
&\abs{\Delta \overline{x}_v}^2 = \frac{1}{2}
\begin{pmatrix}
E\left[\abs{\overline{v}\left(x(t),t\right)}^2\right]
\\
+\frac{1}{\gamma^4}
E\left[\abs{\frac{b^+(x(t),t)-b^-(x(t),t)}{2}}^2\right]
\end{pmatrix}\overline{\tau_v}^2.
\end{align*}

This means that \eqref{l:REL3} reduces to
\begin{align}
\label{l:REL4} \sum_{j=0}^{j=N_{\overline{\tau}}}
\left(c^2\overline{\tau}_v^2-\abs{\Delta
\overline{x}_v}^2\right) =\frac{n\epsilon T}{2\alpha^2M},
\end{align}
where now the summation ranges over $N_{\overline{\tau}}$ rather
than over $N_{\overline{\tau}_v}$.   To obtain an estimate of
the size of the correlation assume that
$\overline{\tau}\approx\overline{\tau_0}$, write
$v^2=\abs{\Delta
\overline{x}_{\overline{\tau}_v}}^2/\overline{\tau}_v^2$, then
substitute \eqref{l:VELC1} into \eqref{l:REL4} to yield
\begin{align*}
\sum_{j=0}^{j=N_{\overline{\tau}_0}}
\left(c^2\overline{\tau}_v^2-\abs{\Delta
\overline{x}_v}^2\right)=\frac{T}{\tau_0}\left(c^2\overline{\tau}_v^2-
v^2\overline{\tau}_v^2\right)=\frac{n\epsilon
T}{2\alpha^2M}=c^2\tau_0T,
\end{align*}
which is equivalent to
\begin{align}
\label{l:REL5} \frac{\overline{\tau}_v^2}{\tau^2_0}=\frac{1}{1-
\frac{v^2}{c^2}}.
\end{align}
As a result then equation \eqref{l:VELC2} shows immediately that
$\rho^2_v\approx v^2/c^2$.

Interpreting $\overline{\tau}$ and $\abs{\Delta
\overline{x}_v}^2$ as differentials equation \eqref{l:REL4}
becomes the well known Minkowski invariant in Relativity Theory
and its solution is the (linear) Lorentz transformation which
relates elapsed time versus motion as perceived in different
reference frames. The following Theorem summarizes the results
and shows the simple linear transformation.
\begin{thm}
\label{l:HAMILT20} Let the main particle in the heatbath be in
energetic equilibrium with the heatbath so that
$E\left[w_2^2\right]=E\left[w_1^2\right]=c^2$ at any collision
point $t$. Now let the forward and backward velocities
$\Delta^+z$ and $\Delta^-z$ be statistically correlated so that
$corr\left(\frac{\Delta^+z}{\tau_2},\frac{\Delta^-z}{\tau_1}\right)
=\rho_{\Delta^+z,\Delta^-z}I$ and let
$\rho^2_v=\left(1-2\alpha^2\right)\rho_{\Delta^+z\Delta^-z}$.
Then condition \eqref{l:REL1} is equivalent to
\begin{align}
\label{l:REL9} \sum_{j=0}^{j=N_{\overline{\tau}}}
\left(c^2\overline{\tau}_v^2-\abs{\Delta
\overline{x}_v}^2\right) =\frac{n\epsilon T}{2\alpha^2M},
\end{align}
where $N_{\overline{\tau}}=T/\overline{\tau}$ is the average
number of collisions in time $T$ assuming inter-collision time
$\tau$ and where
\begin{align*}
&\tau_v=\frac{\tau}{\sqrt{1-\rho^2_v}},
\\
&\abs{\Delta \overline{x}_v}^2 = \frac{1}{2}
\begin{pmatrix}E\left[\abs{\overline{v}\left(x(t),t\right)}^2\right]
\\
+\frac{1}{\gamma^4}
E\left[\abs{\frac{b^+(x(t),t)-b^-(x(t),t)}{2}}^2\right]
\end{pmatrix}
\overline{\tau_v}^2.
\end{align*}
Now if $v=\Delta \overline{x}_{v}/\tau_{v}$ and $v'=\Delta
\overline{x}/\tau$ the solution to equation \eqref{l:REL9}
equals
\begin{align}
\label{l:REL10}
\begin{pmatrix}
\Delta\overline{x}_{v} \\
\overline{\tau}_{v}
\end{pmatrix}
= \frac{1}{\sqrt{1-\frac{\abs{v-v'}^2}{c^2}}}
\begin{pmatrix}
1 & (v-v')^T
\\
\frac{(v-v')^T}{c^2} & 1
\end{pmatrix}
\begin{pmatrix}
\Delta\overline{x}_{\overline{\tau}}\\
\overline{\tau}
\end{pmatrix}.
\end{align}
\end{thm}
\begin{proof}
For the more formal solution to \eqref{l:REL9} let $\Delta
\overline{x}_{\overline{\tau}}$ be the drift associated with the
inter-collision time $\overline{\tau}$ and $\Delta
\overline{x}_v$ be the (larger) drift associated with the
inter-collision time $\overline{\tau}_v$ then the invariance
condition \eqref{l:REL9} suggests that
\begin{align}
\label{l:REL6}
\begin{pmatrix}
\Delta\overline{x}_v \\
\overline{\tau}_v
\end{pmatrix}
=
\begin{pmatrix}
A & B \\
F & E
\end{pmatrix}
\begin{pmatrix}
\Delta\overline{x}_{\overline{\tau}}\\
\overline{\tau}
\end{pmatrix}
=
\begin{pmatrix}
A \Delta \overline{x}_{\overline{\tau}}+B\overline{\tau}
\\
F\Delta \overline{x}_{\overline{\tau}}+E\overline{\tau}
\end{pmatrix}
\end{align}
for a constant matrix $A\in\Real^{n\times n}$,
$B,F^T\in\Real^{n\times 1}$ and $E$ a constant. Substituting
this into \eqref{l:REL9} yields
\begin{align}
\label{l:REL7a}
\begin{split}
c^2\overline{\tau}_v^2-\abs{\Delta \overline{x}_v}^2
&=c^2\left(E\overline{\tau} +F\Delta
\overline{x}_{\overline{\tau}}\right)^2-\abs{A \Delta
\overline{x}_{\overline{\tau}}+B\overline{\tau}}^2
\\
&=\left(c^2E^2-\abs{B}^2\right)\overline{\tau}^2 -\Delta
\overline{x}_{\overline{\tau}}^T\left(A^TA-c^2F^TF\right) \Delta
\overline{x}_{\overline{\tau}},
\end{split}
\end{align}
where $c^2EF^T=AB$ is chosen to avoid mixing terms. Applying
$\overline{\tau}=\overline{\tau}_0$ then $\Delta
\overline{x}_{\overline{\tau}_0}=0$ so that \eqref{l:REL7a}
reduces to
\begin{align*}
\begin{pmatrix}
\Delta\overline{x}_v \\
\overline{\tau}_v
\end{pmatrix}
=
\begin{pmatrix}
B\overline{\tau}
\\
E\overline{\tau}
\end{pmatrix},
\end{align*}
and define
$B/E=v=\Delta\overline{x}_v/\overline{\tau}_v\in\Real^n$.

From $c^2EF^T=AB$ it then follows that $F^T=Av/c^2$.  Equation
\eqref{l:REL7a} now reduces to
\begin{align}
\label{l:REL7b}
\begin{split}
c^2\overline{\tau}_v^2-\Delta \overline{x}_v^2
&=E^2\left(1-\frac{\abs{v}^2}{c^2}\right)c^2\overline{\tau}^2
\\
&-\Delta
\overline{x}_{\overline{\tau}}^T\left(A^TA-c^2v^TF^TFv\right)
\Delta \overline{x}_{\overline{\tau}}
\end{split}
\end{align}
Applying this to the $\overline{\tau}_0$, $\Delta
\overline{x}_{\overline{\tau}}\approx 0$ case shows that
$E=1/\sqrt{1-\frac{v^2}{c^2}}$ so that finally
\begin{align*}
c^2\overline{\tau}_v^2-\abs{\Delta \overline{x}_v}^2
=E^2\left(1-\frac{v^2}{c^2}\right)c^2\overline{\tau}^2
=c^2\overline{\tau}^2.
\end{align*}
Hence equation \eqref{l:REL6} applies to the drift/time pair
$\Delta\overline{x}_v$, $\overline{\tau}_v$ and
$\Delta\overline{x}_0\approx 0$, $\overline{\tau}_0$ hence
\begin{align*}
\begin{pmatrix}
\Delta\overline{x}_v \\
\overline{\tau}_v
\end{pmatrix}
&=\begin{pmatrix}
A(v) & B(v) \\
F(v) & E(v)
\end{pmatrix}
\begin{pmatrix}
\Delta\overline{x}_{\overline{\tau}_0}\\
\overline{\tau}_0
\end{pmatrix}
\\
&=\frac{1}{\sqrt{1-\frac{\abs{v}^2}{c^"}}}
\begin{pmatrix}
1 & -v
\\
\frac{-v}{c^2} & 1
\end{pmatrix}
\begin{pmatrix}
\Delta\overline{x}_{\overline{\tau}_0}\\
\overline{\tau}_0
\end{pmatrix}.
\end{align*}
Moreover equation \eqref{l:REL6} relates the drift/time pair
$\Delta\overline{x}$, $\overline{\tau}$ and
$\Delta\overline{x}_0\approx 0$, $\overline{\tau}_0$ so again
\begin{align*}
\begin{pmatrix}
\Delta\overline{x} \\
\overline{\tau}
\end{pmatrix}
&=
\begin{pmatrix}
A(v') & B(v') \\
F(v') & E(v')
\end{pmatrix}
\begin{pmatrix}
\Delta\overline{x}_{\overline{\tau}_0}\\
\overline{\tau}_0
\end{pmatrix}
\\
&=\frac{1}{\sqrt{1-\frac{\abs{v'}^2}{c^"}}}
\begin{pmatrix}
1 & -v'
\\
\frac{-v'}{c^2} & 1
\end{pmatrix}
\begin{pmatrix}
\Delta\overline{x}_{\overline{\tau}_0}\\
\overline{\tau}_0
\end{pmatrix}.
\end{align*}
Finally from these two equations it is clear that
\begin{align*}
\begin{pmatrix}
\Delta\overline{x}_v \\
\overline{\tau}_v
\end{pmatrix}
=
\begin{pmatrix}
A(v) & B(v) \\
F(v) & E(v)
\end{pmatrix}
\begin{pmatrix}
A(v') & B(v') \\
F(v') & E(v')
\end{pmatrix}^{-1}
\begin{pmatrix}
\Delta\overline{x}_{\overline{\tau}}\\
\overline{\tau}
\end{pmatrix},
\end{align*}
which yields \eqref{l:REL10} so the argument is complete.
\end{proof}

\begin{rem}
The analogy with relativity extends further than just the
Lorentz transformation above.  For example, for the kinetic
energy of the main particle it is possible to write
$E\left[\h_T\right]\thickapprox \frac{1}{2}M_Tv^2\thickapprox
\frac{1}{2}Mv^2$ ignoring the osmotic term, the effect of a
finite mass ratio and the diffusion term in \eqref{l:HAMILT3}.
Expressing the differential $\abs{\Delta
\overline{x}_{\overline{\tau}}}^2$ in terms of
$\overline{\tau}_0$
\begin{align*}
E\left[\h_k\right]&\thickapprox M\frac{\abs{\Delta
\overline{x}_{\overline{\tau}}}^2}{\overline{\tau}_v^2}=
M\left(c^2- \frac{n\epsilon}{2\overline{\tau}_v\alpha^2M}\right)
\\
&=Mc^2\left(1-
\frac{\overline{\tau}_0}{\overline{\tau}_v}\right)
=Mc^2\left(1-\sqrt{1-\frac{v^2}{c^2}}\right),
\end{align*}
and recalling the definition of the relativistic energy
$E_r=Mc^2/\sqrt{1-v^2/c^2}$ this equation can be rewritten as
\begin{align*}
\frac{M}{\sqrt{1-\frac{v^2}{c^2}}}\frac{\abs{\Delta
\overline{x}_{\overline{\tau}}}^2}{\overline{\tau}_v^2}
=\frac{Mc^2}{\sqrt{1-\frac{v^2}{c^2}}} -Mc^2=E_r-Mc^2,
\end{align*}
which yields
\begin{align}
\label{l:HAMILT13} E_r=M_r\frac{\abs{\Delta
\overline{x}_{\overline{\tau}}}^2}{\overline{\tau}_v^2} +Mc^2=
\frac{E\left[\h_k\right]}{\sqrt{1-\frac{v^2}{c^2}}} +Mc^2,
\end{align}
with $M_r=\frac{M}{\sqrt{1-\frac{v^2}{c^2}}}$ the relativistic
mass.
\end{rem}

The remark allows an interesting interpretation of the various
terms $E\left[\h_k\right]$, $c$ and the diffusion term
$\epsilon/(\alpha^2M\overline{\tau})$. Clearly the relativistic
mass $M_r$ in this context is the mass measured as a function of
correlation while the kinetic energy $E\left[\h_k\right]$ refers
to the total kinetic energy of the associated main particle and
is in Relativity Theory referred to as the relativistic
momentum. The second term on the righthand side of
\eqref{l:HAMILT13} is the rest energy of the main particle and
is in the present context the result of the diffusion energy. In
fact, using \eqref{l:VELC1} it is clear that
\begin{align*}
Mc^2= \frac{n\epsilon}{2\alpha^2\overline{\tau}_0},
\end{align*}
interpreting the relativistic mass in terms of the mass ratio
$\gamma$, the average inter-collision time $\overline{\tau}_0$
for the main particle at rest in the heatbath and the variance
per mass ratio $\epsilon$ (with units of action).

The most important conclusion from these observation is that it
is possible to find a simple relationship between the
inter-particle collision time and forward/backward (main)
particle velocity correlation in the form of the Lorentz
transformation, see Lanczos ~\cite{LANCZOS1} for some more
detail. The relativity analogy can be pushed further and the
relativistic momentum and the rest mass energy of the main
particle can be interpreted in terms of the time to collision,
mass ratio and variance per unit mass $\epsilon$. Unfortunately,
the compounded correlation
$\rho^2_v=\left(1-2\alpha^2\right)\rho_{\Delta^+z\Delta^-z}$
destroys the martingale property for the coordinate process of
the main particle and this in turn affects the correlation
structure of the heatbath particle. Theorem \eqref{a:THM9} shows
the heatbath backward and forward velocity are correlated
already so the superimposed correlation will have an additional
effect as shown below.

The remaining part of this Section will briefly discourse on a
correlation model for the impulses
$\frac{1}{\tau_1}\sigma\Delta^+z$ and
$\frac{1}{\tau_1}\sigma\Delta^-z$.  A convenient route is to
assume that
\begin{gather}
\label{l:DRFTEQ2}
\begin{split}
\begin{pmatrix}
\sigma\frac{\Delta^+z}{\tau_2}
\\
\sigma\frac{\Delta^-z}{\tau_1}
\end{pmatrix}
=
\begin{pmatrix}
\sigma_a\Delta_a
\\
\sigma_a\Delta_a
\end{pmatrix}
+
\begin{pmatrix}
\sigma_o\Delta_o
\\
-\sigma_o\Delta_o
\end{pmatrix}
+
\begin{pmatrix}
\sigma_r\frac{\Delta_rz^+}{\tau_2}
\\
\sigma_r\frac{\Delta_rz^-}{\tau_1}
\end{pmatrix},
\end{split}
\end{gather}
where $\Delta_rz^+,\Delta_rz^-$ are independent Gaussian
increments, $\sigma_a=\sigma_a(x,t)$, $\sigma_o=\sigma_o(x,t)$
are $n \times n$ matrices and $\Delta_o\in\Real^n$ and
$\Delta_a\in\Real^n$ independent processes with
$E\left[\Delta_a\Delta^T_a\right]=E\left[\Delta_o\Delta^T_o\right]
=2I/\overline{\tau}$. Proper conditions on the drift and
variance terms will not be specified here.  This representation
is motivated by the fact that by this construction
$\Delta^+x(t,\beta)-\Delta^-x(t,\beta)$ does not depend on
$\sigma_a\Delta_a$ and $\Delta^+x(t,\beta)+\Delta^-x(t,\beta)$
becomes independent of $\sigma_o\Delta_o$.

The form of the correlation can now be summarized in a
straightforward calculation as follows.
\begin{prop}
\label{l:HAMILT14}
If the correlation structure of the main
particle is represented by equation \eqref{l:DRFTEQ2} then the
random variables $\Delta_a$ and $\Delta_o$ are part of both the
past and the future steps of the process or rather relate the
past to the future.  The correlation structure of the heatbath
particle is then represented by
\begin{gather}
\label{l:DRFTEQ3}
\begin{split}
\begin{pmatrix}
\omega\frac{\Delta^+_w}{\tau_2} \\
\omega\frac{\Delta^-_w}{\tau_1}
\end{pmatrix}
= &\frac{1}{\gamma\sin(\theta)}
\begin{pmatrix}
-\cos(\theta) & 1 \\
1 & -\cos(\theta)
\end{pmatrix}
\begin{pmatrix}
\sigma\frac{\Delta^+z}{\tau_2} \\
\sigma\frac{\Delta^-z}{\tau_1}
\end{pmatrix}
\\
= &\sigma_a\Delta_a
\begin{pmatrix}
1
\\
1
\end{pmatrix}
-\frac{\sigma_o\Delta_o}{\gamma^2}
\begin{pmatrix}
1
\\
-1
\end{pmatrix}
\\
& + \sigma_r \frac{1}{\gamma\sin(\theta)}
\begin{pmatrix}
-\cos(\theta) & 1 \\
1 & -\cos(\theta)
\end{pmatrix}
\begin{pmatrix}
\Delta_rz^+
\\
\Delta_rz^-
\end{pmatrix}
\end{split}
\end{gather}
where $\Delta_rz^+$,
$\Delta_rz^-$,$\sigma_a(x,t)$,$\sigma_o(x,t)$,
$\Delta_o\in\Real^n$ and $\Delta_a\in\Real^n$ are defined above.
\end{prop}
\begin{proof}
The first remark in the proposition is seen from the fact that
equation \eqref{l:DRFTEQ2} implies that
\begin{gather}
\begin{split}
2\sigma_a\Delta_a=\sigma\left(\frac{\Delta^+z}{\tau_2}+
\frac{\Delta^-z}{\tau_1}\right)
-\sigma_r\left(\frac{\Delta^+_rz}{\tau_2}+
\frac{\Delta^-_rz}{\tau_1}\right),
\\
2\sigma_o\Delta_o=\sigma\left(\frac{\Delta^+z}{\tau_2}-
\frac{\Delta^-z}{\tau_1}\right)
-\sigma_r\left(\frac{\Delta^+_rz}{\tau_2}-
\frac{\Delta^-_rz}{\tau_1}\right).
\end{split}
\end{gather}
The matrix in the proposition is the result of inverting
equation \eqref{DELTA1b} so \eqref{l:DRFTEQ3} follows from the
fact that the $\Delta_o\in\Real^n$ and $\Delta_a\in\Real^n$
terms are eigenvectors.  In fact,
\begin{align*}
&\frac{1}{\gamma\sin(\theta)}
\begin{pmatrix}
-\cos(\theta) & 1 \\
1 & -\cos(\theta)
\end{pmatrix}
\begin{pmatrix}
\sigma_a\Delta_a
\\
\sigma_a\Delta_a
\end{pmatrix}
=
\begin{pmatrix}
\sigma_a\Delta_a
\\
\sigma_a\Delta_a
\end{pmatrix},
\\
&\frac{1}{\gamma\sin(\theta)}
\begin{pmatrix}
-\cos(\theta) & 1 \\
1 & -\cos(\theta)
\end{pmatrix}
\begin{pmatrix}
\sigma_o\Delta_o
\\
-\sigma_o\Delta_o
\end{pmatrix}
= -\frac{1}{\gamma^2}
\begin{pmatrix}
\sigma_o\Delta_o
\\
-\sigma_o\Delta_o
\end{pmatrix},
\end{align*}
since $\frac{1-\cos(\theta)}{\gamma\sin(\theta)}=1$ and
$\frac{1+\cos(\theta)}{\gamma\sin(\theta)}=1/\gamma^2$.  This
completes the Proposition.
\end{proof}

To show the effect of the $\sigma_o\Delta_o$ and
$\sigma_a\Delta_a$ terms take for example $n=1$, then equating
the variances on both sides of \eqref{l:DRFTEQ2}
\begin{align}
\begin{split}
\frac{2\sigma^2}{\overline{\tau}} =
&\sigma^2var\left(\Delta^+z\right)=
\sigma^2var\left(\Delta^-z\right)=
\\
&\sigma_a^2var\left(\Delta_a\right)+
\sigma_o^2var\left(\Delta_o\right)+
\frac{2\sigma_r^2}{\overline{\tau}}.
\end{split}
\end{align}
Since $\Delta_a$ and $\Delta_o$ are independent
$\sigma^2E\left[\frac{\Delta^+z}{\tau_2}
\frac{\Delta^-z}{\tau_1}\right]
=\sigma^2_avar\left(\Delta_a\right)-\sigma^2_0var\left(\Delta_o\right)$
so that the correlation between between $\Delta^+z$ and
$\Delta^+z$ reduces to
\begin{gather}
\label{l:DRFTEQ4}
\rho_{\Delta^+z\Delta^-z}=\frac{\sigma^2_a
var\left(\Delta_a\right)-\sigma^2_0var\left(\Delta_o\right)}{
\sigma_a^2var\left(\Delta_a\right)+
\sigma_o^2var\left(\Delta_o\right)+
\frac{2\sigma_r^2}{\overline{\tau}}}.
\end{gather}
If it is assumed that
$var\left(\Delta_a\right)=2/\overline{\tau}=
var\left(\Delta_o\right)=2/\overline{\tau}$ then
\eqref{l:DRFTEQ4} reduces further to
\begin{gather}
\rho_{\Delta^+z\Delta^-z}=\frac{\sigma^2_a -\sigma^2_0}{
\sigma_a^2+ \sigma_o^2+ \sigma_r^2}.
\end{gather}

The reason for this construction now becomes clear.  If
$\sigma_a \thickapprox 0$ and $\sigma_o
>>\sigma_r$ then $\rho_{\Delta^+z\Delta^-z} \rightarrow -1$ while
$\rho_{\Delta^+z\Delta^-z} \rightarrow 1$ if $\sigma_o
\thickapprox 0$ and $\sigma_a >>\sigma_r$.  The correlation in
the driving factors in \eqref{l:DRFTEQ2} relates the future to
the past through the collision time $t$.  This introduces a form
of auto-correlation for the random difference process.  For a
positive or negative correlation the process
$x\left(t,\beta\right),t>0$ can not be a martingale and a two
dimensional process must be introduced.

This Section relates correlation in the motion of the main
particle to the inter-particle collision time via the Lorentz
equation to satisfy a Minkowski invariant and suggest that the
rest mass energy is the result of the energy embedded in the
diffusion energy.  Proposition \eqref{l:HAMILT14} shows how to
decompose the backward and forward velocities into a perfect
correlation part and a martingale process. These results were
predicated on the interaction prescription \eqref{l:ENERG2}
which is exact in the one-dimensional case.  In higher
dimensions the equation applies to the center of mass line
projection as remark \eqref{l:ENERG15} describes.  The energy
conservation arguments in the last two Sections still apply
however another level of complexity will be required to describe
the non-simple collision.  This will be addressed in the
following Section.

\section{Non-Simple Collisions / Scattering}

This Section returns to the case where the matrix $\Omega$ in
equation \eqref{l:ENERG2} characterizes all elastic interactions
incorporating the random anti-symmetric matrix $Z=Z(U)$. In this
case the total energy defined in Theorem \eqref{l:THMENERG}
depends on the statistical characteristics of $Z$ and
Proposition \eqref{l:HAMILT12} is derived in the presence of
this collision scattering matrix.  Finally the conservation
condition for the full collision is derived and an example is
presented that combines this result with an electromagnetic
field type Hamiltonian.  The Section concludes with showing that
it is possible to subsume the collision scattering matrix $Z$
into the heatbath making all the results from Section 2 and 3
applicable for an altered heatbath with a different statistical
structure.

The first step is to obtain the total kinetic energy expression
in Theorem \eqref{l:THMENERG} as a function of the pre- and post
collision velocities $v_2, v_1$ of the main particle using
equations \eqref{l:GAMMDEF1} and \eqref{l:COLLMAT1}.  The form
of the total kinetic energy is introduced in the following
result.
\begin{thm}
\label{l:THMENERG1} As in Theorem \eqref{l:THMENERG}, let the
momentum of the main particle and interacting particle be
presented as $p_1=Mv_1$ (post-collision $p_2=Mv_2$) where
$p_1,p_2,v_1,v_2 \in \Real^n$ and $q_1=Mw_1$ (post-collision
$q_2=Mw_2$) with $q_1,q_2,w_1,w_2 \in \Real^n$.  Then the total
kinetic energy $\h_k=\frac{1}{2}(M\abs{v_2}^2+m\abs{w_2}^2)=
\frac{1}{2}(M\abs{v_1}^2+m\abs{w_1}^2)$ is related to the pre -
and post collision momenta of the main particle as follows
\begin{align}
\label{l:DRFTEQ8}
\begin{split}
\frac{8\h_k}{M_T} =& \Delta^+v^T\Delta^+v -2
\Delta^+v^TZ\Delta^-v
\\
&+\frac{1}{\gamma^2}\Delta^-v^T\Delta^-v +
\left(\frac{1+\gamma^2}{\gamma^2}\right)\Delta^-v^TZZ^T\Delta^-v
\\
=& \left(\Delta^+v-Z\Delta^-v\right)^T
\left(\Delta^+v-Z\Delta^-v\right)\\
&+\frac{1}{\gamma^2}\Delta^-v^T\Delta^-v
+\frac{1}{\gamma^2}\Delta^-v^TZZ^T\Delta^-v,
\end{split}
\end{align}
where
\begin{align}
\label{l:DRFTEQ7}
\begin{split}
&\Delta^+v=v_2+v_1, \\
&\Delta^-v=v_2-v_1.
\end{split}
\end{align}
Here $Z=Z(U)$ is the anti-symmetric matrix ($Z^T+Z=0$) such that
$Z=I-2(I+U)^{-1}=I-\gamma\sin(\theta)Q^{-1}$ where the unitary
matrix $U$ and $Q$ are defined in equation \eqref{l:COLLMAT1}
above .
\end{thm}
\begin{proof}
The proof can be found in Appendix B as well.
\end{proof}
As mentioned above the random collision scattering matrix
$Z=Z\left(U\right)$ represents the collection of all possible
elastic collision transitions and therefore contains the center
of mass line information and the angle of impact between $v_1$
and $w_1$. The matrix will be different from one collision to
the next and has a statistical mean
$E\left[Z\right]=\overline{Z}$ and variance $E\left[ZZ^T\right]$
which may be a function of the collision position $x(t)$. To
deduce the expectation of the energy term $\h_k$ use again
\eqref{l:DRFTDEF0} with the definition \eqref{l:DRFTEQ8} to
derive
\begin{align}
\begin{split}
\label{l:DRFTEQ5} \frac{8E\left[\h_k\right]}{M_T} = &
E\left[\Delta^+v^T\Delta^+v\right] -2
E\left[\Delta^+v^TZ\Delta^-v+\right]
\\
&+\frac{1}{\gamma^2}E\left[\Delta^-v^T\Delta^-v\right] +
\left(\frac{1+\gamma^2}{\gamma^2}\right)
E\left[\Delta^-v^TZZ^T\Delta^-v\right]
\\
=&E\left[\Delta^+v^T\Delta^+v\right] -2
E\left[\Delta^+v^T\overline{Z}\Delta^-v\right]
\\
&+\frac{1}{\gamma^2} E\left[\Delta^-v^T\Gamma^z\Delta^-v\right],
\end{split}
\end{align}
where $E[Z]=\overline{Z}$ and $\Gamma^z=I+
(1+\gamma^2)E\left(ZZ^T\right)$.  This is the result of taking
the expectations over the random matrix $Z$ first and then
rearranging the expression.

Alternatively, this expression can be written as
\begin{align*}
\frac{8E\left[\h_k\right]}{M_T}
=&E\left[\left(\Delta^+v-\overline{Z}\Delta^-v\right)^T
\left(\Delta^+v-\overline{Z}\Delta^-v\right)\right]
\\
&+\frac{1}{\gamma^2}E\left[\Delta^-v^T\left(\Gamma^z
-\gamma^2\overline{Z}\overline{Z}\right)\Delta^-v\right].
\end{align*}
Notice that the matrix in the last terms is positive definite
since $\Gamma^z-\gamma^2\overline{Z}\overline{Z}^T
=I+\overline{ZZ^T}+\gamma^2\left(\overline{ZZ^T}
-\gamma^2\overline{Z}\overline{Z^T}\right)=I+\overline{ZZ^T}
+\gamma^2var\left(ZZ^T\right)$ with the obvious definition for
$var\left(ZZ^T\right)=E\left[ZZ^T\right]-
E\left[Z\right]E\left[Z^T\right]\geq 0$.

Again it is assumed that the change in expected energy equals
the change of an appropriate potential $\Phi_p$ so that
$\frac{d}{dt}E\left[\h_T\right]=\frac{d}{dt}E\left[\h_k+\Phi_p\right]=0$.
The following Proposition shows the form of the total energy
\eqref{l:DRFTEQ5} as a function of the backward and forward
velocity.
\begin{prop}
Reducing the expectations in \eqref{l:DRFTEQ5} the total energy
in can be expressed as
\begin{align}
\label{l:ENERG5}
\begin{split}
\frac{E\left[\h_k\right]+E\left[\Phi_p\right]}{M_T}
&=\frac{1}{2} E
\begin{pmatrix}
\left(\frac{b^++b^-}{2}\right)^2-
2\left(\frac{b^++b^-}{2}\right)^T\overline{Z}
\left(\frac{b^+-b^-}{2}\right)
\\
+\frac{1}{\gamma^2}\left(\frac{b^+-b^-}{2}\right)^T\Gamma^z
\left(\frac{b^+-b^-}{2}\right)
\end{pmatrix}
\\
&+\frac{n\sigma^2}{2\overline{\tau}}
+\frac{\sigma^2}{2\overline{\tau}\gamma^2}
E\left[Tr\left(\Gamma^z\right)\right]+\frac{1}{M_T}E\left[\Phi_p\right].
\end{split}
\end{align}
\end{prop}
\begin{proof}
This expression can be easily derived from substituting
\eqref{l:DRFTEQ7} into (\ref{l:DRFTEQ5}).  Let $z_1=
\left(\frac{\Delta^+}{\tau_2} +\frac{\Delta^-}{\tau_1}\right)$
and let $z_2=\left(\frac{\Delta^+}{\tau_2}
-\frac{\Delta^-}{\tau_1}\right)$ then $z_1$ and $z_2$ are
independent and normally distributed.  Hence
\begin{align*}
&E\left[z^T_1z_1\right]=E\left[z^T_2z_2\right]
=\frac{4n}{\overline{\tau}}I,
\\
&E\left[\left(\frac{\Delta^+z}{\tau_2}
+\frac{\Delta^-z}{\tau_1}\right)\left(\frac{\Delta^+z}{\tau_2}
+\frac{\Delta^-z}{\tau_1}\right)^T\right]=\frac{4n}{\overline{\tau}}I,
\\
&E\left[z^T_1z_2\right]=0.
\end{align*}
Moreover, it is easy to see that $E\left[z^T_1Z^Tz_2\right]=0$
and $E\left[z_2^Tz_2+\left(1+\gamma^2\right)z^T_2Z^TZz_2\right]
=\frac{2}{\overline{\tau}}
E\left[Tr\left(\Gamma^z\right)\right]$.
\end{proof}

The pre - and post -collision velocities for the main and
heatbath particles are linearly related via equation
\eqref{l:GAMMDEF1} for the matrices $P,Q,V$ and $G$ defined in
\eqref{l:COLLMAT1}.  Now the equivalent of Theorem
\eqref{l:SCHR1} is introduced to show the conditions for
maintaining a constant total energy.
\begin{thm}
\label{l:HAMILTPRP1}
Let $\rho=e^{\frac{2\gamma\delta
R}{\sigma^2}}=e^{\frac{2\delta R}{\eta}}$ and introduce the
sufficiently smooth functions $A=A(x,t),S=S(x,t),x\in\Real^n$
and constants $\delta, \xi$ to express the backward and forward
drifts $b^+=b^+(x,t),b^-=b^-(x,t),x\in\Real^n,t>0$ as follows
\begin{align*}
&b^+=\xi\left(\nabla S-A\right)+\gamma\delta \nabla R,
\\
&b^-=\xi\left(\nabla S-A\right)-\gamma\delta \nabla R.
\end{align*}
Assume that the potential $\Phi_p$ satisfies the following
property
\begin{align}
\label{l:HAMILT15}
\begin{split}
\frac{d}{dt}E\left[\Phi_p\right] &=E\left[\left(\nabla
S-A\right).\left(\nabla \phi+\xi\dot{A}\right) \right]
\\
&=E\left[\left(S_{x_j}-A_j\right)\left(
\phi_{x_j}+\xi\dot{A}_j\right) \right],
\end{split}
\end{align}
with $\dot{A}=\frac{\partial A}{\partial t}$ using Einstein's
notation of summing all like indices. Let $\h_k$ be defined as
in equation \eqref{l:ENERG5} with $Z$ a random matrix such that
$\Gamma^z=I+(1+\gamma^2)E\left[ZZ^T\right]$ and
$E\left[Z\right]=\overline{Z}$.  Then the total energy
$\h_T=\h_k+\Phi_p$ for the potential in $\Phi_p$ in
\eqref{l:ENERG5} is conserved if
\begin{align}
\label{l:HAMILT5}
\begin{split}
\frac{d}{dt}&\left(\frac{E\left[\h_k\right]
+E\left[\Phi_p\right]}{M_T}\right)
\\
=&\frac{d}{dt}\frac{1}{2} E
\begin{pmatrix}
\xi^2\,|\nabla S-A\,|^2-2\xi\delta\gamma(\nabla
S-A)^T\overline{Z}\nabla R
\\+\delta^2\nabla R^T\Gamma^z\nabla R\end{pmatrix}
+\frac{d}{dt}\frac{1}{M_T}E\left[\Phi_p\right]
\\
=&\xi\int\rho
\begin{pmatrix}
\left(S_{x_p}-A_p\right)\
\begin{pmatrix}
\xi S_{t}
+\frac{\xi^2}{2}\left(S_{x_j}-A_j\right)\left(S_{x_j}-A_j\right)
\\
-\frac{\delta^2}{2}R_{x_j}\Gamma^z_{jk}R_{x_k}
-\frac{\delta\eta}{2}\left(R_{x_j}\Gamma^z_{jk}\right)_{x_k}
\\
+\frac{\phi}{\xi M_T}
\end{pmatrix}_{x_p}
\end{pmatrix}dx
\\
&-\xi\int\rho\left(S_{x_p}-A_p\right)\left(\xi-\frac{1}{M_T}\right)\dot{A_p}dx
+\Xi\left(Z\right)
\\
&+\frac{\sigma^2}{2\overline{\tau}\gamma^2}\frac{d}{dt}
E\left[Tr\left(\Gamma^z\right)\right]=0,
\end{split}
\end{align}
where
\begin{align}
\label{l:HAMIL23}
\begin{split}
\Xi\left(Z\right)&=\frac{\xi}{2}\int\rho
\begin{pmatrix}
\delta^2R_{x_j}\dot{\Gamma}^z_{jk}R_{x_k}
\end{pmatrix}
dx
\\
&+\frac{\eta\xi}{2\delta}\int\rho \left(S_{x_p}-A_{p}\right)
\begin{pmatrix}
\left(\xi\delta\gamma
\left(S_{x_j}-A_j\right)\overline{Z}_{jk}\right)_{x_kx_p}
\end{pmatrix}
dx
\\
&-\xi\delta\gamma\int\rho
\begin{pmatrix}
\left(S_{x_jt}-\dot{A_j}\right)\overline{Z}_{jk}
R_{x_k}+\left(S_{x_j}-A_j\right)\dot{\overline{Z}}_{jk} R_{x_k}
\end{pmatrix}
dx,
\end{split}
\end{align}
and where $Tr\left(\Gamma^z\right)$ denotes the Trace of the
matrix $\Gamma^z$. Here
\begin{align*}
&\dot{\overline{Z}}=\frac{d}{dt}E\left[Z\right],
\\
&\dot{\Gamma}^z=\frac{d}{dt}E\left[ZZ^T\right].
\end{align*}
\end{thm}
\begin{proof}
For a proof consult Appendix C.
\end{proof}

This representation exhibits three sizeable problems finding
solutions for the time invariance of \eqref{l:ENERG5}. First of
all there is the fact that potentials $\Phi_p=\Phi_p(x,t)$ do
not typically admit property like \eqref{l:HAMILT15} as the time
derivative of the potential introduces terms like
$E\left[\frac{\partial \Phi_p}{\partial t}\right]$.  Obviously
time independent potentials satisfy property \eqref{l:HAMILT15}
and as can be seen below Maxwellian type fields have this
property as well. In fact if the $Z$ term can be ignored for a
moment then the following can be shown.

\begin{thm}
Assume that $Z\equiv 0$ and let
$\psi=\psi(x,t)=e^{\frac{R(x,t)+iS(x,t)}{\chi}}$ with
$\rho(x,t)=\abs{\psi(x,t)}^2$ where
\begin{align*}
b^+=\xi\left(\nabla S-A\right)+\gamma\delta \nabla R,
\\
b^-=\xi\left(\nabla S-A\right)-\gamma\delta \nabla R,
\end{align*}
and let $E$ and $B$ satisfy the (magnetically sourceless)
Maxwell equations
\begin{align}
\label{l:MAXW1}
\begin{split}
&\nabla . E=\rho(x,t),
\\
&\nabla \times B-\xi\frac{\partial}{\partial
t}E=\xi\left(\frac{b^++b^-}{2}\right)\rho(x,t),
\\
&\nabla . B=0,
\\
&\nabla \times E+\xi\frac{\partial}{\partial t}B=0,
\end{split}
\end{align}
so that $E=-\nabla\phi-\xi\frac{\partial}{\partial t}A$ and
$B=\nabla \times A$.  Assume that $\Phi_p=\abs{E}^2+\abs{B}^2$
then
\begin{align*}
\frac{d}{dt}\left(\frac{E\left[\h_k\right]
+E\left[\Phi_p\right]}{M_T}\right)=0,
\end{align*}
if and only if
\begin{align}
\label{l:MAXW2}
i\chi\psi_t=-\frac{1}{2M_T}\left(\chi\nabla-iA\right)^2
\psi+\phi(x,t)\psi,
\end{align}
with $\chi=M_T\eta=M_T\sigma^2/\gamma=\left(\gamma
+\frac{1}{\gamma}\right)\epsilon$ and $\delta=\xi=1/M_T$.
\end{thm}
\begin{proof}
The point of the proof is that for equations \eqref{l:MAXW1} it
is true that
\begin{align*}
\frac{d}{dt}E\left[\Phi_p\right]=
&\frac{d}{dt}\int\rho\left[\abs{E}^2+\abs{B}^2\right]
\\
=& -\int\rho\left[\left(\frac{b^++b^-}{2}\right)^TE\right]
\\
=&\int\rho\left[\left(\frac{b^++b^-}{2}\right)^T\left(\nabla
\phi+\xi\dot{A}\right)\right].
\end{align*}
In other words the time change of the energy of the field in
\eqref{l:MAXW1} satisfies \eqref{l:HAMILT15} and this is
combined with proposition \eqref{l:prp22} and equation
\eqref{l:ENERG6} from Appendix C.
\end{proof}

The second issue is the fact that the terms $
\frac{\delta^2}{2}R_{x_j}\Gamma^z_{jk}R_{x_k},kj,k=1,...,n$ and
$\frac{\delta\eta}{2}\left(R_{x_j}\Gamma^z_{jk}\right)_{x_k}
,k=1,...,n$ in Theorem \eqref{l:HAMILT15} above depend on the
random $Z$ matrix which acts here as an arbitrary scaling
factor. If the $\Gamma^z$ matrix is diagonal is is possible to
scale the solution derived for the case where $\overline{Z}$ is
state independent. In fact the example below shows that a simple
scaling applied to the factors in the wave function allows for a
solution.

\subsection{The two-step Scattering Matrix}
\label{l:EXAMP1} This example derives a solution to equation
\eqref{l:HAMILT5} by rescaling the wave function as follows.
Assume that $n=2$, $\xi=\frac{1}{M_T}$, $A\equiv \phi\equiv 0$
and define the anti-symmetric matrices $Z(\nu), Z(-\nu)$ as
follows
\begin{align*}
Z(\nu)=
\begin{pmatrix}
0 & \nu
\\
-\nu & 0
\end{pmatrix},
Z(-\nu)=
\begin{pmatrix}
0 & -\nu
\\
\nu & 0
\end{pmatrix}.
\end{align*}
Now let the probabilities $p(\nu)$ and $p(-\nu)$ be such that
\begin{align*}
&E[Z]=\overline{Z}=p(\nu)Z(\nu)+p(-\nu)Z(-\nu),
\\
&E[ZZ^T]=p(\nu)Z(\nu)Z^T(\nu)+p(-\nu)Z(-\nu)Z^T(-\nu)=\nu^2I,
\end{align*}
so that $\dot{\Gamma}^z=0$, $Tr\left(\dot{\Gamma}^z\right)=0$
with $\overline{Z}$ state independent (not a function of
$x(t)$).
 Then $\Gamma^z=I\left(1+(1+\gamma^2)\nu^2\right)=
I\sigma^2_\nu$ and the solution that preserves the energy
equation \eqref{l:HAMILT5} reduces to
\begin{align}
\label{l:HAMILT22}
\begin{split}
\frac{d}{dt}&\left(\frac{E\left[\h\right]
+E\left[\Phi_p\right]}{M_T}\right)
\\
=&\xi\int\rho
\begin{pmatrix}
S_{x_p}
\begin{pmatrix}
\xi S_{t} +\frac{\xi^2}{2}\abs{\nabla S}^2
\\
-\frac{\delta^2}{2}\abs{\nabla R}^2
-\frac{\sigma^2_\nu\delta^2}{2}\abs{\nabla R}^2
\\
-\frac{\delta\eta}{2}\Delta_xR
-\frac{\sigma^2_\nu\delta\eta}{2}\Delta_xR
\end{pmatrix}_{x_p}
\end{pmatrix}dx
\\
&+\frac{\eta\xi}{2\delta}\int\rho S_{x_p}
\begin{pmatrix}
\left(\xi\delta\gamma S_{x_j}\overline{Z}_{jk}\right)_{x_kx_p}
\end{pmatrix}
dx,
\end{split}
\end{align}
because the second term in equation \eqref{l:HAMILT5} vanishes
\begin{align*}
&\frac{\eta\xi}{2\delta}\int\rho S_{x_p}
\begin{pmatrix}
\left(\xi\delta\gamma S_{x_j}\overline{Z}_{jk}\right)_{x_kx_p}
\end{pmatrix}
dx
\\
&=\frac{\eta\xi^2\nu}{2}\gamma\int\rho S_{x_p}
\left(\frac{\partial}{\partial x_2}
S_{x_1}-\frac{\partial}{\partial x_1} S_{x_2}\right)_{x_p} dx=0,
\end{align*}
and so do the last two terms.  Hence the solution to
\eqref{l:MAXW2} is given by Proposition \eqref{HAMILT22} below
as
\begin{align}
\label{l:SCROD2}
i\chi_\nu\psi_t=-\frac{\chi_\nu^2}{2M_T}\Delta_x
\psi+\phi(x,t)\psi,
\end{align}
where $\psi=\psi(x,t)=e^{\frac{\delta R(x,t)+i\xi
S(x,t)}{\chi}}$ with $\chi_\nu=M_T\eta/\sigma_\nu$, $\xi=1/M_T$
and $\delta=1/\left(\sigma_\nu M_T\right)$.  Proposition
\eqref{l:HAMILT22} in Appendix C has some more details on the
derivation.

This highlights the second feature of equation \eqref{l:HAMILT5}
which is that the effective osmotic term typically scales up due
to the variance of the random matrix $Z$.  This scaling never
disappears if $E\left[ZZ^T\right]>0$ however equation
\eqref{l:HAMILT5} can still be reduced to the case of the simple
collision in Theorem \eqref{l:SCHR1} as example \eqref{l:EXAMP1}
shows.  There is no real quantum mechanical analogy to this
except that some of the terms in equation \eqref{l:HAMILT22} are
similar to the terms in the Bopp-Haag Hamiltonian where the
additional derivatives are introduced to incorporate spin
states, see for instance Nelson~\cite{ENELSON1}.

The third question on equation \eqref{l:HAMILT5} is the effect
of the $\overline{Z}$ terms. The example above shows that these
terms disappear for the case where $\overline{Z}$ does not
depend on the coordinate system. This is entirely due to the
fact that $\overline{Z}$ is anti-symmetric and their
contribution to equation \eqref{l:HAMILT5} remain conveniently
zero even if the matrix $\overline{Z}$ depends on time $t$.  As
the mean collision scattering matrix relates to the average
center of mass line and average angle of collision this quantity
is unlikely to be dependent on the collision coordinate $x(t)$
so this assumption is not unreasonable.

Obviously the presence of the $Z$ matrix changes both the
correlation between $v_2$ and $v_1$ and simultaneously affects
the correlation structure of the heatbath.  If for instance the
main particle path is a martingale for $Z\equiv 0$ then "turning
on" the $Z$ will create a correlation.  If the collision
scattering matrix $Z>0$ and the main particle path is a
martingale then the correlation structure of the heatbath must
change from the case that $Z\equiv 0$.  The following
Proposition generalizes Theorem \eqref{a:THM9} and calculates
the correlation of the heatbath particles for the latter case.
\begin{prop}
\label{l:HAMILT17} Assume that the backward and forward
velocities of the main particle are uncorrelated.  Then the
correlation matrix for the colliding heatbath particle given a
realization of the random matrix $Z$ looks like
\begin{align*}
&E\left[
\begin{pmatrix}
\frac{\Delta^+_w}{\-\tau_2} \\
\frac{\Delta^-_w}{\-\tau_1}
\end{pmatrix}
\begin{pmatrix}
\frac{\Delta^+_w}{\-\tau_2} & \frac{\Delta^-_w}{\-\tau_1}
\end{pmatrix}
\right]
\\
&= \frac{\sigma^2}{\overline{\tau}\alpha^2}
\begin{pmatrix}
1 & -\left(1-2\alpha^2\right) \\
-\left(1-2\alpha^2\right) & 1
\end{pmatrix}
+\Gamma_Z,
\end{align*}
where
\begin{align*}
\overline{\Gamma}_Z=
\frac{2\sigma^2}{\overline{\tau}\gamma\sin^2(\theta)}
\begin{pmatrix}
E\left[ZZ^T\right] & \Omega
\\
\Omega^T & E\left[ZZ^T\right]
\end{pmatrix},
\end{align*}
and where
\begin{align*}
\Omega=\left(1-\cos(\theta)I\right)\overline{Z}+E\left[ZZ^T\right].
\end{align*}
\end{prop}
\begin{proof}
For a straightforward calculation see Appendix D.
\end{proof}

\begin{rem}
Though Proposition \eqref{l:HAMILT17} specifies how the heatbath
particle field must behave to guarantee that the main particle
moves in a Markovian fashion the distribution of $x(t)$ has now
become complicated.  Clearly the distribution for $v_2,w_2$ must
be convolved with the distribution for $v_1,w_1$ and $Z=Z(U)$
and is therefore not readily calculated.
\end{rem}

The final question in this Section is whether there is a
"momentum term" version of Theorem \eqref{l:HAMILT20} and
whether the complexity of equation \eqref{l:HAMILT5} can be
reduced. Interestingly this can indeed be achieved in a
straightforward manner but some changes in assumptions will be
required.  The approach is to absorb the scattering matrix $Z$
into the heatbath and then show that the results from Section 2
and 3 apply for the transformed heatbath.

To implement this approach the $\Delta^+v$ and $\Delta^-v$ must
be rewritten in a more convenient form.  From the proof of
Proposition \eqref{l:HAMILT17} in Appendix \ref{l:APP4} it
becomes clear that
\begin{align*}
\begin{pmatrix}
w_2 \\
w_1
\end{pmatrix}
= \frac{1}{\gamma\sin(\theta)}
\begin{pmatrix}
-\cos(\theta)I-Z & I+Z
\\
I-Z & -\cos(\theta)I+Z
\end{pmatrix}
\begin{pmatrix}
v_2 \\
v_1
\end{pmatrix}.
\end{align*}
This can be simplified by writing
\begin{align*}
\begin{pmatrix}
w_2 \\
w_1
\end{pmatrix}
=
\begin{pmatrix}
Vv_1+Gw_1\\
w_1
\end{pmatrix}
=
\begin{pmatrix}
V & G \\
0 & I
\end{pmatrix}
\begin{pmatrix}
v_1 \\
w_1
\end{pmatrix},
\end{align*}
and then use equation \eqref{e:ENG6} and \eqref{a:THM6} from the
Appendix to obtain
\begin{align*}
\begin{pmatrix}
w_2 \\
w_1
\end{pmatrix}
&= \frac{1}{2}
\begin{pmatrix}
V & G \\
0 & I
\end{pmatrix}
\begin{pmatrix}
I & -I \\
I & \left(2Q^{-1}-I\right)
\end{pmatrix}
\begin{pmatrix}
\Delta^+v \\
\Delta^-v
\end{pmatrix},
\end{align*}
hence with $V_\Gamma=\left(2Q^{-1}-I\right)$ this reduces to
\begin{align*}
\begin{pmatrix}
w_2 \\
w_1
\end{pmatrix}
&
= \frac{1}{2}
\begin{pmatrix}
V & G \\
0 & I
\end{pmatrix}
\begin{pmatrix}
I & -I \\
I & V_\Gamma
\end{pmatrix}
\begin{pmatrix}
\Delta^+v \\
\Delta^-v
\end{pmatrix}
\\
&= \frac{1}{2}
\begin{pmatrix}
V+G & -V+GV_\Gamma \\
I & V_\Gamma
\end{pmatrix}
\begin{pmatrix}
\Delta^+v \\
\Delta^-v
\end{pmatrix}
\\
&=\frac{1}{2}
\begin{pmatrix}
I & V_\Gamma-\frac{2}{\gamma^2}I \\
I & V_\Gamma
\end{pmatrix}
\begin{pmatrix}
\Delta^+v \\
\Delta^-v
\end{pmatrix}.
\end{align*}

Inverting this relation yields
\begin{align*}
\begin{pmatrix}
\Delta^+v \\
\Delta^-v
\end{pmatrix}
= 2
\begin{pmatrix}
\frac{\gamma^2}{2}V_\Gamma & I-\frac{\gamma^2}{2}V_\Gamma
\\
-\frac{\gamma^2}{2}I & \frac{\gamma^2}{2}I
\end{pmatrix}
\begin{pmatrix}
w_2 \\
w_1
\end{pmatrix},
\end{align*}
which means
\begin{align*}
\Delta^+v &=2w_1+\gamma^2V_\Gamma\Delta^-w,\\
\Delta^-v &=-\gamma^2\Delta^-w.
\end{align*}
Now $\gamma^2V_\Gamma=I-\left(1+\gamma^2\right)Z$ so then
\begin{align}
\label{l:ENERG10}
\begin{split}
\Delta^+v &=\Delta^+w-\left(1+\gamma^2\right)Z\Delta^-w,\\
\Delta^-v &=-\gamma^2\Delta^-w,
\end{split}
\end{align}
which can also be written as
\begin{subequations}
\begin{align}
\label{l:ENERG11a}
\Delta^+v &=\Delta^+w+2\W,
\\
\label{l:ENERG11b} \Delta^-v &=-\gamma^2\Delta^-w.
\end{align}
\end{subequations}
Here $m\W$ is the additional momentum such that
$-\W=\frac{1+\gamma^2}{2}Z(w_2-w_1)=\frac{1+\gamma^2}{2}Z\Delta^-w$.

From this the following two Theorems are easily shown.
\begin{thm}
The additional momentum defined in equation \eqref{l:ENERG11a}
above is perpendicular to the $\Delta^-w$, in other words
\begin{subequations}
\begin{align}
\label{l:ENERG12a}
\W^T\Delta^-w
=\frac{1+\gamma^2}{2}\Delta^-w^TZ^T\Delta^-w=0,
\end{align}
and so if $E\left[\abs{w_1}^2\right]=E\left[\abs{w_2}^2\right]$
then
\begin{align}
\label{l:ENERG12b}
E\left[\abs{w_1+\W}^2\right]=E\left[\abs{w_2+\W}^2\right].
\end{align}
Let $w_{\top,1}=w_1+\W$, $w_{\top,2}=w_1+\W$ and define
$\Delta_\top^+v=w_{\top,2}+w_{\top,1}$, $\Delta_\top^-
v=w_{\top,2}-w_{\top,1}=w_2-w_1$ then equation
\eqref{l:ENERG11a} reduces to
\begin{align}
\label{l:ENERG13a} \Delta^+v &=\Delta_\top^+w,
\\
\label{l:ENERG13b} \Delta^-v &=-\gamma^2\Delta_\top^-w,
\end{align}
hence
\begin{align}
\label{l:ENERG14}
&\frac{1}{2}E\left[\abs{\frac{\Delta^+v}{2}}^2+
\frac{1}{\gamma^4}\abs{\frac{\Delta^-v}{2}}^2\right] =c_\top^2,
\end{align}
where $c_\top^2=E\left[\abs{w_{\top,2}}^2\right]=
E\left[\abs{w_{\top,1}}^2\right]$.
\end{subequations}
\end{thm}
\begin{proof}
To prove assertion \eqref{l:ENERG12a} consider
\eqref{l:ENERG11a} and multiply with \eqref{l:ENERG11b} so that
\begin{align*}
\Delta^-v^T\Delta^+v &=\Delta^-v^T\left(\Delta^+w+2\W\right)
\\
&=-\gamma^2\Delta^-w^T\Delta^+w-\gamma^2\Delta^-w^T\W,
\end{align*}
or
\begin{align*}
\abs{v_2}^2-\abs{v_1}^2+\gamma^2\abs{w_2}^2-\gamma^2\abs{w_1}^2=
-\gamma^2\Delta^-w^T\W.
\end{align*}
However the left hand side equals
$\abs{v_2}^2+\gamma^2\abs{w_2}^2-\left(\abs{v_1}^2
+\gamma^2\abs{w_1}^2\right) =\h_k/M-\h_k/M=0$ by energy
conservation hence $\gamma^2\Delta^-w^T\W=0$ and
\eqref{l:ENERG12a} is proved.

Assertion \eqref{l:ENERG12b} now follows since
\begin{align*}
&E\left[\abs{w_1+\W}^2\right]-E\left[\abs{w_2+\W}^2\right]
\\
&=E\left[\abs{w_1}^2\right]-E\left[\abs{w_2}^2\right]
+2E\left[\left(w_1-w_2\right)^T\W\right]
\\
&=-2E\left[\Delta^-w^T\W\right]=0,
\end{align*}
and equation \eqref{l:ENERG14} follows directly from
\eqref{l:ENERG13a} and \eqref{l:ENERG13b} since
\begin{align*}
&\Delta^+v^T\Delta^+v+\frac{1}{\gamma^4}\Delta^-v^T\Delta^-v
\\
&=\Delta^{\top +}w^T\Delta^{\top +}w+\Delta^{\top
-}w^T\Delta^{\top -}w
\\
&=\frac{1}{2}\left(\abs{w^\top_2}^2+\abs{w^\top_1}^2\right).
\end{align*}
This concludes the proof.
\end{proof}

In other words, the momentum constraint on the forward and
backward drift $b^+$ and $b^-$ with $Z\neq 0$ with the heatbath
particles moving at average speed $c$ can be transposed into the
case where $Z\equiv 0$ with the main particle in a heatbath
where the particles move at average speed $c_\top$. Here again
$-m\W=\frac{1+\gamma^2}{2}mZ(w_2-w_1)=
-\frac{1+\gamma^2}{2}mZ\Delta^-w$ is the additional momentum.
Moreover, due to \eqref{l:ENERG12b} and \eqref{l:ENERG14} the
momentum conservation requirement for $w_{\top,1}$ and
$w_{\top,2}$ as described in Theorem \eqref{l:HAMILT20} is
equivalent to a requirement on $w_{1}$ and $w_{2}$.  The sole
difference between the original and the transposed case is that
the correlation between $w_{\top,1}$ and $w_{\top,2}$ is
different from the correlation between $w_{1}$ and $w_{2}$ due
to the fact that $\W$ is a (random) function of $w_{1}$ and
$w_{2}$. The exact expression is calculated in Proposition
\eqref{l:HAMILT17} for the case where the main particle follows
a Markovian path. Any changes in correlation going from $w_1$,
$w_2$ to $w_{\top,1}$, $w_{\top,2}$ suggest that the average
inter-collision time is affected as detailed in Section 3.

The next result completes the identification between a heatbath
$w_2$, $w_1$ in which $Z\neq 0$ and a heatbath $w_{\top,2}$,
$w_{\top,1}$ in which $Z\equiv 0$.  While \eqref{l:ENERG14} is
the transposed heatbath equivalent of the momentum constraint
\eqref{l:DRFT9} it is not immediately obvious that
\eqref{l:DRFT8} has an equivalent as well. The next Theorem
shows that this is the case and as an addendum calculates the
original energy in terms of the original heatbath terms.
\begin{thm}
Let the main particle diffuse in a heatbath where
$w_{\top,1}=w_1+\W$, $w_{\top,2}=w_2+\W$ with
$-\W=\frac{1+\gamma^2}{2}Z(w_2-w_1)=
-\frac{1+\gamma^2}{2}Z\Delta^-w$.  Then the total kinetic energy
$\h_{\top,k}$ equals
\begin{subequations}
\begin{align}
\label{l:HAMILT21a}
\begin{split}
\frac{8\h_{\top,k}}{M}
&=\Delta^+v^T\Delta^+v+\frac{1}{\gamma^2}\Delta^-v^T\Delta^-v
\\
&=\frac{1}{2}\abs{v_1}^2+\frac{1}{2}\abs{w_{\top,1}}^2
\\
&=\frac{1}{2}\abs{v_2}^2+\frac{1}{2}\abs{w_{\top,2}}^2,
\end{split}
\end{align}
and
\begin{align}
\label{l:HAMILT21b}
\begin{split}
\frac{8\h_{\top,k}}{M}
&=\Delta^+v^T\Delta^+v+\frac{1}{\gamma^4}\Delta^-v^T\Delta^-v
\\
&\quad+\gamma^2\Delta^+w^T\Delta^+w
+\gamma^4\Delta^-w^T\Delta^-w
\end{split}
\end{align}
\end{subequations}
\end{thm}
\begin{proof}
No calculations are required to show assertion
\eqref{l:HAMILT21a} as the definition of $\W$ in
\eqref{l:ENERG10} and \eqref{l:ENERG11a} shows that
\begin{align}
\label{l:GAMMDEF25}
\begin{pmatrix}
v_2 \\
v_1
\end{pmatrix}
= \frac{\gamma}{\sin(\theta)}
\begin{pmatrix}
\cos(\theta) & 1
\\
1 & \cos(\theta)
\end{pmatrix}
\begin{pmatrix}
w_{\top,2}\\
w_{\top,1}
\end{pmatrix},
\end{align}
using \eqref{l:GAMMDEF22} and \eqref{l:GAMMDEF24}.  From this it
is easy to work back and derive
\begin{align*}
\begin{pmatrix}
v_2 \\
w_{\top,2}
\end{pmatrix}
= \begin{pmatrix}
\cos(\theta) & \gamma\sin(\theta) \\
\frac{\sin(\theta)}{\gamma} & -\cos(\theta)
\end{pmatrix}
\begin{pmatrix}
v_1 \\
w_{\top,1}
\end{pmatrix},
\end{align*}
which implies in turn that \eqref{l:DRFT8} holds with
$w_{\top,1}$, $w_{\top,2}$ replacing  $w_{1}$ and $w_{2}$. Hence
\begin{align*}
\frac{\h_{\top,k}}{M_T}=\frac{1}{2}
\abs{\frac{v_2+v_1}{2}}^2+\frac{1}{2\gamma^2}
\abs{\frac{v_2-v_1}{2}}^2,
\end{align*}
with
$\h_{\top,k}=\frac{M}{2}\abs{v_1}^2+\frac{m}{2}\abs{w_{\top,1}}^2=
\frac{M}{2}\abs{v_2}^2+\frac{m}{2}\abs{w_{\top,2}}^2$ but this
is exactly assertion \eqref{l:HAMILT21a}.

Equation \eqref{l:HAMILT21b} can be shown from straightforward
calculation.  Equation \eqref{e:ENG4} shows that
\begin{align*}
&\frac{8\h_T}{M_T}-\Delta^+w^T\Delta^+w-\gamma^2\Delta^-w^T\Delta^-w
\\
&=-2\Delta^+w^TZ\Delta^-w +
\left(1+\gamma^2\right)\Delta^-w^TZ^TZ\Delta^-w,
\end{align*}
but from \eqref{l:ENERG10} it follows that
\begin{align*}
&\Delta^+v^T\Delta^+v+\frac{1}{\gamma^2}\Delta^-v^T\Delta^-v
\\
&=\Delta^+w^T\Delta^+w+\gamma^2\Delta^-w^T\Delta^-w
\\
&\quad+\left(1+\gamma^2\right)
\begin{pmatrix}
-2\Delta^+w^TZ\Delta^-w
\\
+\left(1+\gamma^2\right)\Delta^+w^TZZ^T\Delta^+w
\end{pmatrix}
\\
&=\Delta^+w^T\Delta^+w+\gamma^2\Delta^-w^T\Delta^-w
\\
&\quad+\left(1+\gamma^2\right)
\begin{pmatrix}
\frac{8\h_T}{M_T}-\Delta^+w^T\Delta^+w
\\
-\gamma^2\Delta^-w^T\Delta^-w
\end{pmatrix}.
\end{align*}
Finally then
\begin{align*}
&\Delta^+v^T\Delta^+v+\frac{1}{\gamma^4}\Delta^-v^T\Delta^-v
\\
&=\frac{8\h_T}{M} -\gamma^2\Delta^+w^T\Delta^+w
\\
&\quad-\gamma^4\Delta^-w^T\Delta^-w,
\end{align*}
which concludes the proof.
\end{proof}

\section{Conclusions}

The purpose of this paper is to revisit the classical diffusion
problem of a main particle moving through a heatbath propelled
by elastic collisions and introduce interaction energy and
momentum considerations.  The main particle path is modeled as
moving linearly from one random collision to the next with
random inter-collision times $\tau_j,j\geq 0$, represented as
second order Gamma distributions.  This pre-limit microscopic
construction models the motion of the main particle by collision
positions $x(t_j),j \geq 0$, at stopping times $t_j$, $j \geq
0$, with $t_{j+1}-t_j=\tau_j$, $j=0,1,...$.  The derived linear
interpolated path $x\left(t,\beta\right),t>0$ constitutes a
"best estimate" of the main particle position.  An important
model assumption here is the choice of the second (or higher)
order gamma distributions as the particle inter-collision times
$\tau_j,j\geq 0$. If the distance traveled between the
collisions $x(t_j)$ and $x(t_{j+1})$ is modeled as
$b^+(x,t)\tau_j$ and a (subordinate) Gaussian contribution then
the inter-collision momentum and energy of the main particle is
finite with probability one.

Section 1 shows that if the mean inter-collision times decreases
the main particle path approaches the strong solution to the
continuous stochastic differential equation with drift
$b^+(x,t)$.  The drift and the variance term must satisfy
standard requirements which guarantee the existence of a strong
unique solution but one additional time growth condition on the
drift was introduced to control the increasing multitude of
"collision" path contributions. Some constraint must be applied
to the time-variability of the drift because otherwise there is
no guarantee that the drift specified at the collision points
does not deviate too much from the continuous drift function.

Section 2 follows the consequences of the fact that the
collisions with the heatbath particle are elastic and introduces
the canonical solution to the collision energy and momentum
conservation constraint. The pre- and post collision velocities
of the colliding main and heatbath are linearly related via a
matrix containing a random collision scattering matrix $Z=Z(U)$
which carries the center of mass line and impact angle collision
information.  For simple collisions for which $Z\equiv 0, U=I$
the canonical solution implies two constraints on the motion of
the main particle.  The first one relating the total kinetic
energy (main and impacting heatbath particle combined) to the
pre- and post-collision motion of the main particle.  The second
relationship has been referred to as the "momentum constraint"
and looks more like a velocity requirement.  If the total
kinetic energy along the path of the main particle is not
constant (on average) then some energy is being transferred
between the main particle and the heatbath.

The main result in this Section shows that if there is no energy
leakage (on average) between the main particle and the heatbath
hence if the total kinetic energy plus possible potential is
conserved then the probability distribution of the position of
the main particle must be derived from Schr\"{o}dinger's
equation. Planck's constant is then replaced by a variance per
unit of mass term $\epsilon$ and further depends on $\gamma$ the
mass ratio.  The total energy functional may contain only
certain suitable potentials satisfying an energy conservation
property. The derivation relies heavily on the stochastic
mechanics results and on convergence of the collision path
representation to a suitable stochastic process.

The important aspect of this derivation is that a combined
energy constraint for a particle diffusing through a heatbath is
a purely classical problem and Schr\"{o}dinger's equation is
invoked to prevent energy exchange between the main particle and
the heatbath.  The analogy with quantum mechanics however is not
perfect as for instance the forward and backward drifts explode
when the mass ratio $\gamma$ becomes small. The present
derivation moreover only allows certain appropriate potentials
though this includes all time-independent potentials and the
electromagnetic potentials see Section 2 and 4.  Apart from the
presence of the mass ratio $\gamma$ there is an important
conceptual difference.  The total energy functional that is
conserved by the presence of the Schr\"{o}dinger wave function
is the total kinetic energy of both the main and heatbath
particle.  Quantum mechanics associates the wave function only
with the "main" particle and the energy eigenvalues of the wave
function supposedly are the energy states of the main particle
alone. In this paper the energy states obtained from the wave
function relate to the combined main particle and heatbath
particle energy.  Once the wave function solution has been
obtained the main particle and heatbath particle energy
contributions still need to be separated.

Yet another difference between the present formulation and the
Theory of Quantum Mechanics is the curious aspect noted in the
Gaussian Wave Packet example \eqref{l:QM1} where the statistical
characteristics of both the main and the heatbath particles are
affected by the energy constraint.  The kinetic and energy
dispersion terms that are found in the energy of the main
particle can also be found in the heatbath energies so the
heatbath is affected by the presence of the main particle in a
time dependent manner.  This effect decreases if the mass ratio
$\gamma$ becomes smaller but the model can not be structurally
amended. The results in Section 3 suggest in fact that a
different approach may be required which incorporates
correlation between the main particle pre- and post collision
velocities.

Intuition suggests that the diffusion/quantal effect becomes
noticeable if the diffusion per mass term is significant in
comparison to the size of the drift term or the energy
dispersion.  The example of the Gaussian Wave particle in
Section 2 shows that the energy dispersion and the average main
particle momentum increase the combined kinetic energy.  This
suggests that a high temperature heatbath environment that is
not not overly dense is dominated by its energy dispersion and
mean particle drift.  On the other hand if the main particle
moves through a relatively narrow energy band with a relatively
small kinetic motion then the diffusion term constitutes the
larger part of the total kinetic energy.  Ultimately if the
heatbath is very dense then the diffusion term will become the
dominant energy provider.  A very dense heatbath environment
eventually forces the dominant motion of the main particle to be
entirely diffusive.

The example also shows that the heatbath looks like a reflection
of the main particle. The main particle carries almost all the
momentum and dispersion energy with little diffusion energy
while the heatbath has a very large diffusion term and very
little kinetic energy.  If the pre- and post collision
velocities of the main particle are uncorrelated as one would
expect for a Markovian path then the pre- and post collision
velocities for the heatbath particles will be correlated through
the collision point. Both the main and heatbath particle kinetic
energies are time dependent but the first one increases in size
while the heatbath energy decreases proportionally.  For a very
small mass ratio $\gamma$ the time dependence almost disappears
and the heatbath starts to behave as if it has only one velocity
which is reflected by the collision.

The momentum constraint established in Section 2 shows that a
similar symmetric quadratic expression employing the forward and
backward velocities of the main particle can be directly related
to the average velocity of the incident particle.  The average
is calculated as the arithmetic average of forward and backward
heatbath particle velocity.  The contribution of Section 3 is to
show that if the heatbath particles are in energetic equilibrium
with the main particle then the drift of the main particle and
the correlation between the forward and backward velocities of
the colliding heatbath particles must depend on the average
inter-collision time.  The condition is identical to the
geometrical Minskowski invariant employed in Special Relativity
and it is shown that the invariant can be satisfied by applying
the Lorentz transformation to the average collision time and the
squared distance traveled.

The analogy with Relativity Theory can be pushed further to
suggest that the energy rest mass of the main particle equals
its diffusive energy.  In fact the rest mass can be expressed in
terms of the mass ratio $\gamma$, the average inter-collision
time $\overline{\tau}_0$ for the main particle at rest in the
heatbath and the variance per mass ratio $\epsilon$. This
argument only employs the Lorentz transformation as a means of
generating a solution to the Minkowski invariant and is not
necessarily the only solution which balances the mean
inter-collision time and the squared distance traveled.  By
comparison in Relativity Theory the homogeneity of space and the
constancy of the speed of light in all directions leads to a
unique solution.

The third set of results in section 4 focusses on the
"non-simple" solution to the elastic collisions to include the
random collision scattering matrix $Z$.  The path of the main
particle now becomes a "conditionally" Gaussian process in the
sense that the main particle path process remains a Gaussian
process given the realizations of the random collision
scattering matrix $Z$. In general however the additional random
matrix $Z$ alters the main particle coordinate distribution. The
stochastic dynamics for the random matrix involve the center of
mass line distribution and depend on the dimensions and physical
setting rather than on the motion of the main particle.  It is
therefore reasonable to assume that the collision scattering
matrix $Z$ does not depend on the pre- and post velocities of
the main particle.

One of the most obvious effects of the random collision
scattering matrix is the arbitrary change in correlation between
the pre- and post collision heatbath velocities causing in turn
a correlation between the pre- and post-collision velocity of
the main particle. This destroys the Markovian property of the
main particle path process so that additional variables must be
introduced to model its motion. None of the results in Section 2
are then applicable because the fundamental Markovian result
relating the backward and forward drift difference and main
particle position probability density is not valid. However
Section 3 demonstrates how the pre- and post-collision
velocities correlation of the main particle process relates to
the mean inter-particle collision time which can then be
captured by the Lorentz transformation. Unfortunately there are
no results that describe how the probability density of the main
particle position can be calculated in the presence of
correlation.

Theorem \eqref{l:HAMILTPRP1} in Section 4 presents the main
result showing that the conservation of total energy depends on
the mean scattering matrix $\overline{Z}$ and expected
covariance matrix $E\left[ZZ^T\right]$.  A full solution for the
probability density has not been derived but for the case where
$Z\equiv 0$ it was shown again that the probability density must
be obtained employing Schr\"{o}dinger's equation as long as the
potential is time independent or satisfies a Maxwell type set of
equations.  Another example is presented for the case where
$\overline{Z}$ is a function of time only while
$E\left[ZZ^T\right]$ is a constant diagonal matrix. Then it is
possible to obtain the probability for the main particle
position in the form of a wave function for the $Z\equiv 0$ case
but with some of the weighting parameters altered.  In
comparison to the $Z\equiv 0$ wave function this solution looks
as if the mass weighting and the diffusion per unit of mass have
changed to account for the collision scattering matrix.

A very useful result from this Section is that it is possible to
absorb the collision scattering matrix $Z$ into the heatbath.
Specifically it is possible to represent the case with a
non-zero scattering matrix $Z\neq 0$ in a $w_{1}$, $w_{2}$
heatbath with the case where $Z\equiv 0$ with a
$w_{\top,1}=w_1+\W$, $w_{\top,2}=w_2+\W$ heatbath.  The
additional momentum $m\W$ is proportional to the collision
scattering matrix $Z$ and acts as a straight increase in the
energy in the heatbath while it is orthogonal to the momentum
exchange. However the correlation structure for $w_{\top,1}$,
$w_{\top,2}$ is different from the correlation matrix for
$w_{1}$ and $w_{2}$. Therefore the path for a diffusing main
particle can not be Markovian in both the $w_{1}$, $w_{2}$
heatbath and the $w_{\top,1}$, $w_{\top,2}$ heatbath
simultaneously see Proposition \eqref{l:HAMILT17}.

The results in Section 2 do not apply when the main particle
path is not a Markovian process however the results in Section 3
take correlation into account as long as the correlation is
homogeneous.  In fact Theorem \eqref{l:HAMILT18} establishes a
relationship between the mean particle speed
$c^2=\abs{w_1}^2=\abs{w_2}^2$, the compound correlation $\rho_v
$ and the mean inter-particle collision time $\overline{\tau}$.
As a result of \eqref{l:ENERG12b} this exact relationship must
hold for for some
$c_\top^2=\abs{w_{\top,1}}^2=\abs{w_{\top,2}}^2$, $\rho'_v $ and
$\overline{\tau}'$ as well.  The results of Section 2 may still
provide a good Markovian approximation if the $w_{\top,1}$,
$w_{\top,2}$ heatbath renders the pre- and post-collision
velocities of the main particle independent.

From section 3 it is clear that the motion for a main particle
in the presence of correlation between pre- and post-collision
velocities must satisfy a relativistic invariance hence
extrapolating from results in Section 2 and Section 4 a proper
distribution for the main particle position is likely to satisfy
a type of Klein-Gordon equation.  This line of research should
pursue the ideas of Serva~\cite{SERVA1} or Guerra~\cite{GUERRA1}
and will be investigated in Part II of this paper.



\bibliographystyle{phcpc}
\bibliography{xbib}

\appendix

\section{}
\label{l:APP7} \noindent {\bf Proof of Theorem
\eqref{l:THMENERG}} Equations \eqref{l:DRFT8} can of course be
verified by direct substitution however the following simple
argument is more intuitive.  Using $p_1=Mv_1$, $p_2=Mv_2$,
$q_1=Mw_1$ and $q_2=Mw_2$ it follows from equation
\eqref{l:ENERG2} that
\begin{align*}
q_2+q_1
&=p_1 \gamma\sin(\theta)-q_1(\cos(\theta)-1) \\
&=p_1 \gamma\sin(\theta)+q_12\sin\left(\theta/2\right)^2  \\
&=2p_1 \gamma\sin\left(\theta/2\right)\cos\left(\theta/2\right)
+2q_1\sin\left(\theta/2\right)^2 ,
\end{align*}
so
\[
\frac{q_2+q_1}{2} =\sin\left(\theta/2\right)\left(p_1
\gamma\cos\left(\theta/2\right)+q_1\sin\left(\theta/2\right)
\right).
\]

Similarly using momentum conservation $p_2-p_1=-(q_2-q_1)$ and
equation \eqref{l:ENERG2} results in
\begin{align*}
\gamma\frac{q_2-q_1}{2} &=\frac{\gamma}{2}p_1
(1-\cos(\theta))-q_1\frac{1}{2}\sin(\theta)
\\
&=p_1 \gamma\left(\sin\left(\theta/2\right)\right)^2-
q_1\sin\left(\theta/2\right)\cos\left(\theta/2\right)
\\
&=\sin\left(\theta/2\right)\left(\gamma p_1
\sin\left(\theta/2\right)-q_1\cos\left(\theta/2\right)\right),
\end{align*}
and adding the squares results in
\begin{align*}
&\abs{\frac{q_2+q_1}{2}}^2+\gamma^2 \abs{\frac{q_2-q_1}{2}}^2
\\
&=\gamma^2\sin\left(\theta/2\right)^2
\left(\abs{p_1}^2+\frac{\abs{q_1}^2}{\gamma^2}\right)
\\
&=2M\h_k\gamma^2\sin\left(\theta/2\right)^2=
\frac{2M\gamma^4}{1+\gamma^2}\h_k.
\end{align*}

Substituting
\begin{align*}
q_2-q_1 &=p_1-p_2, \\
q_2+q_1 &=\gamma^2(p_1+p_2),
\end{align*}
into this yields
\[
\gamma^4\abs{\frac{p_2+p_1}{2}}^2+\gamma^2 \abs{\frac{p_2-p_1}{2}}^2
= \frac{2M\gamma^4}{1+\gamma^2}\h_k,
\]
hence
\[
\abs{\frac{p_2+p_1}{2}}^2+\frac{1}{\gamma^2}
\abs{\frac{p_2-p_1}{2}}^2 = \frac{2M}{1+\gamma^2}\h_k,
\]
which covers the first two expressions in Theorem
\eqref{l:THMENERG}. Now dividing this equation by $M^2$ yields
\[
\abs{\frac{v_2+v_1}{2}}^2+\frac{1}{\gamma^2}
\abs{\frac{v_2-v_1}{2}}^2 = \frac{2}{M(1+\gamma^2)}\h_k
=\frac{2}{M_T}\h_k,
\]
proving equation (\ref{l:DRFT8}).

Direct substitution can be employed again to obtain equation
(\ref{l:DRFT9}) but this equation can be derived more
organically from the pre- and post-collision total kinetic
energy and equation (\ref{l:DRFT8}).

So for a direct derivation write
\begin{subequations}
\begin{align}
\label{l:DRFT10a}
\h_k &=\frac{M}{2}\abs{v_2}^2+\gamma^2\abs{w_2}^2
\\
\label{l:DRFT10b}
&=\frac{M(1+\gamma^2)}{2}
\left(\abs{\frac{v_2+v_1}{2}}^2+\frac{1}{2\gamma^2}
\abs{\frac{v_2-v_1}{2}}^2 \right)
\\
\label{l:DRFT10c}
&=\frac{M}{2}\left(\abs{v_1}^2+\gamma^2\abs{w_1}^2\right),
\end{align}
\end{subequations}
so that
\begin{align*}
& M(1+\gamma^2)
\left(\abs{\frac{v_2+v_1}{2}}^2+\frac{1}{2\gamma^2}
\abs{\frac{v_2-v_1}{2}}^2 \right)\\
& =\frac{M}{2}(\abs{v_2}^2+\abs{v_1}^2)+M\gamma^2 \abs{w_1}^2+
\frac{M\gamma^2}{2}\left(\abs{w_2}^2-\abs{w_1}^2\right),
\end{align*}
which is achieved by adding \eqref{l:DRFT10a} and \eqref{l:DRFT10c}
and equating that to twice the middle term \eqref{l:DRFT10b}.

Carrying the $v_2$, $v_1$ terms to the left hand side then
yields
\begin{align*}
\frac{M(1+\gamma^2)}{4}
&\left(\left(\abs{v_2}^2+2v^T_2v_1+\abs{v_1}^2\right)+\frac{1}{\gamma^2}
\left(\abs{v_2}^2-2v^T_2v_1+\abs{v_1}^2\right) \right)
\\
&-\frac{M}{2}(\abs{v_2}^2+\abs{v_1}^2)
\\
=\frac{M(1+\gamma^2)}{4} &\left(
\left(\abs{v_2}^2+\abs{v_1}^2\right)\left(1+\frac{1}{\gamma^2}\right)
+2v^T_2v_1\left(1-\frac{1}{\gamma^2}\right) \right)
\\
&-\frac{M}{2}(\abs{v_2}^2+\abs{v_1}^2)
\\
&=M\gamma^2 \abs{w_1}^2+
\frac{M\gamma^2}{2}\left(\abs{w_2}^2-\abs{w_1}^2\right).
\end{align*}

Dividing both sides of this expression by
$\frac{M(1+\gamma^2)}{4}$ yields
\begin{align*}
&
\left(\abs{v_2}^2+\abs{v_1}^2\right)\left(1+\frac{1}{\gamma^2}\right)
+2v^T_2v_1\left(1-\frac{1}{\gamma^2}\right)
\\
&-\frac{2}{1+\gamma^2}(\abs{v_2}^2+\abs{v_1}^2)
\\
&=\frac{4\gamma^2}{1+\gamma^2} \abs{w_1}^2+
\frac{2\gamma^2}{1+\gamma^2}\left(\abs{w_2}^2-\abs{w_1}^2\right),
\end{align*}
or
\begin{align*}
&
\left(\abs{v_2}^2+\abs{v_1}^2\right)\left(1+\frac{1}{\gamma^2}-
\frac{2}{1+\gamma^2}\right)
+2v^T_2v_1\left(1-\frac{1}{\gamma^2}\right)
\\
&=\frac{4\gamma^2}{1+\gamma^2} \abs{w_1}^2+
\frac{2\gamma^2}{1+\gamma^2}\left(\abs{w_2}^2-\abs{w_1}^2\right).
\end{align*}
Now the constant in this equation can be written as
\begin{align*}
\left(1+\frac{1}{\gamma^2}- \frac{2}{1+\gamma^2}\right)
=\frac{1+\gamma^4}{\gamma^2\left(1+\gamma^2\right)},
\end{align*}
so that
\begin{align*}
&\left(\abs{v_2}^2+\abs{v_1}^2\right)
\left(\frac{1+\gamma^4}{\gamma^2\left(1+\gamma^2\right)} \right)
+2v^T_2v_1\left(1-\frac{1}{\gamma^2}\right)
\\
&=\frac{4\gamma^2}{1+\gamma^2} \abs{w_1}^2+
\frac{2\gamma^2}{1+\gamma^2}\left(\abs{w_2}^2-\abs{w_1}^2\right).
\end{align*}
Dividing by $\frac{1+\gamma^4} {\gamma^2\left(1+\gamma^2\right)}
$ yields
\begin{align}
\label{l:App2}
\begin{split}
&\left( \left(\abs{v_2}^2+\abs{v_1}^2\right)
-2v^T_2v_1\left(\frac{1-\gamma^4}{1+\gamma^4}\right)\right)
\\
&=\frac{4\gamma^4}{1+\gamma^4} \abs{w_1}^2+
\frac{2\gamma^4}{1+\gamma^4}\left(\abs{w_2}^2-\abs{w_1}^2\right),
\end{split}
\end{align}
and if $\theta^2=\frac{1-\gamma^4}{1+\gamma^4}$ then
$1-\theta^2=\frac{2\gamma^4}{1+\gamma^4}$ and so \eqref{l:App2}
reduces to
\begin{align*}
&\left( \left(\abs{v_2}^2+\abs{v_1}^2\right)
-2v^T_2v_1\theta^2\right)\\
&=(1-\theta^2)\left(2
\abs{w_1}^2+\left(\abs{w_2}^2-\abs{w_1}^2\right)\right).
\end{align*}
Some further rewriting shows that
\begin{align*}
&\frac{(1-\theta^2)}{2}\abs{v_2+v_1}^2+
\frac{(1+\theta^2)}{2}\abs{v_2-v_1}^2\\
&=(1-\theta^2)\left(2 \abs{w_1}^2+\abs{w_2}^2-\abs{w_1}^2\right),
\end{align*}
or
\begin{align*}
&\abs{v_2+v_1}^2+
\frac{(1+\theta^2)}{(1-\theta^2)}\abs{v_2-v_1}^2\\
&=2\left(2 w^2_1+\left(w^2_2-w^2_1\right)\right),
\end{align*}
so that finally
\begin{align*}
&\abs{\frac{v_2+v_1}{2}}^2+
\frac{1}{\gamma^4}\abs{\frac{v_2-v_1}{2}}^2\\
&=\frac{1}{2}\left(\abs{w_1}^2+\abs{w_1}^2\right),
\end{align*}
since $\frac{1+\theta^2}{1-\theta^2}=\frac{1}{\gamma^4}$. This
concludes the proof.

\section{}
\label{l:APP6}
\noindent {\bf Proof of Theorem
\eqref{l:ENERG16}.}  This Theorem can be generalized slightly to
the case where $M$ and $m$ are matrices. So if the collision
matrix can be written as $\Gamma=\left(
\begin{smallmatrix}
P & Q \\
V & G
\end{smallmatrix}\right)$
then the solution to equations
\eqref{l:COLLMAT3a}-\eqref{l:COLLMAT3c} equal
\begin{align}
\label{e:AVQ1}
\begin{split}
&P=X(m,M)+\left(M+Mm^{-1}M\right)^{-\frac{1}{2}}UY(m,M)^{\frac{1}{2}},
\\
&G=X(M,m)+\left(m+mM^{-1}m\right)^{-\frac{1}{2}}UY(M,m)^{\frac{1}{2}}
,
\\
&V=m^{-1}M(I-P),
\\
&Q=M^{-1}m(I-S),
\end{split}
\end{align}
with
\begin{align*}
X(m,M)=&(M+m)^{-1}M,
\\
Y(m,M)=&M-Mm^{-1}M
\\
&+X(m,M)^T\left(M+Mm^{-1}M\right)X(m,M).
\end{align*}
This reduces to \eqref{l:COLLMAT1} if $M$ and $m$ become
diagonal.  Notice that solution \eqref{l:COLLMAT1} can also be
written as
\begin{align*}
&P=\cos\left(\theta/2\right)^2\left(I-\gamma^2U\right),
Q=\sin\left(\theta/2\right)^2\left(I+U\right), \\
&V=\cos\left(\theta/2\right)^2\left(I+U\right) ,
G=\sin\left(\theta/2\right)^2\left(I-\frac{1}{\gamma^2}U\right),\\
&U^TU=I.
\end{align*}

To prove that \eqref{l:COLLMAT1} and \eqref{e:AVQ1} above are
solutions to equations \eqref{l:COLLMAT3a} - \eqref{l:COLLMAT3c}
notice that \eqref{l:COLLMAT3a} is equivalent to
$V=m^{-1}M(I-P)$. Substituting that into $P^TMP+V^TmV=M$ yields
\begin{align*}
&P^TMP+(I-P)^TMm^{-1}M(I-P)=M,
\end{align*}
or
\begin{align*}
&P^T(M+Mm^{-1}M)P-P^TMm^{-1}M
\\
&-Mm^{-1}MP+Mm^{-1}M=M,
\end{align*}
so that finally
\begin{align}
\label{e:QE3} &(P-X)^T(M+Mm^{-1}M)(P-X)=Y,
\end{align}
with
\begin{align*}
X(m,M)=&(M+Mm^{-1}M)^{-1}Mm^{-1}M=(m+M)^{-1}M, \\
Y(m,M)=&M-Mm^{-1}M
\\
&+X(m,M)^T(M+Mm^{-1}M)^{-1}X(m,M).
\end{align*}

The solution to equation \eqref{e:QE3} equals
\begin{align}
\label{e:QE4}
&P=X(m,M)+(M+Mm^{-1}M)^{-\frac{1}{2}}UY(m,M)^{\frac{1}{2}},
\end{align}
and $V$ follows from $V=m^{-1}M(I-P)$ so that
\begin{align}
\label{e:QE4a}
\begin{split}
&V=m^{-1}M\left(I-X(m,M)-(M+Mm^{-1}M)^
{-\frac{1}{2}}UY(m,M)^{\frac{1}{2}}\right)
\\ &=m^{-1}M\left((m+M)^{-1}m-(M+Mm^{-1}M)^{-\frac{1}{2}}UY(m,M)^
{\frac{1}{2}}\right),
\end{split}
\end{align}

Specifically if $M=MI$,$m=mI$ (i.e. the matrices $m,M$ equal the
constant masses $m,M$ times the unit matrix $I$) then
\eqref{e:QE3} simplifies to
\begin{align*}
P^TP-\frac{\sin(\theta)}{2\gamma}\left(P^T+P\right)+cos(\theta)I=0.
\end{align*}
Then
\begin{align*}
\left(P-\frac{sin(\theta)}{2\gamma}\right)^T
\left(P-\frac{sin(\theta)}{2\gamma}\right)-\gamma^2\frac{sin^2(\theta)}{4}I=0,
\end{align*}
since
\begin{align*}
&\frac{sin^2(\theta)}{4\gamma^2}-\gamma^2\frac{sin^2(\theta)}{4}
\\
&=\left(\frac{1}{1+\gamma^2}\right)^2-
\frac{\gamma^4}{\left(1+\gamma^2\right)^2}=cos(\theta)
\end{align*}
Equation \eqref{e:QE4} then simplifies to
\begin{align*}
P=\frac{sin(\theta)}{2\gamma}\left(I-\gamma^2U\right),
\end{align*}
for some unitary matrix $U$ and so equation \eqref{e:QE4a} for
the matrix $V$ reduces to
\begin{align*}
V=\frac{sin(\theta)}{2\gamma}\left(I+U\right).
\end{align*}

Using the same approach the matrix $G$ can be calculated as
\begin{align}
\label{e:QE5} &G=X(M,m)+(m+mM^{-1}m)^{-\frac{1}{2}}U_\top
Y(M,m)^{\frac{1}{2}},
\end{align}
for another arbitrary unitary matrix $U_\top$.  Again, if
$M=MI$,$m=mI$ (i.e. the matrices $m,M$ equal the constant masses
$m,M$ times the unit matrix $I$), then
\begin{align}
\label{e:QE6}
G&=\frac{\gamma\sin(\theta)}{2}\left(I-\frac{1}{\gamma^2}V\right),
\\
\label{e:QE6a}
Q&=\frac{\gamma\sin(\theta)}{2}\left(I+V\right).
\end{align}
Finally from the fourth equation
\begin{align*}
0=\frac{\sin^2(\theta)}{4}\left(1+\gamma^2\right)\left(I-U^TU_\top\right),
\end{align*}
from which follows $UU_\top=I$ so that $U=U_\top$ ($U$ and
$U_\top$ are unitary) and the proof is complete.

\bigskip
\noindent {\bf Proof of Theorem \eqref{l:THMENERG1}.} Using
\eqref{l:GAMMDEF1} and denoting $\Delta^+v=v_2+v_1$,
$\Delta^-v=v_2-v_1$ it is clear that
\begin{align*}
\begin{pmatrix}
\Delta^+v \\
\Delta^-v
\end{pmatrix}
&=
\begin{pmatrix}
v_2+v_1 \\
v_2-v_1
\end{pmatrix}
=
\begin{pmatrix}
P+I & Q \\
P-I & Q
\end{pmatrix}
\begin{pmatrix}
v_1 \\
w_1
\end{pmatrix}
\\
&=\begin{pmatrix}
P+I & Q \\
-Q & Q
\end{pmatrix}
\begin{pmatrix}
v_1 \\
w_1
\end{pmatrix},
\end{align*}
so inverting yields
\begin{align}
\label{e:ENG6}
\begin{pmatrix}
v_1 \\
w_1
\end{pmatrix}
&=
\begin{pmatrix}
P+I & Q \\
-Q & Q
\end{pmatrix}^{-1}
\begin{pmatrix}
\Delta^+v \\
\Delta^-v
\end{pmatrix},
\end{align}
which can be entered into the total kinetic energy expression
$\h_k=v^2_1 + w^2_1=v^2_2 + w^2_2$ once the inverse of this
matrix has been determined.  Since $P+Q=I$ it is clear that
$P+Q+I=2I$ so some manipulation shows that
\begin{align}
\label{a:THM6}
\begin{pmatrix}
P+I & Q \\
-Q & Q
\end{pmatrix}^{-1}
=\frac{1}{2}
\begin{pmatrix}
I & -I \\
I & \left(2Q^{-1}-I\right)
\end{pmatrix}.
\end{align}
Hence
\begin{align*}
&\begin{pmatrix}
P^T+I & -Q^T \\
Q^T & Q^T
\end{pmatrix}^{-1}
\begin{pmatrix}
I & 0 \\
0 & \gamma^2
\end{pmatrix}
\begin{pmatrix}
P+I & Q \\
-Q & Q
\end{pmatrix}^{-1}
\\
&=\begin{pmatrix}
I & I \\
-I & \left(2Q^{-T}-I\right)
\end{pmatrix}
\begin{pmatrix}
I & 0 \\
0 & \gamma^2
\end{pmatrix}
\begin{pmatrix}
I & -I \\
I & \left(2Q^{-1}-I\right)
\end{pmatrix}
\\
&=\begin{pmatrix}
\left(1+\gamma^2\right)I & -I+\gamma^2V_\Gamma \\
-I+\gamma^2V_\Gamma^T & I+\gamma^2V_\Gamma^TV_\Gamma
\end{pmatrix}
,
\end{align*}
where $V_\Gamma=\left(2Q^{-1}-I\right)$.

As a result
\begin{align}
\label{a:THM7}
\begin{split}
\frac{8\h_k}{M} =&
\begin{pmatrix}
v^T_1 & w^T_1
\end{pmatrix}
\begin{pmatrix}
I & 0 \\
0 & \gamma^2
\end{pmatrix}
\begin{pmatrix}
v_1 \\ w_1
\end{pmatrix}
\\
=&
\begin{pmatrix} \Delta^+v^T & \Delta^-v^T
\end{pmatrix}
\begin{pmatrix}
I & \gamma^2I \\
-I & \gamma^2V_\Gamma^T
\end{pmatrix}
\begin{pmatrix}
I & -I \\
I & V_\Gamma
\end{pmatrix}
\begin{pmatrix}
\Delta^+v \\ \Delta^-v
\end{pmatrix}
\\
=&\begin{pmatrix} \Delta^+v^T & \Delta^-v^T
\end{pmatrix}
\begin{pmatrix}
\left(1+\gamma^2\right)I & -I+\gamma^2V_\Gamma \\
-I+\gamma^2V_\Gamma^T & I+\gamma^2V_\Gamma^TV_\Gamma
\end{pmatrix}
\begin{pmatrix}
\Delta^+v
\\\Delta^-v
\end{pmatrix}
\\
=&\left(1+\gamma^2\right)\Delta^+v^T\Delta^+v -2
\Delta^+v^T\left(I-\gamma^2V_\Gamma\right)\Delta^-v
\\
&+\Delta^-v^T\left(I+\gamma^2V_\Gamma^TV_\Gamma\right)\Delta^-v
\end{split}
\end{align}
To further simplify the appearance of this expression the
following Lemma is required.

\begin{lem}
\label{e:LEM1} Let $Z$ be the collision scattering matrix
$Z=I-2(I+U)^{-1}=I-\gamma\sin(\theta)Q^{-1}$ with $U$ the
unitary matrix and $Q$ defined in equation \eqref{e:QE6a}. Then
it is true that
\begin{subequations}
\begin{align}
\label{e:QE10a}
&Q^{-1}=\frac{1+\gamma^2}{\gamma^2}\left(I+U\right)^{-1},
\\
\label{e:QE10b}
&Q^{-T}+Q^{-1}=\frac{2}{\gamma sin(\theta)}I,
\end{align}
\end{subequations} and
\begin{subequations}
\begin{align}
\label{e:QE7a}
&Z^{-T}+Z^{-1}=0,
\\
\label{e:QE7b} &\frac{1-\gamma^2V_\Gamma}{1+\gamma^2}=Z,
\\
\label{e:QE7c} &\frac{1+\gamma^2V_\Gamma^TV_\Gamma}{1+\gamma^2}
=\frac{1}{\gamma^2}I +\frac{1+\gamma^2}{\gamma^2}Z^TZ.
\end{align}
\end{subequations}
\end{lem}
\begin{proof}
Using result \eqref{l:COLLMAT1} it is clear that
\begin{align*}
U=\frac{2}{\gamma
\sin(\theta)}Q-I=\frac{1+\gamma^2}{\gamma^2}Q-I
\end{align*}
for the unitary matrix $U$.  Hence,
\begin{align*}
I=&U^TU=\left(\frac{2}{\gamma \sin(\theta)}Q-I\right)^T
\left(\frac{2}{\gamma \sin(\theta)}Q-I\right)
\\
&=\frac{4Q^TQ}{\gamma^2 \sin^2(\theta)}-\frac{2}{\gamma
\sin(\theta)}\left(Q^T+Q\right)+I,
\end{align*}
so that
\begin{align*}
\frac{2Q^TQ}{\gamma \sin(\theta)}=\left(Q^T+Q\right).
\end{align*}
Multiplying left with matrix $Q^{-T}$ and right with matrix $Q^{-1}$
it follows that
\begin{align*}
\frac{2}{\gamma
\sin(\theta)}I=\left(Q^{-T}+Q^{-1}\right),
\end{align*}
which demonstrates equation \eqref{e:QE10b}.

Substituting definition $Q^{-1}=(I-Z)/\left(\gamma
\sin(\theta)\right)$ into \eqref{e:QE10b} yields
\begin{align*}
&\frac{2}{\gamma \sin(\theta)}I=\left(\frac{(I-Z)^T}{\gamma
\sin(\theta)}+\frac{(I-Z)}{\gamma \sin(\theta)}\right)
\\
&=\frac{2}{\gamma \sin(\theta)}I-\frac{1}{\gamma
\sin(\theta)}\left(Z^T+Z\right),
\end{align*}
from which follow $Z^T+Z=0$ proving \eqref{e:QE7a} and equation
\eqref{e:QE7b} follows from
\begin{align*}
\left(I-\gamma^2V_\Gamma\right)&=I-\gamma^2\
\left(\frac{2\left(1+\gamma^2\right)}{\gamma^2}(I+U)^{-1}-I\right)
\\
&=\left(I+\gamma^2\right)\left(I-2(I+U)^{-1}\right)
\\
&=
\left(I+\gamma^2\right)\left(I-\gamma\sin(\theta)Q^{-1}\right)=
\left(1+\gamma^2\right)Z.
\end{align*}
Finally to demonstrate \eqref{e:QE7c} use the definition
$V_\Gamma=2Q^{-1}-I$ again to show that
\begin{align*}
V_\Gamma&=2\left(\frac{I-Z}{\gamma\sin(\theta)}\right)-I=\frac{2}{\gamma\sin(\theta)}I-
I-2\frac{Z}{\gamma\sin(\theta)}
\\
&=\frac{1}{\gamma^2} \left(I-\left(1+\gamma^2\right)Z\right),
\end{align*}
so that
\begin{align*}
I+\gamma^2V_\Gamma^TV_\Gamma&=1+\frac{1}{\gamma^2}\left(I-
\left(1+\gamma^2\right)Z^T\right)\left(I-\left(1+
\gamma^2\right)Z\right)
\\
&=\frac{\left(1+\gamma^2\right)}{\gamma^2}I+\frac{\left(1+\gamma^2\right)^2}{\gamma^2}
\left(Z^TZ\right)
\\
&=\frac{\left(1+\gamma^2\right)}{\gamma^2}\left(I+\left(1+\gamma^2\right)
\left(Z^TZ\right)\right),
\end{align*}
which concludes the Lemma.
\end{proof}
Applying \eqref{e:QE7a}-\eqref{e:QE7c} to \eqref{a:THM7} yields
\begin{align*}
\frac{8\h_k}{M_T}=& \Delta^+v^T\Delta^+v -2
\Delta^+v^T\frac{\left(I-\gamma^2V_\Gamma\right)}
{\left(1+\gamma^2\right)}\Delta^-v
\\
&+\Delta^-v^T\frac{\left(I+\gamma^2V_\Gamma^TV_\Gamma\right)}
{\left(1+\gamma^2\right)}\Delta^-v
\\
=&\Delta^+v^T\Delta^+v -2 \Delta^+v^TZ\Delta^-v+
\frac{1}{\gamma^2}\Delta^-v^T\Delta^-v
\\
&+\left(\frac{1+\gamma^2}{\gamma^2}\right)\Delta^-v^TZ^TZ\Delta^-v,
\end{align*}
which is equivalent to equation \eqref{l:DRFTEQ8} and this
concludes the proof.

\bigskip
The total kinetic energy $\h_k$ can also be expressed in terms
of the velocities of the heatbath particle and the same random
matrix $Z$ as is shown in the following Theorem.
\begin{thm}
The combined energy of the main and colliding heatbath particle
$\h_k$ can be expressed as
\begin{align}
\label{e:ENG4}
\begin{split}
\frac{8\h_k}{M_T} =&\Delta^+w^T\Delta^+w -2
\Delta^+w^TZ\Delta^-w
\\
&+ \gamma^2\Delta^-w^T\Delta^-w +
\left(1+\gamma^2\right)\Delta^-w^TZ^TZ\Delta^-w,
\end{split}
\end{align}
where
\begin{align*}
&\Delta^+w=w_2+w_1, \\
&\Delta^-w=w_2-w_1.
\end{align*}
As previously the anti-symmetric matrix $Z$ is defined as
$Z=I-2(I+U)^{-1}=I-\gamma\sin(\theta)Q^{-1}$ with $U$ a unitary
matrix and $Q$ as defined in \eqref{l:COLLMAT1}.
\end{thm}
\begin{proof}
Using  \eqref{l:THMENERG1} again it is clear that
\begin{align*}
\begin{pmatrix}
\Delta^+w \\
\Delta^-w
\end{pmatrix}
=&
\begin{pmatrix}
w_2+w_1 \\
w_2-w_1
\end{pmatrix}
\\
=&
\begin{pmatrix}
V & G+I \\
V & G-I
\end{pmatrix}
\begin{pmatrix}
v_1 \\
w_1
\end{pmatrix}
\\
=&
\begin{pmatrix}
V & G+I \\
V & -V
\end{pmatrix}
\begin{pmatrix}
v_1 \\
w_1
\end{pmatrix}
\end{align*}
so that
\begin{align*}
\begin{pmatrix}
v_1 \\
w_1
\end{pmatrix}
=\begin{pmatrix}
V &  G+I\\
V & -V
\end{pmatrix}^{-1}
\begin{pmatrix}
\Delta^+v \\
\Delta^-v
\end{pmatrix}.
\end{align*}
As before use $V+G=I$ so that $V+G+I=2I$ to show that the
inverse can be calculated as follows
\begin{align*}
\begin{pmatrix}
V & G+I \\
V & -V
\end{pmatrix}^{-1}
=\frac{1}{2}
\begin{pmatrix}
I & \left(2V^{-1}-I\right) \\
I & -I
\end{pmatrix}
=\frac{1}{2}
\begin{pmatrix}
I & X \\
I & -I
\end{pmatrix}.
\end{align*}
with $X=\left(2V^{-1}-I\right)$ so that
\begin{align}
\label{a:THM8}
\begin{split}
&\begin{pmatrix}
V^T & V^T \\
G^T+I & -V^T
\end{pmatrix}^{-1}
\begin{pmatrix}
I & 0 \\
0 & \gamma^2
\end{pmatrix}
\begin{pmatrix}
V & G+I \\
V & -V
\end{pmatrix}^{-1}
\\
&=
\begin{pmatrix}
I & I \\
X^T & -I
\end{pmatrix}
\begin{pmatrix}
I & 0 \\
0 & \gamma^2
\end{pmatrix}
\begin{pmatrix}
I & X \\
I & -I
\end{pmatrix}
\\
&=\begin{pmatrix}
\left(1+\gamma^2\right)I & X-\gamma^2I \\
X^T-\gamma^2I & X^TX+\gamma^2I
\end{pmatrix}.
\end{split}
\end{align}
Hence
\begin{align}
\label{e:ENG5}
\begin{split}
&\frac{8\h_k}{M} =\begin{pmatrix} v_1^T & w_1^T
\end{pmatrix}
\begin{pmatrix}
I & 0 \\
0 & \gamma^2I
\end{pmatrix}
\begin{pmatrix}
v_1^T
\\
w_1^T
\end{pmatrix}
\\
&=\begin{pmatrix} \Delta^+w^T & \Delta^-w^T
\end{pmatrix}
\begin{pmatrix}
\left(1+\gamma^2\right)I & X-\gamma^2I \\
X^T-\gamma^2I & X^TX+\gamma^2I
\end{pmatrix}
\begin{pmatrix}
\Delta^+w
\\
\Delta^-w
\end{pmatrix}
\\
&=\left(1+\gamma^2\right)\Delta^+w^T\Delta^+w -2
\Delta^+w^T\left(\gamma^2I-X\right)\Delta^-w
\\
&+\Delta^-w^T\left(X^TX+\gamma^2I\right)\Delta^-w
\end{split}
\end{align}
To further simplify the appearance of this expression the
following Lemma is required.

\begin{lem}
Using the same definition as in Lemma \eqref{e:LEM1} let
$V^{-1}=\gamma^2Q^{-1}=\frac{2\gamma}{sin(\theta)}\left(I+U\right)^{-1}
=\gamma(I-Z)/\left( \sin(\theta)\right)$ with $U$ the unitary
matrix and using equation (\ref{e:QE6}) it follows that
\begin{subequations}
\begin{align}
\label{e:QE9a}
&V^{-1}=\left(1+\gamma^2\right)\left(I+U\right)^{-1},
\\
\label{e:QE9b}
&V^{-T}+V^{-1}=\frac{2\gamma}{sin(\theta)}I,
\end{align}
and
\begin{align}
\label{e:QE8a}
&\frac{\gamma^2I-X}{1+\gamma^2}=Z,
\\
\label{e:QE8b}
&\frac{X^TX+\gamma^2I}{1+\gamma^2}=\left(1+\gamma^2\right)Z^TZ
+\gamma^2I.
\end{align}
\end{subequations}
\end{lem}
\begin{proof}
Using $V^{-1}=\gamma^2Q^{-1}$, equations \eqref{e:QE9a} and
\eqref{e:QE9b} follow immediately from \eqref{e:QE10a} and
\eqref{e:QE10b}. Now $X=2V^{-1}-I=2\gamma^2Q^{-1}-I$, hence
\begin{align*}
\left(X-\gamma^2I\right)&= 2\left(1+\gamma^2\right)(I+U)^{-1}
-\left(1+\gamma^2\right)I
\\
&=\left(1+\gamma^2\right)\left(2(I+U)^{-1}-I\right)
\\
&= -\left(1+\gamma^2\right)Z,
\end{align*}
so that $X=\gamma^2I-\left(1+\gamma^2\right)Z$ which proves
\eqref{e:QE8a}. Then finally
\begin{align*}
&X^TX+\gamma^2I
\\
&=\left(\gamma^2I-\left(1+\gamma^2\right)Z\right)^T
\left(\gamma^2I-\left(1+\gamma^2\right)Z\right)+\gamma^2I
\\
&=\left(1+\gamma^2\right)^2Z^TZ+\gamma^2\left(1+\gamma^2\right)I,
\end{align*}
since $Z^T+Z=0$.  This concludes the proof to the Lemma.
\end{proof}

\bigskip
Applying expressions \eqref{e:QE9a}-\eqref{e:QE8b} to
\eqref{e:ENG5} yields
\begin{align*}
\frac{8\h_k}{M_T} =&\Delta^+w^T\Delta^+w -2
\Delta^+w^TZ\Delta^-w
\\
&+ \gamma^2\Delta^-w^T\Delta^-w +
\left(1+\gamma^2\right)\Delta^-w^TZ^TZ\Delta^-w,
\end{align*}
which is equivalent to equation \eqref{e:ENG4} and this
concludes the proof of the Theorem.
\end{proof}

\section{}
\label{l:APP5}
\noindent {\bf Proof of Theorem
\eqref{l:HAMILTPRP1}.} This argument is a variation on the
proofs presented by E. Nelson~\cite{ENELSON1} and E.
Carlen~\cite{ECARL1}.  To prove Theorem \eqref{l:SCHR1} express
the probability of the particle position as
$\rho=\rho(x,t),x\in\Real^n$ for an appropriate (smoothly)
differentiable function $R=R(x,t),x\in\Real^n$ such that
$\rho=e^{\frac{2\gamma\delta R}{\sigma^2}}=e^{\frac{2\delta
R}{\eta}}$. Then introduce the sufficiently smooth functions
$A=A(x,t),S=S(x,t),x\in\Real^n$ and constants $\delta, \xi$ to
express the backward and forward drifts
$b^+=b^+(x,t),b^-=b^-(x,t),x\in\Real^n,t>0$ as follows
\begin{subequations}
\begin{align}
\label{l:APP3a}
&2\gamma\delta\nabla
R=\sigma^2\frac{\nabla\rho}{\rho}
=b^+-b^-,
\\
\label{l:APP3b} &2\xi\left(\nabla S-A\right)=b^++b^-,
\end{align}
\end{subequations}
or equivalently
\begin{gather}
\label{l:APP1}
\begin{split}
b^+=\xi\left(\nabla S-A\right)+\gamma\delta \nabla R,
\\
b^-=\xi\left(\nabla S-A\right)-\gamma\delta \nabla R.
\end{split}
\end{gather}
Equation \eqref{l:APP3a} is a consequence of equation
\eqref{l:DRFT2c} and the fact that $\rho=e^{\frac{2\gamma\delta
R}{\sigma^2}}$ while equation \eqref{l:APP3b} supplies a
definition for the functions $A=A(x,t),S=S(x,t),x\in\Real^n$.

Using this definition equation \eqref{l:ENERG5} becomes
equivalent to
\begin{align}
\label{l:HAMILT7}
\begin{split}
&\frac{E\left[\h_k+\Phi_p\right]}{M_T}
\\
&= \frac{1}{2} E\left[
\begin{matrix}
\xi^2\,\abs{\nabla S-A}^2-2\xi\delta\gamma\left(\nabla S
-A\right)^T\overline{Z}\nabla R
\\
+\delta^2{\nabla
R^T}\Gamma^z{\nabla R}
\end{matrix}
\right]+
\\
& +\frac{n\sigma^2}{2\overline{\tau}}
+\frac{\sigma^2}{2\overline{\tau}\gamma^2}
E\left[Tr\left(\Gamma^z\right)\right]+\frac{1}{M_T}E\left[\Phi_p\right].
\end{split}
\end{align}
where $\Gamma^z=I+(1+\gamma^2)E\left[ZZ^T\right]$ and where $Z$
is the collision scattering matrix such that
$E\left[Z\right]=\overline{Z}$. The time derivative of this
functional depends on the continuity equation \eqref{l:DRFT3c}
which demands that
\begin{align*}
\rho_t=-\nabla .\left(\frac{b^++b^-}{2}\rho
\right)=-\xi\nabla.\left(\left(\nabla S-A\right) \rho\right),
\end{align*}
where $\eta=\sigma^2/\gamma$ and $\rho=e^{\frac{2\gamma\delta
R}{\sigma^2}}=e^{\frac{2\delta R}{\eta}}$.

From this follows
\begin{align}
\label{l:HAMILT6}
\begin{split}
&R_t=-\frac{\eta\xi}{2\delta}\left(S_{x_jx_j} -A_{jx_j}\right)- \xi
\left(S_{x_j}-A_j\right)R_{x_j},
\\
&R_{tx_k}=\left(-\frac{\eta\xi}{2\delta}\left(S_{x_jx_j}
-A_{jx_j}\right)- \xi \left(S_{x_j}-A_j\right)R_{x_j}\right)_{x_k},
\end{split}
\end{align}
using Einstein's convention for the summation of indices.  This
abbreviates $S_{x_jx_j}:=\Delta_x S=\left(\frac{\partial^2
S}{\partial x^2_1}+...+\frac{\partial^2 S}{\partial
x^2_n}\right)$ and $A_{jx_j}:=\sum_j
A_{jx_j}=A_{1x_1}+...A_{nx_n}$.

Taking the time derivative of the kinetic part of the
Hamiltonian \eqref{l:HAMILT5} using \eqref{l:HAMILT6} and
ignoring the trace term results in
\begin{align*}
&\frac{d}{dt}\left(\frac{1}{2} E\left[\frac{\h_T}{M_T}\right]\right)
\\
=&\frac{1}{2}\int\rho_t
\begin{pmatrix}
\xi^2\left(S_{x_j}-A_j\right)\left(S_{x_j}-A_j\right)
-2\xi\delta\gamma\left(S_{x_j}-A_j\right)Z_{jk} R_{x_k}
\\
+\delta^2R_{x_j}\Gamma^z_{jk}R_{x_k}
\end{pmatrix}
dx
\\
&-\xi\delta\gamma\int\rho
\begin{pmatrix}
\left(S_{x_jt}-\dot{A_j}\right)\overline{Z}_{jk}
R_{x_k}+\left(S_{x_j}-A_j\right)\dot{\overline{Z}}_{jk}R_{x_k}
\\
+\left(S_{x_j}-A_j\right)\overline{Z}_{jk}R_{tx_k}
\end{pmatrix}
dx
\\
&+\int\rho
\begin{pmatrix}
\xi^2\left(S_{x_j}-A_j\right)\left(S_{x_jt}-\dot{A_j}\right)+
\\
\delta^2R_{x_j}\Gamma^z_{jk}R_{tx_k}
+\frac{1}{2}\delta^2R_{x_j}\dot{\Gamma}^z_{jk}R_{x_k}
\end{pmatrix}
dx,
\end{align*}
so with rearranging this reduces to
\begin{align*}
&\frac{d}{dt}\left(\frac{1}{2} E\left[\frac{\h_T}{M_T}\right]\right)
\\
=&\frac{1}{2}\int\rho_t
\begin{pmatrix}
\xi^2\left(S_{x_j}-A_j\right)\left(S_{x_j}-A_j\right)
-2\xi\delta\gamma\left(S_{x_j}-A_j\right)\overline{Z}_{jk}
R_{x_k}
\\
+\delta^2R_{x_j}\Gamma^z_{jk}R_{x_k}
\end{pmatrix}
dx && \text{(a)}
\\
&-\xi\delta\gamma\int\rho
\begin{pmatrix}
\left(S_{x_jt}-\dot{A_j}\right)\overline{Z}_{jk}
R_{x_k}+\left(S_{x_j}-A_j\right)\dot{\overline{Z}}_{jk} R_{x_k}
\end{pmatrix}
dx && \text{(b)}
\\
&+\int\rho
\begin{pmatrix}
\xi^2\left(S_{x_j}-A_j\right)\left(S_{x_jt}-\dot{A_j}\right)
\\
+\left(\delta^2R_{x_j}\Gamma^z_{jk}-\xi\delta\gamma
\left(S_{x_j}-A_j\right)\overline{Z}_{jk}\right)R_{tx_k}
\\
+\frac{1}{2}\delta^2R_{x_j}\dot{\Gamma}^z_{jk}R_{x_k}
\end{pmatrix}
dx. && \text{(c)}
\end{align*}
The first term becomes
\begin{align*}
\text{(a)}
=&\frac{1}{2}\int\frac{2\delta}{\eta}\rho R_t
\begin{pmatrix}
\xi^2\left(S_{x_j}-A_j\right)\left(S_{x_j}-A_j\right)
-2\xi\delta\gamma\left(S_{x_j}-A_j\right)\overline{Z}_{jk}
R_{x_k}
\\
+\delta^2R_{x_j}\Gamma^z_{jk}R_{x_k}
\end{pmatrix}
dx
\\
=&\frac{\delta}{\eta}\int\rho
\begin{pmatrix}
\left( -\frac{\eta\xi}{2\delta}\left(S_{x_px_p}
-A_{px_p}\right)- \xi \left(S_{x_p}-A_p\right)R_{x_p}\right)
\\
\begin{pmatrix}
\xi^2\left(S_{x_j}-A_j\right)\left(S_{x_j}-A_j\right)
\\
-2\xi\delta\gamma\left(S_{x_j}-A_j\right)\overline{Z}_{jk}
R_{x_k}
\\
+\delta^2R_{x_j}\Gamma^z_{jk}R_{x_k}
\end{pmatrix}
\end{pmatrix}
dx
\\
=&-\frac{\xi}{2}\int \rho
\begin{pmatrix}
\left(S_{x_px_p} -A_{px_p}\right)
\begin{pmatrix}\xi^2\left(S_{x_j}-A_j\right)\left(S_{x_j}-A_j\right)
\\
-2\xi\delta\gamma\left(S_{x_j}-A_j\right)\overline{Z}_{jk}
R_{x_k}
\\
+\delta^2R_{x_j}\Gamma^z_{jk}R_{x_k}\end{pmatrix}
\end{pmatrix}
dx
\\
&-\frac{\xi\delta}{\eta}\int\rho
\begin{pmatrix}
\left(S_{x_p}-A_p\right)R_{x_p}
\begin{pmatrix}
(\xi^2\left(S_{x_j}-A_j\right)\left(S_{x_j}-A_j\right)
\\
-2\xi\delta\gamma\left(S_{x_j}-A_j\right)\overline{Z}_{jk}
R_{x_k}
\\
+\delta^2R_{x_j}\Gamma^z_{jk}R_{x_k}
\end{pmatrix}
\end{pmatrix}
dx,
\end{align*}
so with one partial integral this reduces to
\begin{align*}
\text{(a)}=&\frac{\xi}{2}\int \rho
\begin{pmatrix}
\left(S_{x_p} -A_{p}\right)
\begin{pmatrix}\xi^2\left(S_{x_j}-A_j\right)\left(S_{x_j}-A_j\right)
\\
-2\xi\delta\gamma\left(S_{x_j}-A_j\right)\overline{Z}_{jk}
R_{x_k}
\\
+\delta^2R_{x_j}\Gamma^z_{jk}R_{x_k}\end{pmatrix}_{x_p}
\end{pmatrix}
dx
\\
&+\frac{\xi}{2}\int \rho_{x_p}
\begin{pmatrix}
\left(S_{x_p} -A_{p}\right)
\begin{pmatrix}\xi^2\left(S_{x_j}-A_j\right)\left(S_{x_j}-A_j\right)
\\
-2\xi\delta\gamma\left(S_{x_j}-A_j\right)\overline{Z}_{jk}
R_{x_k}
\\
+\delta^2R_{x_j}\Gamma^z_{jk}R_{x_k}\end{pmatrix}
\end{pmatrix}
dx
\\
&-\frac{\xi\delta}{\eta}\int\rho
\begin{pmatrix}
\left(S_{x_p}-A_p\right)R_{x_p}
\begin{pmatrix}
\xi^2\left(S_{x_j}-A_j\right)\left(S_{x_j}-A_j\right)
\\
-2\xi\delta\gamma\left(S_{x_j}-A_j\right)\overline{Z}_{jk}
R_{x_k}
\\
+\delta^2R_{x_j}\Gamma^z_{jk}R_{x_k}\end{pmatrix}
\end{pmatrix}
dx,
\end{align*}
or
\begin{align*}
\text{(a)}=&\frac{\xi}{2}\int \rho
\begin{pmatrix}
\left(S_{x_p} -A_{p}\right)
\begin{pmatrix}\xi^2\left(S_{x_j}-A_j\right)\left(S_{x_j}-A_j\right)
\\
-2\xi\delta\gamma\left(S_{x_j}-A_j\right)\overline{Z}_{jk}
R_{x_k}
\\
+\delta^2R_{x_j}\Gamma^z_{jk}R_{x_k}\end{pmatrix}_{x_p}
\end{pmatrix}
dx
\\
&+\frac{\xi}{2}\int \frac{2\delta}{\eta}\rho R_{x_p}
\begin{pmatrix}
\left(S_{x_p} -A_{p}\right)
\begin{pmatrix}\xi^2\left(S_{x_j}-A_j\right)\left(S_{x_j}-A_j\right)
\\
-2\xi\delta\gamma\left(S_{x_j}-A_j\right)\overline{Z}_{jk}
R_{x_k}
\\
+\delta^2R_{x_j}\Gamma^z_{jk}R_{x_k}\end{pmatrix}
\end{pmatrix}
dx
\\
&-\frac{\xi\delta}{\eta}\int\rho
\begin{pmatrix}
\left(S_{x_p}-A_p\right)R_{x_p}
\begin{pmatrix}
\xi^2\left(S_{x_j}-A_j\right)\left(S_{x_j}-A_j\right)
\\
-2\xi\delta\gamma\left(S_{x_j}-A_j\right)\overline{Z}_{jk}
R_{x_k}
\\
+\delta^2R_{x_j}\Gamma^z_{jk}R_{x_k}\end{pmatrix}
\end{pmatrix}
dx,
\end{align*}
so that finally
\begin{align*}
\text{(a)}=&\frac{\xi}{2}\int \rho
\begin{pmatrix}
\left(S_{x_p} -A_{p}\right)
\begin{pmatrix}\xi^2\left(S_{x_j}-A_j\right)\left(S_{x_j}-A_j\right)
\\
-2\xi\delta\gamma\left(S_{x_j}-A_j\right)\overline{Z}_{jk}
R_{x_k}
\\
+\delta^2R_{x_j}\Gamma^z_{jk}R_{x_k}\end{pmatrix}_{x_p}
\end{pmatrix}
dx.
\end{align*}

Now the third term becomes
\begin{align*}
\text{(c)}=
&\int\rho
\begin{pmatrix}
\xi^2\left(S_{x_j}-A_j\right)\left(S_{x_jt}-\dot{A_j}\right)
+\frac{1}{2}\delta^2R_{x_j}\dot{\Gamma}^z_{jk}R_{x_k}
\\
+\begin{pmatrix}
\delta^2R_{x_j}\Gamma^z_{jk}
\\
-\xi\delta\gamma \left(S_{x_j}-A_j\right)\overline{Z}_{jk}
\end{pmatrix}
\begin{pmatrix}
-\frac{\eta\xi}{2\delta}\left(S_{x_px_p} -A_{px_p}\right)
\\
-\xi \left(S_{x_p}-A_p\right)R_{x_p}
\end{pmatrix}_{x_k}
\end{pmatrix}
dx
\\
=&\int\rho
\begin{pmatrix}
\xi^2\left(S_{x_j}-A_j\right)\left(S_{x_jt}-\dot{A_j}\right)
+\frac{1}{2}\delta^2R_{x_j}\dot{\Gamma}^z_{jk}R_{x_k}
\\
-\begin{pmatrix}
\delta^2R_{x_j}\Gamma^z_{jk}
\\
-\xi\delta\gamma \left(S_{x_j}-A_j\right)\overline{Z}_{jk}
\end{pmatrix}_{x_k}
\begin{pmatrix}
-\frac{\eta\xi}{2\delta}\left(S_{x_px_p} -A_{px_p}\right)
\\
-\xi \left(S_{x_p}-A_p\right)R_{x_p}
\end{pmatrix}
\\
-\frac{2\delta}{\eta}R_{x_k}
\begin{pmatrix}
\delta^2R_{x_j}\Gamma^z_{jk}
\\
-\xi\delta\gamma \left(S_{x_j}-A_j\right)\overline{Z}_{jk}
\end{pmatrix}
\begin{pmatrix}
-\frac{\eta\xi}{2\delta}\left(S_{x_px_p} -A_{px_p}\right)
\\
-\xi \left(S_{x_p}-A_p\right)R_{x_p}
\end{pmatrix}
\end{pmatrix}
dx
\\
=&\int\rho
\begin{pmatrix}
\xi^2\left(S_{x_j}-A_j\right)\left(S_{x_jt}-\dot{A_j}\right)
+\frac{1}{2}\delta^2R_{x_j}\dot{\Gamma}^z_{jk}R_{x_k}
\\
+\frac{\eta\xi}{2\delta}
\left(\delta^2R_{x_j}\Gamma^z_{jk}-\xi\delta\gamma
\left(S_{x_j}-A_j\right)\overline{Z}_{jk}\right)_{x_k}
\left(S_{x_px_p}-A_{px_p}\right)
\\
+\xi\left(\delta^2R_{x_j}\Gamma^z_{jk}-\xi\delta\gamma
\left(S_{x_j}-A_j\right)\overline{Z}_{jk}\right)_{x_k}
\left(S_{x_p}-A_p\right)R_{x_p}
\\
+\xi R_{x_k} \left(\delta^2R_{x_j}\Gamma^z_{jk}-\xi\delta\gamma
\left(S_{x_j}-A_j\right)\overline{Z}_{jk}\right)
\left(S_{x_px_p}-A_{px_p}\right)
\\
+\frac{2\xi\delta}{\eta}R_{x_k}
\left(\delta^2R_{x_j}\Gamma^z_{jk}-\xi\delta\gamma
\left(S_{x_j}-A_j\right)\overline{Z}_{jk}\right)
\left(S_{x_p}-A_p\right)R_{x_p}
\end{pmatrix}
dx.
\end{align*}
This can be written as
\begin{align*}
\text{(c)}=
&\int\rho
\begin{pmatrix}
\xi^2\left(S_{x_j}-A_j\right)\left(S_{x_jt}-\dot{A_j}\right)
+\frac{1}{2}\delta^2R_{x_j}\dot{\Gamma}^z_{jk}R_{x_k}
\\
+\xi\left(\delta^2R_{x_j}\Gamma^z_{jk}-\xi\delta\gamma
\left(S_{x_j}-A_j\right)\overline{Z}_{jk}\right)_{x_k}
\left(S_{x_p}-A_p\right)R_{x_p}
\\
+\frac{2\xi\delta}{\eta}R_{x_k}
\left(\delta^2R_{x_j}\Gamma^z_{jk}-\xi\delta\gamma
\left(S_{x_j}-A_j\right)\overline{Z}_{jk}\right)
\left(S_{x_p}-A_p\right)R_{x_p}
\end{pmatrix}
dx.
\\
+&\int\rho
\begin{pmatrix}
\frac{\eta\xi}{2\delta}
\left(\delta^2R_{x_j}\Gamma^z_{jk}-\xi\delta\gamma
\left(S_{x_j}-A_j\right)\overline{Z}_{jk}\right)_{x_k}
\left(S_{x_px_p}-A_{px_p}\right)
\\
+\xi R_{x_k} \left(\delta^2R_{x_j}\Gamma^z_{jk}-\xi\delta\gamma
\left(S_{x_j}-A_j\right)\overline{Z}_{jk}\right)
\left(S_{x_px_p}-A_{px_p}\right)
\end{pmatrix}
dx,
\end{align*}
and with one partial integral again this becomes
\begin{align*}
\text{(c)}=
&\int\rho
\begin{pmatrix}
\xi^2\left(S_{x_j}-A_j\right)\left(S_{x_jt}-\dot{A_j}\right)
+\frac{1}{2}\delta^2R_{x_j}\dot{\Gamma}^z_{jk}R_{x_k}
\\
+\xi\left(\delta^2R_{x_j}\Gamma^z_{jk}-\xi\delta\gamma
\left(S_{x_j}-A_j\right)\overline{Z}_{jk}\right)_{x_k}
\left(S_{x_p}-A_p\right)R_{x_p}
\\
+\frac{2\xi\delta}{\eta}R_{x_k}
\left(\delta^2R_{x_j}\Gamma^z_{jk}-\xi\delta\gamma
\left(S_{x_j}-A_j\right)\overline{Z}_{jk}\right)
\left(S_{x_p}-A_p\right)R_{x_p}
\end{pmatrix}.
dx
\\
+&\int\rho
\begin{pmatrix}
-\xi R_{x_k} \left(\delta^2R_{x_j}\Gamma^z_{jk}-\xi\delta\gamma
\left(S_{x_j}-A_j\right)\overline{Z}_{jk}\right)_{x_p}
\left(S_{x_p}-A_{p}\right)
\\
-\xi R_{x_kx_p}
\left(\delta^2R_{x_j}\Gamma^z_{jk}-\xi\delta\gamma
\left(S_{x_j}-A_j\right)\overline{Z}_{jk}\right)
\left(S_{x_p}-A_{p}\right)
\\
-\frac{2\xi\delta}{\eta} R_{x_k}R_{x_p}
\left(\delta^2R_{x_j}\Gamma^z_{jk}-\xi\delta\gamma
\left(S_{x_j}-A_j\right)\overline{Z}_{jk}\right)
\left(S_{x_p}-A_{p}\right)
\end{pmatrix}
dx
\\
+&\int\rho
\begin{pmatrix}
-\frac{\eta\xi}{2\delta}
\left(\delta^2R_{x_j}\Gamma^z_{jk}-\xi\delta\gamma
\left(S_{x_j}-A_j\right)\overline{Z}_{jk}\right)_{x_kx_p}
\left(S_{x_p}-A_{p}\right)
\\
-\xi R_{x_p} \left(\delta^2R_{x_j}\Gamma^z_{jk}-\xi\delta\gamma
\left(S_{x_j}-A_j\right)\overline{Z}_{jk}\right)_{x_k}
\left(S_{x_p}-A_{p}\right)
\end{pmatrix}
dx.
\end{align*}
So then combining the $a)$, $b)$ and $c)$ terms results in
\begin{align*}
&\text{(a)}+\text{(b)}+\text{(c)}
\\
&=\int\rho
\begin{pmatrix}
\xi^2\left(S_{x_j}-A_j\right)\left(S_{x_jt}-\dot{A_j}\right)
+\frac{1}{2}\delta^2R_{x_j}\dot{\Gamma}^z_{jk}R_{x_k}
\\
+\xi\left(\delta^2R_{x_j}\Gamma^z_{jk}-\xi\delta\gamma
\left(S_{x_j}-A_j\right)\overline{Z}_{jk}\right)_{x_k}
\left(S_{x_p}-A_p\right)R_{x_p}
\\
+\frac{2\xi\delta}{\eta}R_{x_k}
\left(\delta^2R_{x_j}\Gamma^z_{jk}-\xi\delta\gamma
\left(S_{x_j}-A_j\right)\overline{Z}_{jk}\right)
\left(S_{x_p}-A_p\right)R_{x_p}
\end{pmatrix}
dx
\\
&+\int\rho
\begin{pmatrix}
-\xi R_{x_k} \left(\delta^2R_{x_j}\Gamma^z_{jk}-\xi\delta\gamma
\left(S_{x_j}-A_j\right)\overline{Z}_{jk}\right)_{x_p}
\left(S_{x_p}-A_{p}\right)
\\
-\xi R_{x_kx_p}
\left(\delta^2R_{x_j}\Gamma^z_{jk}-\xi\delta\gamma
\left(S_{x_j}-A_j\right)\overline{Z}_{jk}\right)
\left(S_{x_p}-A_{p}\right)
\\
-\frac{2\xi\delta}{\eta} R_{x_k}R_{x_p}
\left(\delta^2R_{x_j}\Gamma^z_{jk}-\xi\delta\gamma
\left(S_{x_j}-A_j\right)\overline{Z}_{jk}\right)
\left(S_{x_p}-A_{p}\right)
\end{pmatrix}
dx
\\
&+\int\rho
\begin{pmatrix}
-\frac{\eta\xi}{2\delta}
\left(\delta^2R_{x_j}\Gamma^z_{jk}-\xi\delta\gamma
\left(S_{x_j}-A_j\right)\overline{Z}_{jk}\right)_{x_kx_p}
\left(S_{x_p}-A_{p}\right)
\\
-\xi R_{x_p} \left(\delta^2R_{x_j}\Gamma^z_{jk}-\xi\delta\gamma
\left(S_{x_j}-A_j\right)\overline{Z}_{jk}\right)_{x_k}
\left(S_{x_p}-A_{p}\right)
\end{pmatrix}
dx
\\
&-\xi\delta\gamma\int\rho
\begin{pmatrix}
\left(S_{x_jt}-\dot{A_j}\right)\overline{Z}_{jk}
R_{x_k}+\left(S_{x_j}-A_j\right)\dot{\overline{Z}}_{jk} R_{x_k}
\end{pmatrix}
dx
\\
&+\frac{1}{2}\int_{-\infty}^{\infty}\xi\rho
\begin{pmatrix}
\left(S_{x_p} -A_{p}\right)
\begin{pmatrix}\xi^2\left(S_{x_j}-A_j\right)\left(S_{x_j}-A_j\right)
\\
-2\xi\delta\gamma\left(S_{x_j}-A_j\right)\overline{Z}_{jk}
R_{x_k}
\\
+\delta^2R_{x_j}\Gamma^z_{jk}R_{x_k}\end{pmatrix}_{x_p}
\end{pmatrix}
dx,
\end{align*}
which reduces to
\begin{align*}
&\text{(a)}+\text{(b)}+\text{(c)}
\\
&\int\rho
\begin{pmatrix}
\xi^2\left(S_{x_j}-A_j\right)\left(S_{x_jt}-\dot{A_j}\right)
+\frac{1}{2}\delta^2R_{x_j}\dot{\Gamma}^z_{jk}R_{x_k}
\end{pmatrix}
dx
\\
&-\xi\int\rho
\left(S_{x_p}-A_{p}\right)\left(R_{x_k}
\left(\delta^2R_{x_j}\Gamma^z_{jk}-\xi\delta\gamma
\left(S_{x_j}-A_j\right)\overline{Z}_{jk}\right)\right)_{x_p} dx
\\
&-\frac{\eta\xi}{2\delta}\int\rho \left(S_{x_p}-A_{p}\right)
\left(\delta^2R_{x_j}\Gamma^z_{jk}-\xi\delta\gamma
\left(S_{x_j}-A_j\right)\overline{Z}_{jk}\right)_{x_kx_p} dx
\\
&-\xi\delta\gamma\int\rho
\begin{pmatrix}
\left(S_{x_jt}-\dot{A_j}\right)\overline{Z}_{jk}
R_{x_k}+\left(S_{x_j}-A_j\right)\dot{\overline{Z}}_{jk} R_{x_k}
\end{pmatrix}
dx
\\
&+\frac{\xi}{2}\int_{-\infty}^{\infty}\rho
\left(S_{x_p}
-A_{p}\right)
\begin{pmatrix}
\xi^2\left(S_{x_j}-A_j\right)\left(S_{x_j}-A_j\right)
\\
-2\xi\delta\gamma\left(S_{x_j}-A_j\right)\overline{Z}_{jk}
R_{x_k}
\\
+\delta^2R_{x_j}\Gamma^z_{jk}R_{x_k}
\end{pmatrix}_{x_p}
dx.
\end{align*}
Gathering terms finally shows that
\begin{align}
\label{l:ENERG6}
\begin{split}
\frac{d}{dt}&\frac{1}{2} E
\begin{pmatrix}
\xi^2\,|\nabla S-A\,|^2-2\xi\delta\gamma(\nabla
S-A)^T\overline{Z}\nabla R
\\
+\delta^2\nabla R^T\Gamma^z\nabla R\end{pmatrix}
\\
=&\xi\int\rho
\left(S_{x_p}-A_p\right)\
\begin{pmatrix}
\xi S_{t}
+\frac{\xi^2}{2}\left(S_{x_j}-A_j\right)\left(S_{x_j}-A_j\right)
\\
-\frac{\delta^2}{2}R_{x_j}\Gamma^z_{jk}R_{x_k}
-\frac{\delta\eta}{2}\left(R_{x_j}\Gamma^z_{jk}\right)_{x_k}
\end{pmatrix}_{x_p}dx
\\
&-\xi^2\int\rho\left(S_{x_p}-A_p\right)\dot{A_p}+\frac{1}{2}\int\rho
\begin{pmatrix}
\delta^2R_{x_j}\dot{\Gamma}^z_{jk}R_{x_k}
\end{pmatrix}
dx
\\
&+\frac{\eta\xi}{2\delta}\int\rho \left(S_{x_p}-A_{p}\right)
\begin{pmatrix}
\left(\xi\delta\gamma
\left(S_{x_j}-A_j\right)\overline{Z}_{jk}\right)_{x_kx_p}
\end{pmatrix}
dx
\\
&-\xi\delta\gamma\int\rho
\begin{pmatrix}
\left(S_{x_jt}-\dot{A_j}\right)\overline{Z}_{jk}
R_{x_k}+\left(S_{x_j}-A_j\right)\dot{\overline{Z}}_{jk} R_{x_k}
\end{pmatrix}
dx.
\end{split}
\end{align}

Now the static potential term time derivative
$\frac{1}{M_T}\frac{d}{dt}E\left[\Phi_p\right]$ contains the
differential $\frac{\xi}{M_T}E\left[\left(\nabla
S-A\right)^T\dot{A}\right]$.  The assumption on the potential
turns it into a full differential and the $A$ term disappears if
$\xi=\frac{1}{M_T}$.  Once the term with the Trace of the
$\Gamma^z$ matrix is reintroduced the Theorem is proved.

\bigskip
Therefore a sufficient condition for the Hamiltonian in
proposition \eqref{l:HAMILTPRP1} to become time-independent the
integrand in \eqref{l:HAMILT5} is to equal zero.  The following
proposition shows that this is equivalent to demanding that the
probability for the main particle position $\rho(x,t)$ is
derived from a wave function satisfying Schr\"{o}dinger's
equation.

\begin{prop}
\label{l:prp22} Assume that $Z\equiv 0$ so that
$E\left[ZZ^T\right]\equiv 0$, $E[Z]\equiv 0$, $\Gamma^z=I$ and
let the potential $\Phi_p$ satisfy \eqref{l:HAMILT15}. Define
the wave function
\begin{gather*}
\psi=\psi(x,t)=e^{\frac{\delta R(x,t)+i\xi S(x,t)}{\eta}},
\end{gather*}
with $\rho(x,t)=\,|\psi(x,t)\,|^2$.  Then equation
\begin{align}
\label{l:HAMILT8}
\xi^2(S_{t}-A_t)-\frac{\delta^2}{2}\abs{\nabla
R}^2 +\frac{\xi^2}{2} \abs{\nabla S-A}^2
-\frac{\eta\delta}{2}\Delta_x R +\frac{1}{\xi M_T}\phi=0,
\end{align}
makes the energy terms invariant.  Here
$\Delta_x=\left(\frac{\partial^2}{\partial
x^2_1},...,\frac{\partial^2}{\partial x^2_n}\right)$.  Now this
equation is equivalent to
\begin{align*}
i\chi\psi_t=-\frac{1}{2M_T}\left(\chi\nabla-iA\right)^2
\psi+\phi(x,t)\psi.
\end{align*}
with $\chi=M_T\eta$ and $\delta=\xi=1/M_T$ so that
$\psi=\psi(x,t)=e^{\frac{R(x,t)+iS(x,t)}{\chi}}$.
\end{prop}
\begin{proof}
The proof involves a straightforward verification of equation
\eqref{l:SCROD2} by brute force.  Taking derivatives
\begin{align*}
&\frac{\eta}{i}\psi_t=\psi\left( -i\delta R_t+\xi S_t\right),
\\
&\frac{\eta}{2}\psi_{x_j}=\frac{1}{2}\psi\left( \delta
R_{x_j}+i\xi S_{x_j}\right),
\\
&\frac{\eta^2}{2}\psi_{x_jx_j}=\frac{1}{2}\psi\left( \delta
R_{x_j}+i\xi S_{x_j}\right)^2 + \frac{\eta}{2}\psi\left( \delta
R_{x_jx_j}+i\xi S_{x_jx_j}\right),
\\
&i\xi\eta A_j\psi_{x_j}=\psi\left(-i\delta\xi A_jR_{x_j} + A_j
\xi^2 S_{x_j} \right),
\end{align*}
which combined becomes
\begin{align*}
\frac{\eta}{i}\psi_t&-\frac{\eta^2}{2}\psi_{x_jx_j}+i\xi\eta
A_j\psi_{x_j}
+\frac{1}{2}\abs{A}^2\xi^2\psi+\frac{\xi\eta}{2}iA_{jx_j}\psi
\\
&=-i\psi\left( \delta R_t-\delta\xi A_jR_{x_j}+\delta\xi
R_{x_j}S_{x_j}-\frac{\eta\xi}{2}A_{jx_j}+\frac{\eta\xi}{2}S_{xx}\right)
\\
&+\left( \xi S_t-\frac{\delta^2}{2}\abs{\nabla R}^2 +
\frac{\xi^2}{2}\abs{\nabla S-A}^2
-\frac{\eta\delta}{2}R_{x_jx_j}\right)
\\
&=-\frac{\phi}{\xi M_T}\psi.
\end{align*}
Now
\begin{align*}
-\frac{1}{2}&\left(\eta\frac{\partial}{\partial x_j}-iA_j\xi
\right)^2 \psi=-\frac{1}{2}\left(\eta\frac{\partial}{\partial
x_j}-iA_j\xi \right)\left( \eta\psi_{x_j} -iA_j\xi\psi\right)
\\
=&-\frac{1}{2}\left(\eta^2\psi_{x_j}-\eta iA_j\xi \psi
\right)_{x_j} +\frac{1}{2}iA_j\xi\left( \eta\psi_{x_j}
-iA_j\xi\psi\right)
\\
=&-\frac{1}{2}\left(\eta^2\psi_{x_jx_j}-\eta iA_j\xi \psi_{x_j}
-\eta i\xi A_{jx_j}\psi\right) +\frac{1}{2}iA_j\xi\eta\psi_{x_j}
+\frac{1}{2}\abs{A}^2\xi^2\psi
\\
=&-\frac{1}{2}\eta^2\psi_{x_jx_j} +\frac{\eta}{2} i\xi
A_{jx_j}\psi +iA_j\xi\eta\psi_{x_j}
+\frac{1}{2}\abs{A}^2\xi^2\psi,
\end{align*}
so indeed
\begin{align*}
i\eta\psi_t=-\frac{1}{2}\left(\eta\nabla-iA\xi \right)^2
\psi+\frac{\phi}{\xi M_T}\psi.
\end{align*}

Finally, to adjust for the masses, let $\chi=M_T\eta$ and let
$\xi=1/M_T$, then
\begin{gather*}
\psi=\psi(x,t)=e^{\frac{R(x,t)+iS(x,t)}{\chi}},
\end{gather*}
and
\begin{align*}
i\chi\psi_t=-\frac{1}{2M_T}\left(\chi\nabla-iA\right)^2
\psi+\phi(x,t)\psi.
\end{align*}
This proves (\ref{l:HAMILT6}).  Also quite clearly
\begin{align*}
b^+(x,t)=\frac{1}{M_T}\left(\nabla S-A+\gamma R_x \right), \\
b^-(x,t)=\frac{1}{M_T}\left(\nabla S-A-\gamma R_x \right),
\end{align*}
which proves equation (\ref{l:DRFT15}).
\end{proof}

An immediate conclusion from quantum mechanics is that the mean
acceleration of the motion is guided by the potential $\phi$.
\begin{prop}
Some straightforward derivatives show that
\begin{align*}
\frac{d^2}{dt^2}E[x(t)]=-\frac{1}{M_T}\int\rho\left(\nabla\phi\right)
dx.
\end{align*}
\end{prop}
\begin{proof}
In addition,
\begin{align*}
\frac{d}{dt}E[x(t)]=-\xi\int x\nabla.\left((\nabla S-A)\rho
\right)dx=\xi\int\rho\nabla (S-A)dx,
\end{align*}
while
\begin{align*}
\frac{d^2}{dt^2}E[x(t)] &=\xi^2\int\rho \nabla
(S-A).\left(\Delta_xS-\nabla.A\right)dx+\xi\int\rho (\nabla
\dot{S}-\dot{A})dx
\\
&=\int\rho \nabla\left(\xi S_t+\frac{\xi^2}{2}\abs{(\nabla
S-A)}^2\right) dx
\\
&=\int\rho
\nabla\left(\frac{\delta^2}{2}\abs{R}^2+\frac{\eta\delta}{2}\Delta_x
R- \frac{\phi}{M_T}\right) dx
\\
&=-\frac{1}{M_T}\int\rho\left(\nabla \phi\right) dx.
\end{align*}
\end{proof}

Now the example \eqref{l:EXAMP1} can be slightly extended in the
form of the following Proposition.
\begin{prop}
\label{HAMILT22}
Consider the two-dimensional case n=2 and
assume that
\begin{align*}
Z^\pm=
\begin{pmatrix}
0 & \pm\nu
\\
\mp\nu & 0
\end{pmatrix},
\end{align*}
so that $E\left[ZZ^T\right]=\nu^2I$ and
$\Gamma^z=I+(1+\gamma^2)E\left[ZZ^T\right]=
\left(1+(1+\gamma^2)\nu^2\right)I=\sigma^2_\nu I$.  Then define
the wave function
\begin{gather*}
\psi=\psi(x,t)=e^{\frac{\delta R(x,t)+i\xi S(x,t)}{\eta}},
\end{gather*}
with $\rho(x,t)=\,|\psi(x,t)\,|^2$.  Then equation
\begin{align}
\label{l:HAMILT9}
\xi^2(S_{t}-A_t)-\frac{\delta^2\sigma^2_\nu}{2} \abs{\nabla R}^2
+\frac{\xi^2}{2} \abs{\nabla S-A}^2
-\frac{\eta\delta\sigma^2_\nu}{2}\Delta_x R +\frac{1}{\xi
M_T}\phi=0,
\end{align}
makes the energy terms invariant.  Here
$\Delta_x=\left(\frac{\partial^2}{\partial
x^2_1},...,\frac{\partial^2}{\partial x^2_n}\right)$.  Now this
equation is equivalent to
\begin{align}
\label{l:SCROD3}
i\chi\psi_t=-\frac{1}{2M_T}\left(\chi\nabla-iA\right)^2
\psi+\phi(x,t)\psi
\end{align}
with $\chi=M_T\eta/\sigma_\nu$, $\xi=1/M_T$ and
$\delta=1/\left(\sigma_\nu M_T\right)$.
\end{prop}
\begin{proof}
Let the wave function $\psi=\psi(x,t,\delta,\xi,\chi)$ satisfy
equation (\ref{l:HAMILT8}). Then it is clear that the wave
function $\psi'=\psi(x,t,\frac{\delta}{\sigma_\nu},\xi,
\frac{\chi}{\sigma_\nu})$ satisfies equation (\ref{l:HAMILT9})
above.
\end{proof}

\section{}
\label{l:APP4}
\noindent {\bf Proof of Proposition
\eqref{l:HAMILT17}.} Move all the main particle terms in
equation \eqref{l:GAMMDEF1} to the left and all heatbath
particle terms to the righthand side (reference also equation
\eqref{l:COLLMAT1}) to obtain
\begin{align}
\label{l:GAMMDEF21}
\begin{pmatrix}
I & -P \\
0 & -R
\end{pmatrix}
\begin{pmatrix}
v_2 \\
v_1
\end{pmatrix}
=
\begin{pmatrix}
0 & Q \\
-I & S
\end{pmatrix}
\begin{pmatrix}
w_2 \\
w_1
\end{pmatrix},
\end{align}
and using remark \eqref{l:REM1} this can be reduced to
\begin{align*}
\begin{pmatrix}
v_2 \\
v_1
\end{pmatrix}
&=
\begin{pmatrix}
I & -P \\
0 & -R
\end{pmatrix}^{-1}
\begin{pmatrix}
0 & Q \\
-I & S
\end{pmatrix}
\begin{pmatrix}
w_2 \\
w_1
\end{pmatrix}
\\
&=
\begin{pmatrix}
I & -PR^{-1} \\
0 & -R^{-1}
\end{pmatrix}
\begin{pmatrix}
0 & Q \\
-I & S
\end{pmatrix}
\begin{pmatrix}
w_2 \\
w_1
\end{pmatrix}
\\
&=
\begin{pmatrix}
I & -\gamma^2\left(Q^{-1}-I\right) \\
0 & -\gamma^2Q^{-1}
\end{pmatrix}
\begin{pmatrix}
0 & Q \\
-I & \left(I-\frac{Q}{\gamma^2}\right)
\end{pmatrix}
\begin{pmatrix}
w_2 \\
w_1
\end{pmatrix}
\\
&=
\begin{pmatrix}
\gamma^2\left(Q^{-1}-I\right) & -\gamma^2Q^{-1}+\left(1+\gamma^2\right)I \\
\gamma^2Q^{-1} & -\left(\gamma^2Q^{-1}-I\right)
\end{pmatrix}
\begin{pmatrix}
w_2 \\
w_1
\end{pmatrix}.
\end{align*}
Using the collision scattering matrix
$\left(I-Z\right)=2\frac{\gamma^2}{\left(1+\gamma^2\right)}Q^{-1}$
the expression above reduces to
\begin{align}
\label{l:GAMMDEF23}
\begin{split}
\begin{pmatrix}
v_2 \\
v_1
\end{pmatrix}
= &\frac{\gamma}{\sin(\theta)}
\begin{pmatrix}
\cos(\theta) & 1
\\
1 & \cos(\theta)
\end{pmatrix}
\begin{pmatrix}
w_2 \\
w_1
\end{pmatrix}
\\
&+ \frac{\gamma}{\sin(\theta)}
\begin{pmatrix}
-Z & Z
\\
-Z & Z
\end{pmatrix}
\begin{pmatrix}
w_2 \\
w_1
\end{pmatrix}
\end{split}
\end{align}
or
\begin{align}
\label{l:GAMMDEF22}
\begin{pmatrix}
v_2 \\
v_1
\end{pmatrix}
= \frac{\gamma}{\sin(\theta)}
\begin{pmatrix}
\cos(\theta) & 1
\\
1 & \cos(\theta)
\end{pmatrix}
\begin{pmatrix}
w_2 \\
w_1
\end{pmatrix}
+
\begin{pmatrix}
\W
\\
\W
\end{pmatrix},
\end{align}
where
$-m\W=\frac{1+\gamma^2}{2}mZ(w_2-w_1)=\frac{1+\gamma^2}{2}mZ\Delta^-w$
is the additional momentum transfer from the heatbath to the main
particle. Notice the rather interesting fact that this allows
\eqref{l:GAMMDEF22} to be rewritten as follows
\begin{align}
\label{l:GAMMDEF24}
\begin{pmatrix}
v_2 \\
v_1
\end{pmatrix}
= \frac{\gamma}{\sin(\theta)}
\begin{pmatrix}
\cos(\theta) & 1
\\
1 & \cos(\theta)
\end{pmatrix}
\begin{pmatrix}
w_2 +\W\\
w_1+\W
\end{pmatrix},
\end{align}
since the vector $\left(\begin{smallmatrix}
\W \\
\W
\end{smallmatrix}\right)
$ is an eigenvector of the matrix $\frac{\gamma}{\sin(\theta)}
\left(
\begin{smallmatrix}
\cos(\theta) & 1 \\
1 & \cos(\theta)
\end{smallmatrix}
\right)$.  This is in fact equivalent to equation
\eqref{l:GAMMDEF25}.

Now returning to \eqref{l:GAMMDEF23}
\begin{align*}
\begin{pmatrix}
v_2 \\
v_1
\end{pmatrix}
= \frac{\gamma}{sin(\theta)}
\begin{pmatrix}
\cos(\theta)I-Z & I+Z
\\
I-Z & \cos(\theta)I+Z
\end{pmatrix}
\begin{pmatrix}
w_2 \\
w_1
\end{pmatrix},
\end{align*}
so that
\begin{align*}
\begin{pmatrix}
w_2 \\
w_1
\end{pmatrix}
= &\frac{\sin(\theta)}{\gamma}
\begin{pmatrix}
\cos(\theta)I-Z & I+Z
\\
I-Z & \cos(\theta)I+Z
\end{pmatrix}^{-1}
\begin{pmatrix}
v_2 \\
v_1
\end{pmatrix}
\\
=& \frac{1}{\gamma\sin(\theta)}
\begin{pmatrix}
-\cos(\theta)I-Z & I+Z
\\
I-Z & -\cos(\theta)I+Z
\end{pmatrix}
\begin{pmatrix}
v_2 \\
v_1
\end{pmatrix}.
\end{align*}
Then
\begin{align*}
E&\left[
\begin{pmatrix}
\frac{\Delta^+_w}{\-\tau_2} \\
\frac{\Delta^-_w}{\-\tau_1}
\end{pmatrix}
\begin{pmatrix}
\frac{\Delta^+_w}{\-\tau_2} & \frac{\Delta^-_w}{\-\tau_1}
\end{pmatrix}
\right]
\\
=& \frac{2\sigma^2}{\overline{\tau}\gamma\sin^2(\theta)} E\left[
\begin{matrix}
\begin{pmatrix} -\cos(\theta)I-Z & I+Z
\\
I-Z & -\cos(\theta)I+Z
\end{pmatrix}.
\\
\qquad
\begin{pmatrix} -\cos(\theta)I-Z^T & I-Z^T
\\
I+Z^T & -\cos(\theta)I+Z^T
\end{pmatrix}
\end{matrix}
\right]
\\
=&\frac{2\sigma^2}{\overline{\tau}\gamma\sin^2(\theta)}
\begin{pmatrix}
\left(1+\cos^2(\theta)\right)I & -2\cos(\theta)I
\\
-2\cos(\theta)I & \left(1+\cos^2(\theta)\right)I
\end{pmatrix}
\\
&+\frac{2\sigma^2}{\overline{\tau}\gamma\sin^2(\theta)}
\begin{pmatrix}
E\left[ZZ^T\right] & \Omega
\\
\Omega^T & E\left[ZZ^T\right]
\end{pmatrix},
\end{align*}
where
\begin{align*}
\Omega=\left(1-\cos(\theta)I\right)\overline{Z}+E\left[ZZ^T\right].
\end{align*}
Recombining the $\cos(\theta)^2$ and $\sin(\theta)^2$ terms in
the first matrix (refer to Theorem \eqref{a:THM9}) this reduces
to
\begin{align*}
E&\left[
\begin{pmatrix}
\frac{\Delta^+_w}{\-\tau_2} \\
\frac{\Delta^-_w}{\-\tau_1}
\end{pmatrix}
\begin{pmatrix}
\frac{\Delta^+_w}{\-\tau_2} & \frac{\Delta^-_w}{\-\tau_1}
\end{pmatrix}
\right]
\\
=& \frac{\sigma^2}{\overline{\tau}\alpha^2}
\begin{pmatrix}
1 & -\left(1-2\alpha^2\right) \\
-\left(1-2\alpha^2\right) & 1
\end{pmatrix}
+\Gamma_Z,
\end{align*}
with
\begin{align*}
\Gamma_Z=\frac{2\sigma^2}{\overline{\tau}\gamma\sin^2(\theta)}
\begin{pmatrix}
E\left[ZZ^T\right] & \Omega
\\
\Omega^T & E\left[ZZ^T\right]
\end{pmatrix}.
\end{align*}
Clearly here $\Gamma_Z$ is positive definite and symmetric.  This
concludes the proof.

\end{document}